%% file: RR-9590.tex
\documentclass[twoside,leqno]{article}

\pdfoutput=1

\usepackage{RR}
\usepackage[a4paper]{geometry}
\usepackage{a4wide}

\usepackage[english]{babel}
\usepackage[numbers]{natbib}

\usepackage{caption}
\input{commands}

\addto\captionsenglish{%
  }

\newcommand{\oftypeJ}{~\coqe{:}~}
\newcommand{\oftypeRR}{ of type }
\newcommand{\equalsRR}{ equals }
\newcommand{\inRR}{ in~}
\newcommand{\thedefofJ}{}

\RRdate{June 2025}

\RRversion{2\thanks{%
  Shortened version, with modified presentation and greater focus on
  novelties.}}
\RRdater{April 2026}

\RRauthor{%
  Sylvie Boldo%
  \thanks[lmf]{{\LMF} \texttt{\{sylvie.boldo,houda.mouhcine\}@inria.fr}}
  \and François Clément%
  \thanks[serena]{{\SERENA} \texttt{francois.clement@inria.fr}}
  \and Vincent Martin%
  \thanks{{\LMAC} \texttt{vincent.martin@utc.fr}}
  \and Micaela Mayero%
  \thanksref{lmf}\thanks[lipn]{{\LIPN}\goodbreak
    \texttt{mayero@lipn.univ-paris13.fr}}
  \and Houda Mouhcine\thanksref{lmf}\thanksref{lipn}\thanksref{serena}
}
\authorhead{\AuthorsList}

\RRtitle{\TitreRR}
\RRetitle{\TitleRR}
\titlehead{\TitleRR}

\RRnote{\Funding}

\RRresume{\Resume}
\RRabstract{\Abstract}

\RRmotcle{\Motscles}
\RRkeyword{\Keywords}

\RRprojets{Toccata and Serena}

\RCSaclay

\begin{document}

\RRNo{9590}
\makeRR

\tableofcontents
\listoffigures

\input{body}

\section*{Acknowledgments}
\Acknowledgments

\bibliographystyle{plainnat}
\bibliography{biblio-RR}

\end{document}

%% file: commands.tex
\usepackage[utf8]{inputenc}
\usepackage[OT1]{fontenc} 
\usepackage{hyperref}
\usepackage{amssymb, amsmath, amsthm, mathtools, bm}
\usepackage{amsfonts, bbold, dsfont}
\usepackage{enumerate} 

\usepackage[nottoc]{tocbibind}

\newcommand{\Notation}[1]{\textcolor{magenta}{\fbox{#1}}}

\usepackage{listings, lstcoq}
\newcommand{\code}[1]{\texttt{#1}}

\newcommand{\iconlk}{\includegraphics[width=0.5em]{icon_lk.jpg}}

\newcommand{\cnaURL}{%
  https://lipn.univ-paris13.fr/rocqdoc-num-analysis/2.2/NumAnalysis}
\newcommand{\genlk}[2]{%
  \hspace{0.1em}\raisebox{0.5ex}{\href{\cnaURL.#1.html#2}{\iconlk}}}
\bgroup\catcode`\#=12\gdef\hash{#}\egroup
  \newcommand{\lk}[2]{\genlk{#1}{\hash{#2}}}
\newcommand{\coqelk}[2]{\mbox{\coqe{#2}\lk{#1}{#2}}}
\newcommand{\filelk}[3]{\mbox{\soft{#3}\genlk{#1.#2}{}}}

\newcommand{\LLKAlg}[2]{#2\lk{Algebra.#1}{#2}}
\newcommand{\LLKH}[1]{\LLKAlg{Hierarchy_compl}{#1}}
\newcommand{\LLKM}[2]{\LLKAlg{Monoid.#1}{#2}}
\newcommand{\LLKG}[2]{\LLKAlg{Group.#1}{#2}}
\newcommand{\LLKR}[2]{\LLKAlg{Ring.#1}{#2}}
\newcommand{\LLKMS}[2]{\LLKAlg{ModuleSpace.#1}{#2}}
\newcommand{\LLKAS}[2]{\LLKAlg{AffineSpace.#1}{#2}}

\newcommand{\flkLogic}{\filelk{Subset.Logic}{logic_compl}{logic\_compl}}
\newcommand{\FLKSUB}[2]{\filelk{Subset.Subsets}{#1}{#2}}
\newcommand{\flkFun}{\FLKSUB{Function}{Function}}
\newcommand{\flkFunS}{\FLKSUB{Function_sub}{Function\_sub}}

\newcommand{\FLKFF}[2]{\filelk{Subset.Subsets.FF}{#1}{#2}}
\newcommand{\flkOrd}{\FLKFF{ord_compl}{ord\_compl}}
\newcommand{\flkFF}{\FLKFF{Finite_family}{Finite\_family}}

\newcommand{\flkH}{\filelk{Algebra}{Hierarchy_compl}{Hierarchy\_compl}}
\newcommand{\FLKALG}[3]{\filelk{Algebra.#1}{#2}{#3}}
\newcommand{\FLKMON}[2]{\FLKALG{Monoid}{#1}{#2}}
\newcommand{\flkSum}{\FLKMON{Monoid_sum}{Monoid\_sum}}
\newcommand{\flkProd}{\FLKMON{MonoidMult}{MonoidMult}}
\newcommand{\flkComp}{\FLKMON{MonoidComp}{MonoidComp}}
\newcommand{\flkConcat}{\FLKMON{ConcatnF}{ConcatnF}}
\newcommand{\FLKRG}[2]{\FLKALG{Ring}{#1}{#2}}
\newcommand{\flkRing}{\FLKRG{Ring_compl}{Ring\_compl}}
\newcommand{\flkKron}{\FLKRG{Ring_kron}{Ring\_kron}}
\newcommand{\FLKMS}[2]{\FLKALG{ModuleSpace}{#1}{#2}}
\newcommand{\flkMS}{\FLKMS{ModuleSpace_compl}{ModuleSpace\_compl}}
\newcommand{\flkLM}{\FLKMS{ModuleSpace_lin_map}{ModuleSpace\_lin\_map}}

\newcommand{\lkNat}[1]{\lk{Subset.Numbers.nat_compl}{#1}}
\newcommand{\LKSUB}[2]{\lk{Subset.Subsets.#1}{#2}}

\newcommand{\lkSdec}[1]{\LKSUB{Subset_dec}{#1}}

\newcommand{\lkFunS}[1]{\LKSUB{Function_sub}{#1}}
\newcommand{\LKFF}[2]{\lk{Subset.Subsets.FF.#1}{#2}}
\newcommand{\lkOrd}[1]{\LKFF{ord_compl}{#1}}
\newcommand{\lkFF}[1]{\LKFF{Finite_family}{#1}}
\newcommand{\LKALG}[2]{\lk{Algebra.#1}{#2}}
\newcommand{\LKMON}[2]{\LKALG{Monoid.#1}{#2}}
\newcommand{\lkMonS}[1]{\LKMON{Monoid_sub}{#1}}
\newcommand{\lkProd}[1]{\LKMON{MonoidMult}{#1}}
\newcommand{\lkConcat}[1]{\LKMON{ConcatnF}{#1}}
\newcommand{\lkMO}[1]{\LKMON{Monomial_order}{#1}}
\newcommand{\LKAS}[2]{\LKALG{AffineSpace.#1}{#2}}
\newcommand{\lkAS}[1]{\LKAS{AffineSpace_def}{#1}}
\newcommand{\lkASB}[1]{\LKAS{AffineSpace_baryc}{#1}}
\newcommand{\lkAM}[1]{\LKAS{AffineSpace_aff_map}{#1}}
\newcommand{\lkASS}[1]{\LKAS{AffineSpace_sub}{#1}}
\newcommand{\LKFD}[2]{\LKALG{Finite_dim.Finite_dim_#1}{#2}}
\newcommand{\lkFDLMR}[1]{\LKALG{Finite_dim.Finite_dim_MS_lin_map_R}{#1}}
\newcommand{\lkFDAS}[1]{\LKFD{AS_def}{#1}}
\newcommand{\lkFDAI}[1]{\LKFD{AS_aff_indep}{#1}}
\newcommand{\lkBinom}[1]{\LKALG{binomial_compl}{#1}}
\newcommand{\LKFEM}[2]{\lk{FEM.#1}{#2}}
\newcommand{\lkGeom}[1]{\LKFEM{geometry}{#1}}
\newcommand{\lkMI}[1]{\LKFEM{multi_index}{#1}}
\newcommand{\lkPdk}[1]{\LKFEM{poly_Pdk}{#1}}
\newcommand{\lkLik}[1]{\LKFEM{poly_LagP1k}{#1}}
\newcommand{\lkLdiref}[1]{\LKFEM{poly_LagPd1_ref}{#1}}
\newcommand{\lkTgeom}[1]{\LKFEM{geom_transf_affine}{#1}}
\newcommand{\lkLdi}[1]{\LKFEM{poly_LagPd1}{#1}}
\newcommand{\lkNoderef}[1]{\LKFEM{LagP_node_ref}{#1}}
\newcommand{\lkNode}[1]{\LKFEM{LagP_node}{#1}}
\newcommand{\lkFE}[1]{\LKFEM{FE}{#1}}
\newcommand{\lkTFE}[1]{\LKFEM{FE_transf}{#1}}
\newcommand{\lkFELagP}[1]{\LKFEM{FE_LagP}{#1}}

\newcommand{\coqelkLogic}[1]{\coqelk{Subset.Logic.logic_compl}{#1}}
\newcommand{\COQELKSUB}[2]{\coqelk{Subset.Subsets.#1}{#2}}
\newcommand{\coqelkFun}[1]{\COQELKSUB{Function}{#1}}
\newcommand{\coqelkFunS}[1]{\COQELKSUB{Function_sub}{#1}}
\newcommand{\COQELKFF}[2]{\coqelk{Subset.Subsets.FF.#1}{#2}}
\newcommand{\coqelkOrd}[1]{\COQELKFF{ord_compl}{#1}}
\newcommand{\coqelkFF}[1]{\COQELKFF{Finite_family}{#1}}
\newcommand{\COQELKALG}[2]{\coqelk{Algebra.#1}{#2}}
\newcommand{\coqelkH}[1]{\COQELKALG{Hierarchy_compl}{#1}}
\newcommand{\COQELKMON}[2]{\COQELKALG{Monoid.#1}{#2}}
\newcommand{\coqelkMonF}[1]{\COQELKMON{Monoid_FF}{#1}}
\newcommand{\coqelkMonT}[1]{\COQELKMON{Monoid_FT}{#1}}
\newcommand{\coqelkSum}[1]{\COQELKMON{Monoid_sum}{#1}}
\newcommand{\coqelkMonS}[1]{\COQELKMON{Monoid_sub}{#1}}
\newcommand{\coqelkProd}[1]{\COQELKMON{MonoidMult}{#1}}
\newcommand{\coqelkComp}[1]{\COQELKMON{MonoidComp}{#1}}
\newcommand{\coqelkConcat}[1]{\COQELKMON{ConcatnF}{#1}}
\newcommand{\coqelkMO}[1]{\COQELKMON{Monomial_order}{#1}}
\newcommand{\COQELKRG}[2]{\COQELKALG{Ring.#1}{#2}}
\newcommand{\coqelkRing}[1]{\COQELKRG{Ring_compl}{#1}}
\newcommand{\coqelkRingF}[1]{\COQELKRG{Ring_FF_FT}{#1}}
\newcommand{\coqelkRChar}[1]{\COQELKRG{Ring_charac}{#1}}
\newcommand{\coqelkKron}[1]{\COQELKRG{Ring_kron}{#1}}
\newcommand{\COQELKMS}[2]{\COQELKALG{ModuleSpace.#1}{#2}}
\newcommand{\coqelkMSF}[1]{\COQELKMS{ModuleSpace_FF_FT}{#1}}
\newcommand{\coqelkLC}[1]{\COQELKMS{ModuleSpace_lin_comb}{#1}}
\newcommand{\coqelkLM}[1]{\COQELKMS{ModuleSpace_lin_map}{#1}}
\newcommand{\coqelkMSS}[1]{\COQELKMS{ModuleSpace_sub}{#1}}
\newcommand{\coqelkMSR}[1]{\COQELKMS{ModuleSpace_R_compl}{#1}}
\newcommand{\COQELKAS}[2]{\COQELKALG{AffineSpace.#1}{#2}}
\newcommand{\coqelkAS}[1]{\COQELKAS{AffineSpace_def}{#1}}
\newcommand{\coqelkASF}[1]{\COQELKAS{AffineSpace_FF}{#1}}
\newcommand{\coqelkASB}[1]{\COQELKAS{AffineSpace_baryc}{#1}}
\newcommand{\coqelkAM}[1]{\COQELKAS{AffineSpace_aff_map}{#1}}
\newcommand{\coqelkASS}[1]{\COQELKAS{AffineSpace_sub}{#1}}
\newcommand{\COQELKFD}[2]{\COQELKALG{Finite_dim.Finite_dim_#1}{#2}}
\newcommand{\coqelkFDMS}[1]{\COQELKFD{MS_def}{#1}}
\newcommand{\coqelkFDLG}[1]{\COQELKFD{MS_lin_gen}{#1}}
\newcommand{\coqelkFDB}[1]{\COQELKFD{MS_basis}{#1}}
\newcommand{\coqelkFDLM}[1]{\COQELKFD{MS_lin_map}{#1}}

\newcommand{\coqelkFDBR}[1]{\COQELKFD{MS_basis_R}{#1}}
\newcommand{\coqelkFDLMR}[1]{\COQELKFD{MS_lin_map_R}{#1}}
\newcommand{\coqelkFDAS}[1]{\COQELKFD{AS_def}{#1}}
\newcommand{\coqelkFDASp}[1]{\COQELKFD{AS_aff_span}{#1}}
\newcommand{\coqelkFDAI}[1]{\COQELKFD{AS_aff_indep}{#1}}
\newcommand{\coqelkFDAB}[1]{\COQELKFD{AS_aff_basis}{#1}}
\newcommand{\coqelkFDAM}[1]{\COQELKFD{AS_aff_map}{#1}}
\newcommand{\coqelkBinom}[1]{\COQELKALG{binomial_compl}{#1}}
\newcommand{\COQELKFEM}[2]{\coqelk{FEM.#1}{#2}}
\newcommand{\coqelkMonom}[1]{\COQELKFEM{monomial}{#1}}
\newcommand{\coqelkMI}[1]{\COQELKFEM{multi_index}{#1}}
\newcommand{\coqelkPdk}[1]{\COQELKFEM{poly_Pdk}{#1}}
\newcommand{\coqelkLik}[1]{\COQELKFEM{poly_LagP1k}{#1}}
\newcommand{\coqelkLdiref}[1]{\COQELKFEM{poly_LagPd1_ref}{#1}}
\newcommand{\coqelkTgeom}[1]{\COQELKFEM{geom_transf_affine}{#1}}
\newcommand{\coqelkLdi}[1]{\COQELKFEM{poly_LagPd1}{#1}}
\newcommand{\coqelkNoderef}[1]{\COQELKFEM{LagP_node_ref}{#1}}
\newcommand{\coqelkNode}[1]{\COQELKFEM{LagP_node}{#1}}
\newcommand{\coqelkFE}[1]{\COQELKFEM{FE}{#1}}
\newcommand{\coqelkTFE}[1]{\COQELKFEM{FE_transf}{#1}}
\newcommand{\coqelkFELagP}[1]{\COQELKFEM{FE_LagP}{#1}}

\usepackage[framemethod=tikz]{mdframed}

\usepackage{ifthen, tikz, circuitikz}
\usetikzlibrary{arrows.meta,calc,patterns,patterns.meta,%
  positioning,shadings,shapes}
\tikzset{math3d/.style={x= {(-0.55cm,-0.55cm)}, z={(0cm,1cm)},y={(1cm,0cm)}}}

\theoremstyle{plain}
\mdfdefinestyle{thsty}{linecolor=black, roundcorner=5pt, linewidth=1pt}
\newmdtheoremenv[style=thsty]{TextbookTheorem}{Textbook Theorem}
\mdtheorem[style=thsty]{Theorem}{Theorem}
\mdtheorem[style=thsty]{Lemma}[Theorem]{Lemma}
\mdtheorem[style=thsty]{Definition}[Theorem]{Definition}
\mdtheorem[style=thsty]{Remark}[Theorem]{Remark}

\usepackage{color}

\definecolor{tanIII}{RGB}{205,135,63}
\definecolor{yellowII}{RGB}{238,238,0}
\definecolor{darkOliveGreenIII}{RGB}{162,205,90}
\definecolor{turquoiseII}{RGB}{0,229,238}
\definecolor{turquoiseOO}{RGB}{176,221,232}

\definecolor{darkred}{rgb}{0.7,0.2,0.2}
\definecolor{darkgreen}{rgb}{0.2,0.7,0.2}
\definecolor{darkblue}{rgb}{0.2,0.2,0.7}
\definecolor{lightred}{RGB}{255,225,225}
\definecolor{lightgreen}{RGB}{200,255,200}
\definecolor{lightblue}{RGB}{225,225,255}

\definecolor{darkgray}{gray}{0.45}
\definecolor{gray}{gray}{0.60}
\definecolor{lightgray}{gray}{0.75}
\definecolor{lightlightgray}{gray}{0.90}

\def\iscolor{1}  
\if\iscolor1

\newcommand{\colorB}{yellowII}
\newcommand{\colorC}{darkOliveGreenIII}
\newcommand{\colorD}{turquoiseII}

\else

\newcommand{\colorB}{gray}
\newcommand{\colorC}{lightgray}
\newcommand{\colorD}{lightlightgray}

\fi

\newcommand{\AuthorsList}{%
  S. Boldo, F. Clément, V. Martin, M. Mayero, H. Mouhcine}

\newcommand{\LMForgdiv}{Laboratoire Méthodes Formelles}
\newcommand{\LMForgname}{Université Paris-Saclay, CNRS, ENS Paris-Saclay, Inria}
\newcommand{\LMFcity}{Gif-sur-Yvette}
\newcommand{\LMFpostcode}{91190}
\newcommand{\LMFcountry}{France}
\newcommand{\LMF}{%
  {\LMForgname}, {\LMForgdiv},
   {\LMFpostcode} {\LMFcity}, {\LMFcountry}.}

\newcommand{\SERENAorgname}{Centre Inria de Paris}
\newcommand{\SERENAstreet}{48 rue Barrault, CS 61534}
\newcommand{\SERENAcity}{Paris Cedex}
\newcommand{\SERENApostcode}{75647}
\newcommand{\SERENAcountry}{France}
\newcommand{\CERMICSorgdiv}{CERMICS}
\newcommand{\CERMICSorgname}{École des Ponts}
\newcommand{\CERMICScity}{Marne-la-Vallée}
\newcommand{\CERMICSpostcode}{77455}
\newcommand{\CERMICScountry}{France}
\newcommand{\SERENA}{%
  a. {\SERENAorgname}, {\SERENAstreet},
     {\SERENApostcode} {\SERENAcity}, {\SERENAcountry}.\protect\\
  b. {\CERMICSorgdiv}, {\CERMICSorgname},
     {\CERMICSpostcode} {\CERMICScity}, {\CERMICScountry}.}

\newcommand{\LMACorgdiv}{LMAC}
\newcommand{\LMACorgname}{Université de technologie de Compiègne}
\newcommand{\LMACcity}{Compiègne}
\newcommand{\LMACpostcode}{60203}
\newcommand{\LMACcountry}{France}
\newcommand{\LMAC}{%
  {\LMACorgname}, {\LMACorgdiv},
  {\LMACpostcode} {\LMACcity}, {\LMACcountry}.}

\newcommand{\LIPNorgdiv}{LIPN}
\newcommand{\LIPNorgname}{Université Paris 13 - USPN, CNRS UMR 7030}
\newcommand{\LIPNcity}{Villetaneuse}
\newcommand{\LIPNpostcode}{93430}
\newcommand{\LIPNcountry}{France}
\newcommand{\LIPN}{%
  {\LIPNorgdiv}, {\LIPNorgname},
  {\LIPNpostcode} {\LIPNcity}, {\LIPNcountry}.}

\newcommand{\Acknowledgments}{%
  The authors thank Alexandre Ern, Guillaume Melquiond, and Pierre Rousselin
  for fruitful discussions about proof details, canonical structures, and
  notations.
  The authors are also thankful to David Hamelin for his help in tuning the CI
  and about the {\Nix} package manager.
}

\newcommand{\Funding}{%
  This work was partly supported by the European Research Council (ERC) under
  the European Union's Horizon 2020 Research and Innovation Programme – Grant
  Agreement n$^\circ$810367.
}

\newcommand{\Title}{%
  A Rocq Formalization of Simplicial Lagrange Finite Elements}
\newcommand{\TitleRR}{%
  \Title}
\newcommand{\TitreRR}{%
  Une formalisation en Rocq des éléments finis de Lagrange simpliciaux}

\newcommand{\Abstract}{%
  Formalization of mathematics is a major topic, that includes in particular
  numerical analysis, towards proofs of scientific computing programs.
  The present study is about the finite element method, a popular method to
  numerically solve partial differential equations.
  In the long-term goal of proving its correctness, we focus here on the formal
  definition of what is a finite element.
  Mathematically, a finite element describes what happens in a cell of a mesh.
  It notably includes the geometry of the cell, the polynomial approximation
  space, and a finite set of linear forms that computationally characterizes
  the polynomials.
  Formally, we design a finite element as a record in the Rocq proof assistant
  with both values (such as the vertices of the cell) and proofs of
  validity (such as the dimension of the approximation space).
  The decisive validity proof is unisolvence, that makes the previous
  characterization unique.
  We then instantiate this record with the most popular and useful, the
  simplicial Lagrange finite elements for evenly distributed nodes, for any
  dimension and any polynomial degree, including the difficult unisolvence
  proof.
  These proofs require many results (definitions, lemmas, canonical structures)
  about finite families, affine spaces, multivariate polynomials, in the
  context of finite or infinite-dimensional spaces.
}
\newcommand{\Resume}{%
  La formalisation des mathématiques est un sujet important, qui inclut en
  particulier l'analyse numérique, vers des preuves de programmes de calcul
  scientifique.
  La présente étude porte sur la méthode des éléments finis, un outil populaire
  pour la résolution numérique d'équations aux dérivées partielles.
  En vue de l'objectif à long terme de la preuve de sa correction, nous nous
  focalisons ici sur la définition formelle de ce qu'est un élément fini.
  Mathématiquement, a élément fini décrit ce qu'il se passe dans une cellule
  du maillage.
  Il inclut notamment la géométrie de la maille, l'espace d'approximation
  polynomiale et un nombre fini de formes linéaires qui caractérisent les
  polynômes d'un point de vue calculatoire.
  Formellement, nous concevons un élément fini comme un enregistrement dans
  l'assistant de preuve Rocq avec à la fois des valeurs (comme les sommets de
  la maille) et des preuves de validité (comme la dimension de l'espace
  d'approximation).
  La preuve de validité décisive est l'unisolvance, qui rend unique la
  caractérisation précédente.
  Puis nous instancions cet enregistrement avec les plus populaires et utiles,
  les éléments finis simpliciaux de Lagrange pour des n{\oe}uds uniformément
  répartis, pour toutes les dimensions et tous les degrés de polynômes,
  incluant la preuve difficile d'unisolvance.
  Ces preuves exigent de nombreux résultats (définitions, lemmes, structures
  canoniques) sur les familles finies, les espaces affines, les polynômes
  multivariés, dans le cadre d'espaces de dimension finie ou infinie.
 }

\newcommand{\Keywords}{%
  Formalization of mathematics,
  Rocq,
  Affine Space,
  Numerical Analysis,
  Finite Element Method,
  Lagrange Finite Element
}
\newcommand{\Motscles}{%
  Formalisation des mathématiques,
  Rocq,
  Espace affine,
  Analyse Numérique,
  Méthode des éléments finis,
  Éléments finis de Lagrange
}

\newcommand{\aka}{a.k.a.}

\newcommand{\eg}{e.g.}

\newcommand{\ie}{i.e.}

\newcommand{\perse}{\emph{per se}}

\newcommand{\multiindex}{multi-index}
\newcommand{\Multiindices}{Multi-indices}
\newcommand{\multiindices}{multi-indices}
\newcommand{\multivariate}{multivariate}
\newcommand{\nonaffine}{non-affine}
\newcommand{\noncancellation}{non\-cancellation}

\newcommand{\nonconstant}{non\-constant}

\newcommand{\nonconvex}{non\-convex}

\newcommand{\nondegenera}{non\-degenera}
\newcommand{\nondegeneracy}{{\nondegenera}cy}
\newcommand{\nondegenerate}{{\nondegenera}te}
\newcommand{\nonempty}{non\-empty}

\newcommand{\nonexpert}{non\-expert}

\newcommand{\noninvertible}{non\-invertible}

\newcommand{\nonnegativ}{non\-negativ}
\newcommand{\nonnegative}{{\nonnegativ}e}

\newcommand{\nonnodal}{non-nodal}
\newcommand{\nonproof}{non\-proof}
\newcommand{\nonquadratic}{non\-quadratic}

\newcommand{\nonzero}{non\-zero}
\newcommand{\Nonzero}{Non\-zero}

\newcommand{\subfamilies}{sub\-families}
\newcommand{\subnode}{sub\-node}
\newcommand{\Subnode}{Sub\-node}
\newcommand{\substructure}{sub\-structure}
\newcommand{\Substructure}{Sub\-structure}
\newcommand{\subvert}{sub\-vert}
\newcommand{\Subvert}{Sub\-vert}

\newcommand{\subvertices}{{\subvert}ices}
\newcommand{\Subvertices}{{\Subvert}ices}

\newcommand{\myskip}{\bigskip} 

\newcommand{\AppSection}{Section} 

\newcommand{\soft}[1]{\textsf{#1}}
\newcommand{\Coq}{\soft{Coq}}
\newcommand{\Rocq}{\soft{Rocq}}
  \newcommand{\Coquelicot}{\soft{Coquelicot}}
  \newcommand{\CoqNumAnalysis}{\soft{coq-num-analysis}}
  \newcommand{\RocqNumAnalysis}{\soft{rocq-num-analysis}}
  \newcommand{\RNASubset}{\RocqNumAnalysis\soft{-subset}}
  \newcommand{\RNAAlgebra}{\RocqNumAnalysis\soft{-algebra}}
  \newcommand{\RNALebesgue}{\RocqNumAnalysis\soft{-lebesgue}}
  \newcommand{\RNALM}{\RocqNumAnalysis\soft{-lm}}
  \newcommand{\RNAFEM}{\RocqNumAnalysis\soft{-fem}}
  \newcommand{\CoRN}{\soft{C-CoRN}}
  \newcommand{\Flocq}{\soft{Flocq}}
  \newcommand{\HB}{\soft{Hierarchy Builder}}
  \newcommand{\MathComp}{\soft{MathComp}}
  \newcommand{\MathCompAnalysis}{\MathComp\soft{-Analysis}}
  \newcommand{\SSReflect}{\soft{SSReflect}}
  \newcommand{\SSRfun}{\soft{ssrfun}}
  \newcommand{\Reals}{\soft{Reals}}
\newcommand{\Cpp}{\soft{C++}}
  \newcommand{\Felisce}{\soft{FELiScE}}
  \newcommand{\FreeFEM}{\soft{FreeFEM++}}
  \newcommand{\XLiFE}{\soft{XLiFE++}}
\newcommand{\GIT}{\soft{GIT}}
\newcommand{\HOL}{\soft{HOL}}
\newcommand{\IsabelleHOL}{\soft{Isabelle/HOL}}

\newcommand{\Lean}{\soft{Lean}}
  \newcommand{\mathlib}{\soft{mathlib}}
  \newcommand{\SciLean}{\soft{SciLean}}
    \newcommand{\SciLeanURL}{https://github.com/lecopivo/SciLean/tree/v4.20.1}
    \newcommand{\SciLeanFEURL}{\SciLeanURL/SciLean/Modules/FiniteElement/Mesh}
    \newcommand{\SciLeanPRURL}{\SciLeanFEURL/PrismRepr.lean}
    \newcommand{\SciLeanTMURL}{\SciLeanFEURL/TriangularMesh.lean}
    \newcommand{\SciLeanODEURL}{\SciLeanURL/SciLean/Analysis/ODE}
\newcommand{\Mizar}{\soft{Mizar}}
\newcommand{\Nix}{\soft{Nix}}
\newcommand{\Opam}{\soft{Opam}}

\newcommand{\thethm}[1]{the #1 theorem}
\newcommand{\Thethm}[1]{The #1 theorem}

\newcommand{\BNB}{Banach--Ne\v{c}as--Babu\v{s}ka}

\newcommand{\LM}{Lax--Milgram}
\newcommand{\theLMt}{\thethm{\LM}}
\newcommand{\TheLMt}{\Thethm{\LM}}
\newcommand{\Cea}{Céa}

\newcommand{\RT}{\mathbb{RT}}
\newcommand{\BDM}{\mathbb{BDM}}
\newcommand{\BDFM}{\mathbb{BDFM}}

\newcommand{\dps}{\displaystyle}
\newcommand{\minitab}[2][l]{\begin{tabular}{@{}#1@{}}#2\end{tabular}}

\renewcommand{\leq}{\leqslant}
\renewcommand{\emptyset}{\varnothing}

\newcommand{\fhi}{\varphi}
\newcommand{\st}{{\,|\,}}

\newcommand{\restr}[2]{{#1}_{|{#2}}}
\newcommand{\makespace}[1]{\quad#1\quad}
\newcommand{\Equiv}{\Longleftrightarrow}
\newcommand{\EQUIV}{\makespace{\Equiv}}
\newcommand{\Implies}{\Longrightarrow}
\newcommand{\IMPLIES}{\makespace{\Implies}}
\newcommand{\Conj}{\land}

\newcommand{\CONJ}{\makespace{\Conj}}

\newcommand{\AND}{\makespace{\text{and}}}

\newcommand{\eqdef}{\stackrel{\text{def.}}{=}}

\newcommand{\equivdef}{\stackrel{\text{def.}}{\Equiv}}
\newcommand{\EQUIVDEF}{\makespace{\equivdef}}
\newcommand{\nin}{\not\in}

\newcommand{\PropPP}{\Pi}

\newcommand{\Span}[1]{{\text{span}(#1)}}
\newcommand{\card}{\text{card}}

\newcommand{\len}[1]{\left|#1\right|}

\newcommand{\Arrow}[2]{#1\to#2}
\newcommand{\ArOmR}{\Arrow{\Omega}{\R}}

\newcommand{\ArRdR}{\Arrow{\Rd}{\R}}
\newcommand{\ArRdRd}{\Arrow{\Rd}{\Rd}}
\newcommand{\ArRdmiRd}{\Arrow{\Rdmi}{\Rd}}
\newcommand{\ArRdRdpi}{\Arrow{\Rd}{\Rdpi}}
\newcommand{\ArNdmiNd}{\Arrow{\Ndmi}{\Nd}}

\newcommand{\N}{\mathbb{N}}
\newcommand{\Nstar}{\N^{\star}}
\newcommand{\Nd}{\N^d}
\newcommand{\Ndmi}{\N^{d-1}}

\newcommand{\matP}{\mathbb{P}}
\newcommand{\hmatP}{\hat{\matP}}
\newcommand{\Q}{\mathbb{Q}}

\newcommand{\R}{\mathbb{R}}
\newcommand{\Rstar}{\R^\star}
\newcommand{\Rd}{\R^d}
\newcommand{\Rdmi}{\R^{d-1}}
\newcommand{\Rdpi}{\R^{d+1}}

\newcommand{\Rbar}{\overline{\R}}

\newcommand{\matPgen}[2]{\matP^{#1}_{#2}}
\newcommand{\matPdk}{\matPgen{d}{k}}
\newcommand{\matPdmik}{\matPgen{d-1}{k}}
\newcommand{\matPdkmi}{\matPgen{d}{k-1}}
\newcommand{\matPdpik}{\matPgen{d+1}{k}}
\newcommand{\matPdkpi}{\matPgen{d}{k+1}}
\newcommand{\matPdpikpi}{\matPgen{d+1}{k+1}}
\newcommand{\matPdo}{\matPgen{d}{0}}
\newcommand{\matPdi}{\matPgen{d}{1}}

\newcommand{\matPik}{\matPgen{1}{k}}

\newcommand{\calA}{\mathcal{A}}

\newcommand{\calC}{\mathcal{C}}
\newcommand{\calF}{\mathcal{F}}
\newcommand{\calH}{\mathcal{H}}

\newcommand{\calL}{\mathcal{L}}

\newcommand{\calS}{\mathcal{S}}

\newcommand{\calT}{\mathcal{T}}
\newcommand{\calTh}{\calT_h}

\newcommand{\calAgen}[2]{\calA^{#1}_{#2}}
\newcommand{\calAdk}{\calAgen{d}{k}}
\newcommand{\calAdpik}{\calAgen{d+1}{k}}
\newcommand{\calAdmik}{\calAgen{d-1}{k}}

\newcommand{\calAdkmi}{\calAgen{d}{k-1}}
\newcommand{\calAdpikpi}{\calAgen{d+1}{k+1}}

\newcommand{\calAdiii}{\calAgen{d}{3}}

\newcommand{\calAdkigen}[1]{\calAgen{d}{k,#1}}
\newcommand{\calAdko}{\calAdkigen{0}}
\newcommand{\calAdki}{\calAdkigen{i}}

\newcommand{\InjAdk}[1]{\mathbb{I}^d_{k,#1}}
\newcommand{\InjAdko}{\InjAdk{0}}
\newcommand{\InjAdki}{\InjAdk{i}}

\newcommand{\calCdk}{{\calC}^{d}_{k}}

\newcommand{\calCdl}{{\calC}^{d}_{l}}
\newcommand{\calCd}[1]{\calC^d_{#1}}
\newcommand{\calCdo}{\calCd{0}}
\newcommand{\calCdi}{\calCd{1}}
\newcommand{\calCdii}{\calCd{2}}
\newcommand{\calCdiii}{\calCd{3}}

\newcommand{\FRdgen}[2]{\calF(\R{#1},\R{#2})}
\newcommand{\FRdgeni}[1]{\FRdgen{^{#1}}{}}
\newcommand{\FRdRq}{\FRdgen{^d}{^q}}
\newcommand{\FRd}{\FRdgeni{d}}

\newcommand{\FRo}{\FRdgeni{0}}
\newcommand{\FRdpgen}[1]{\Arrow{\FRdgeni{#1}}{\R}}
\newcommand{\FRdp}{\FRdpgen{d}}
\newcommand{\FRop}{\FRdpgen{0}}

\newcommand{\hHpl}{\hat{\calH}}
\newcommand{\Hplgen}[1]{\calH^{#1}}
\newcommand{\Hplv}{\Hplgen{\vv}}
\newcommand{\InjhHd}[1]{{\hat{I}_{#1}}}
\newcommand{\InjhHdo}{{\InjhHd{0}}}
\newcommand{\InjhHdi}{{\InjhHd{i}}}
\newcommand{\Injgen}[2]{I^{#1}_{#2}}
\newcommand{\Injv}[1]{\Injgen{\vv}{#1}}
\newcommand{\Injvo}{\Injv{0}}
\newcommand{\Injvi}{\Injv{i}}

\newcommand{\txx}{\tilde{\xx}}
\newcommand{\tpo}{\tilde{p}_0}

\newcommand{\ttimes}{{\boldsymbol{\times}}}
\newcommand{\aalpha}{{\boldsymbol{\alpha}}}
\newcommand{\aalphap}{{\aalpha^\prime}}
\newcommand{\alphap}{\alpha^\prime}
\newcommand{\bbeta}{{\boldsymbol{\beta}}}
\newcommand{\bbetap}{{\bbeta^\prime}}
\newcommand{\ggamma}{{\boldsymbol{\gamma}}}
\newcommand{\ddelta}{{\boldsymbol{\delta}}}
\newcommand{\xxi}{{\boldsymbol{\xi}}}
\newcommand{\SSigma}{{\boldsymbol{\Sigma}}}
\renewcommand{\aa}{{\bf a}}
\renewcommand{\AA}{{\bf A}}

\newcommand{\HH}{{\bf H}}
\newcommand{\LL}{{\bf L}}
\newcommand{\nn}{{\bf n}}

\newcommand{\uu}{{\bf u}}

\newcommand{\UU}{{\bf U}}
\newcommand{\vv}{{\bf v}}
\newcommand{\ww}{{\bf w}}
\newcommand{\xx}{{\bf x}}

\newcommand{\XX}{{\bf X}}
\newcommand{\zzero}{{\bf 0}}
\newcommand{\oone}{{\bf 1}}

\newcommand{\hxxi}{{\hat{\xxi}}}
\newcommand{\haa}{{\hat{\aa}}}
\newcommand{\hvv}{{\hat{\vv}}}
\newcommand{\hxx}{\hat{\xx}}
\newcommand{\hxxp}{\hat{\xx}^\prime}

\newcommand{\hp}{\hat{p}}
\newcommand{\hq}{\hat{q}}
\newcommand{\hx}{\hat{x}}
\newcommand{\hxp}{\hat{x}^\prime}

\newcommand{\nvtx}{{n_{\text{vtx}}}}
\newcommand{\ndof}{{n_{\text{dof}}}}
\newcommand{\Ndof}{{N_{\text{dof}}}}

\newcommand{\bcSymbol}{{\ell}}
\newcommand{\hbcSymbol}{{\hat{\bcSymbol}}}
\newcommand{\bcGEN}[2]{{\boldsymbol{#1}^{#2}}}
\newcommand{\bc}[1]{{\bcGEN{\bcSymbol}{#1}}}
\newcommand{\hbc}{\bcGEN{\hbcSymbol}{}}
\newcommand{\bciGEN}[3]{{{#1}^{#2}_{#3}}}
\newcommand{\bci}[2]{{\bciGEN{\bcSymbol}{#1}{#2}}}

\newcommand{\Tgeom}[1]{T^{#1}}
\newcommand{\Tgeomv}{\Tgeom{\vv}}
\newcommand{\Tgeomhv}{\Tgeom{\hvv}}
\newcommand{\Tgeomw}{\Tgeom{\ww}}
\newcommand{\Tgeomvinv}{\left(\Tgeomv\right)^{-1}}

\newcommand{\Tgeomuhv}{\Tgeom{\uhvv}}
\newcommand{\TgeomPermut}[1]{\Tgeom{\pi(#1)}}
\newcommand{\TgeomPermutv}{\TgeomPermut{\vv}}
\newcommand{\TgeomPermutvinv}{\left(\TgeomPermutv\right)^{-1}}
\newcommand{\TgeomTransp}[3]{\Tgeom{\tau_{#1}^{#2}(#3)}}
\newcommand{\TgeomTranspo}[1]{\TgeomTransp{d}{0}{#1}}
\newcommand{\TgeomTranspov}{\TgeomTranspo{\vv}}
\newcommand{\taudo}{\tau_d^0}
\newcommand{\TgeomTranspi}[1]{\TgeomTransp{d}{i}{#1}}
\newcommand{\TgeomTranspiv}{\TgeomTranspi{\vv}}
\newcommand{\taudi}{\tau_d^i}
\newcommand{\Scal}[1]{\left(#1\cdot\right)}
\newcommand{\Scalinvk}{\Scal{\frac{1}{k}}}
\newcommand{\Scalinvkmi}{\Scal{\frac{1}{k-1}}}
\newcommand{\Scalratio}{\Scal{\frac{k-1}{k}}}

\newcommand{\JacTgeom}[1]{\boldsymbol{J}^{#1}}
\newcommand{\JacTgeomv}{\JacTgeom{\vv}}

\newcommand{\FElagPgen}[3]{\prescript{L\!a\!g}{}{#1}^{#2}_{#3}}
\newcommand{\FElagP}[2]{\FElagPgen{\matP}{#1}{#2}}
\newcommand{\FElagPdk}{\FElagP{d}{k}}
\newcommand{\FElagPdo}{\FElagP{d}{0}}

\newcommand{\FElagPik}{\FElagP{1}{k}}
\newcommand{\FElagPref}[2]{\FElagPgen{\hmatP}{#1}{#2}}
\newcommand{\FElagPdkref}{\FElagPref{d}{k}}
\newcommand{\FElagQ}[2]{\FElagPgen{\Q}{#1}{#2}}

\newcommand{\Lii}{L^2}
\newcommand{\lii}[1]{\Lii(#1)}
\newcommand{\LiiOm}{\lii{\Omega}}
\newcommand{\Ho}{H^0}
\newcommand{\ho}[1]{\Ho(#1)}
\newcommand{\HoOm}{\ho{\Omega}}
\newcommand{\Hi}{H^1}
\newcommand{\hi}[1]{\Hi(#1)}
\newcommand{\HiOm}{\hi{\Omega}}
\newcommand{\Hio}{\Hi_0}
\newcommand{\hio}[1]{\Hio(#1)}
\newcommand{\HioOm}{\hio{\Omega}}
\newcommand{\Hdiv}{\HH(\div)}
\newcommand{\Hcurl}{\HH(\curl)}

\newcommand{\Closure}[1]{\overline{#1}}
\newcommand{\Omegab}{\Closure{\Omega}}
\newcommand{\homega}{\hat{\omega}}

\newcommand{\hK}{\hat{K}}
\newcommand{\hKd}{\hK^d}

\newcommand{\PhiSSigma}{\Phi_\SSigma}
\newcommand{\Interior}[1]{\stackrel{\circ}{#1}}

\newcommand{\txtin}[1]{#1^\text{in}}
\newcommand{\txtout}[1]{#1^\text{transf}}
\newcommand{\Kin}{\txtin{K}}
\newcommand{\Kout}{\txtout{K}}
\newcommand{\Pin}{\txtin{P}}
\newcommand{\Pout}{\txtout{P}}

\newcommand{\pin}{\txtin{p}}

\newcommand{\sigmain}{\txtin{\sigma}}
\newcommand{\sigmaout}{\txtout{\sigma}}
\newcommand{\vvin}{\txtin{\vv}}
\newcommand{\vvout}{\txtout{\vv}}
\newcommand{\xxin}{\txtin{\xx}}
\newcommand{\xxout}{\txtout{\xx}}

\newcommand{\Blf}{a}
\newcommand{\blf}[2]{{\Blf (#1, #2)}}
\newcommand{\Lf}{l}
\newcommand{\lf}[1]{{\Lf (#1)}}

\newcommand{\Vh}{V_h}
\newcommand{\uh}{u_h}
\newcommand{\vh}{v_h}

\newcommand{\p}{\partial}
\newcommand{\nnabla}{\boldsymbol{\nabla}}
\renewcommand{\div}{\text{div}}
\newcommand{\curl}{\text{curl}}

\newcommand{\norm}[2]{\left\|#2\right\|_{#1}}
\newcommand{\normo}[1]{\norm{0}{#1}}
\newcommand{\normi}[1]{\norm{1}{#1}}
\newcommand{\normio}[1]{\norm{1,0}{#1}}
\newcommand{\normV}[1]{\norm{V}{#1}}

\newcommand{\uvv}{{\underline{\vv}}}
\newcommand{\uhvv}{{\underline{\hvv}}}
\newcommand{\uaa}{{\underline{\aa}}}
\newcommand{\uhaa}{{\underline{\haa}}}

\newcommand{\vect}[1]{\overrightarrow{#1}}

\newcommand{\lto}[1]{<^{\text{#1}}}
\newcommand{\ltgrsymlex}{\lto{grsymlex}}
\newcommand{\ltlex}{\lto{lex}}

\newcommand{\curLgen}[1]{\calL^{#1}}
\newcommand{\curLLgen}[1]{\boldsymbol{\calL}^{#1}}
\newcommand{\curLdiv}{\curLgen{\vv}}
\newcommand{\curLLdiv}{\curLLgen{\vv}}
\newcommand{\lagLik}{\mathbb{L}}
\newcommand{\refLgen}{\hat{\calL}}
\newcommand{\refLLgen}{{\,\,\hat{\!\!\boldsymbol{\calL}}}}
\newcommand{\refLdi}{\refLgen}
\newcommand{\refLLdi}{\refLLgen}



%% file: body.tex
\section{Introduction}
\label{sec:Intro}

\subsection{General Introduction}
\label{sec:Intro:gen}

Our long-term goal is to provide greater guarantees for numerical programs
simulating physical systems, as they are widely used, {\eg} in healthcare,
civil engineering, aeronautics, climatology, and astrophysics.
In this article we deal with one of the most popular method for solving Partial
Differential Equations~(PDE), the finite element method~(FEM).
We focus on the formalization of finite elements~(FE), and more specifically on
the construction of the most commonly used, the simplicial Lagrange~FE.
This is a necessary first step towards a formal proof of the whole~FEM.
More details about the~FEM can be found in Section~\ref{sec:FE:Maths} and
Appendix~\ref{sec:Maths:FEM}.
Informally, the~FEM is used to approximate the solution of~PDE.
These equations are used to model a wide range of problems, including fluid
mechanics (Navier--Stokes equations), electromagnetism (Maxwell equations),
heat equation (Fourier), mechanics (elasticity), quantum mechanics (Schrödinger
and Heisenberg), weather forecasting and aeronautics, among others.

\begin{figure}[ht]
  \centering
  \includegraphics[width=4cm]{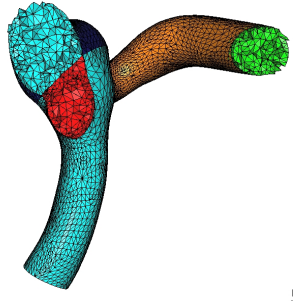}
  \caption[3D mesh of a cerebral aneurysm]{%
    3D mesh of a cerebral aneurysm (colors only help to distinguish zones).}
  \label{fig:aneurysm}
\end{figure}

\begin{figure}[ht]
  \centering
  \resizebox{0.5\linewidth}{!}{\input{fig_Lag_2to3}}
  \caption[Reference Lagrange finite elements in~2D and~3D]{%
    Reference Lagrange finite elements~$\FElagP{d}{k}$ in dimension~$d=2$ (top)
    and~$d=3$ (bottom), with degrees of approximation~$k=1$ (left) and~$k=2$
    (right).
    The geometric elements are $d$-simplices.
    The nodes (depicted by blue dots) are only the vertices when~$k=1$, and
    when~$k=2$, nodes are added in the middle of edges.
    Their coordinates being real numbers, they are denoted with a point suffix,
    such as in ``0.'' and ``1.''.}
  \label{fig:Lag:geom}
\end{figure}
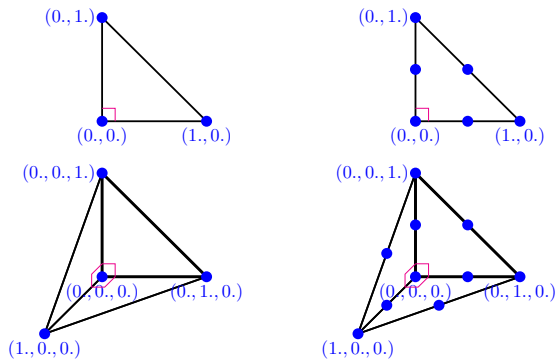

Exact solutions of these~PDE are generally not computable.
To be able to compute an approximate solution, the idea is to discretize, {\ie}
to cut the problem to be solved into small pieces.
As described in Section~\ref{sec:ref-cur-fe-geomap} (and with more mathematical
details in \AppSection~\ref{sec:mesh}), we create a mesh consisting of
``simple'' geometric elements covering the total area or volume, see
Figure~\ref{fig:aneurysm}, on which we know how to perform the calculations.
We are now moving from a continuous problem to a discrete one.
These ``computational'' elements are called ``finite elements''.
To perform approximate calculations on these different elements, a change of
coordinate frame is necessary.
This involves changing the representation of space from some ``current
element'' to the ``reference element'', where calculations are simpler (see
Figure~\ref{fig:Lag:geom}); for example in~2D, one can take the right triangle,
with sides of length~1 as a reference element.
The link between the reference and current elements is also used to establish
approximation properties of the FE~\cite[Chap.~11]{eg:FE1:21}, which are out of
scope of this paper.
The technique of using~FE to approximate the solution to the overall problem
is the ``finite element method''.
For example, Figure~\ref{fig:aneurysm} represents an application of~FEM to
healthcare.
This is a~3D mesh of a cerebral aneurysm used to calculate blood flow inside
the artery (using the Navier--Stokes equation) and wall mechanics (shell
equation, not depicted here), with fluid structure interaction between wall and
blood, see for instance~\cite{mf-jfg-vm-stent:08}.

As explained in Section~\ref{sec:fe-triple}, a~FE is not just a geometric
shape, it is a triple~$(K,P,\SSigma)$.
Simply said, $K$~represents the geometry, $P$~represents the set of
approximation polynomials, and~$\SSigma$ is a way to characterize the
polynomials, for instance by the values taken at some nodes (the blue points in
Figure~\ref{fig:Lag:geom}).
Moreover, a ``correct''~FE needs to satisfy the unisolvence property, which
means that the above characterization is unique.
There are several types of~FE, such as Lagrange, Raviart--Thomas (more in
\AppSection~\ref{sec:FE:zoo}), the choice depending on the equations to be
approximated.
Representations of simplicial Lagrange~FE are given in
Figure~\ref{fig:Lag:geom} for space dimensions~$d=2$ and~$3$, and for affine
($k=1$) and quadratic ($k=2$) polynomials (see Section~\ref{sec:FELagP} for a
formal description of Lagrange~FE).

\myskip

There are many programs for performing approximation calculations based on
the~FEM, and many~FE libraries are available, such as
\FreeFEM,\footnote{\url{https://freefem.org}}
\Felisce,\footnote{\url{https://gitlab.inria.fr/felisce/felisce/}} and
\XLiFE.\footnote{\url{https://xlifepp.pages.math.cnrs.fr/}}
As in any program, errors are possible, either unexpected (bugs) or expected
such as method errors, as well as errors due to calculations made by the
processor, in particular due to floating-point arithmetic.
To make these programs safe, our aim is to formally prove the correctness of
their implementation, in terms of both method and floating-point errors.
To formally prove the correctness of such a program, we need to start by
formalizing and proving the necessary mathematical properties in a proof
assistant.
The proof tool we have chosen is \Rocq,\footnote{\url{https://rocq-prover.org}}
previously called \Coq.
First, {\Rocq} provides a library of floating-point numbers~\cite{BolMel11}.
Moreover, since the 2000s, this tool has enabled us to formalize and prove a~1D
wave equation resolution program~\cite{BCF13}, and also a number of real
analysis properties necessary for our goal of proving critical numerical
programs, such as {\theLMt}~\cite{BCF17}, the Lebesgue
integral~\cite{BCF22,BCM23}, and the Bochner integral~\cite{BCL22}.
These formalizations and the present work are prerequisites to the
formalization of the~FEM.

\myskip

It should also be noted that our formalization is intended for both experts and
{\nonexpert}s in {\Rocq}, including numerical analysts and
students~\cite{BCH24,KMR24}.
We pay particular attention to the readability of statements, which informs
some of our decisions regarding formalization and the use of existing
libraries.
By ``readability'', we mean above all syntax of the statements of the lemmas,
as well as striking the right balance in terms of the abstraction of the
mathematical concepts used (see Section~\ref{sec:Libs}).

\myskip

In Section~\ref{sec:FE:Maths}, we start by describing the main mathematical
notions for the~FE.
In Section~\ref{sec:Libs}, we briefly introduce the {\Rocq} libraries we use,
such as {\Coquelicot} and {\MathComp}.
In Section~\ref{sec:Algebra}, we begin to describe our developments in {\Rocq},
starting with complements on subsets and algebra.
Section~\ref{sec:AffineAlgebra} presents a corpus of results dedicated to
affine algebra.
Section~\ref{sec:FE:Coq} presents the definition of an~FE as a {\Rocq} record.
In Section~\ref{sec:geometry}, we present several geometric transformations
({\ie} related to~$K$), including some that allow us to transform one~FE into
another.
In Section~\ref{sec:poly-space-Pdk}, we focus on a polynomial approximation
space ($P$).
Section~\ref{sec:FELagP} is dedicated to the application of our formalization
to Lagrange~FE (with the corresponding~$\SSigma$), including the proof of
unisolvence, before concluding and describing our perspectives in
Section~\ref{sec:ConclPersp}.
Finally, Appendix~\ref{sec:Maths:FEM} is devoted to a brief mathematical
presentation of the~FEM.

{\bf
  Sections~\ref{sec:FE:Coq} to \ref{sec:FELagP} represent the core of the
  article.
  One can skip reading Sections~\ref{sec:Libs} to \ref{sec:AffineAlgebra} if
  one is familiar with {\Rocq} and trusts the naming of the {\Rocq} definitions
  and lemmas.}

\myskip

The library, now called \RocqNumAnalysis, is available as {\Opam}%
\footnote{\url{https://rocq-prover.org/p/rocq-num-analysis/2.2.0}}
packages, and the {\GIT} repository is available at
\begin{center}
  \url{https://lipn.univ-paris13.fr/rocq-num-analysis/tree/2.2.0/}
\end{center}
The {\Rocq} code and documentation is also browsable at
\begin{center}
  \url{https://lipn.univ-paris13.fr/rocqdoc-num-analysis/2.2/}
\end{center}
The pen-and-paper proofs corresponding to the present work, developed for their
formalization, are detailed in~\cite{CM24}.

Note also that throughout the paper, annotations such as in ``\coqelkFE{FE}''
contain a clickable link towards the corresponding point in our {\Rocq}
formalization, be it a definition (as here), a lemma, or an entire file.
Moreover, annotations in pink boxes such as \Notation{[m..n)} and
\Notation{\coqe{n.+1}} indicate the definition or first mention of notations,
be they mathematical or formal.

\subsection{State of the Art}
\label{sec:Intro:soa}

To our knowledge, finite elements have not yet been formalized elsewhere.
However, the ongoing {\SciLean} project is quite related to our concerns, and
there have also been some developments in the specific areas of geometrical
aspects such as mesh construction, polynomial interpolation, and approximate
solution to differential equations.

\myskip

{\SciLean}%
\footnote{\url{\SciLeanURL}}
is a scientific computing library written in {\Lean}~\cite{Lean_Ref}.
The goal is to mix in the same framework the implementation of scientific
computing algorithms together with the underlying mathematics, and the proofs
of correctness of the algorithms.
However, as the authors write: ``This library is in an {\em early stage of
development}.''
It provides effective computations but with many admitted proofs.

The {\SciLean} library includes a {\tt FiniteElement/Mesh} directory%
\footnote{\url{\SciLeanFEURL}}
that provides a unified representation of various geometric shapes, called
\emph{prism}, such as $n$-simplices, $n$-cubes, and pyramids through
induction.\footnote{\url{\SciLeanPRURL}}
Then, the usual concepts are specified by induction on the prism geometry, such
as barycentric coordinates, Lagrange basis polynomials, and Lagrange nodes.
And this goes up to the construction of a triangular mesh.%
\footnote{\url{\SciLeanTMURL}}
However, there does not seem to be any specification of reference element and
geometric transformation (see Section~\ref{sec:geometry}), or of unisolvence
and finite element (see Sections~\ref{sec:FE:Coq} and~\ref{sec:FELagP}).

\myskip

About geometry, we can mention the following formalization conducted in
{\IsabelleHOL}~\cite{Isabelle_Ref}.
L.I.~Meikle and J.D.~Fleuriot~\cite{MeiFle:10:geom} focused on enhancing
two-dimensional planar geometry, such as convex hulls and Delaunay
triangulations.
This includes basic geometric definitions such as points, and orientation of
points (collinearity, betweenness).

Additionally, Y.~Bertot and J.-F.~Dufourd~\cite{DufBer:10:tri}, focused on the
formal verification of an algorithm for constructing Delaunay triangulations of
a mesh using the {\Rocq} proof assistant.
This algorithm was implemented to construct triangles such that none of the
input points lie inside the circumcircle of any triangle.
In a more recent work, Y.~Bertot extended his earlier work
in~\cite{Ber:18:mesh} by addressing a broader class of triangulation
algorithms, that guarantees that the resulting mesh satisfies important
geometric properties, such as covering the convex hull and maintaining boundary
edges.
In our development, we do not deal with the building of the mesh.
Instead, we specifically focus on a given mesh cell as the convex hull of its
affinely independent vertices, and subsequently construct a~FE on this cell.

\myskip

In {\IsabelleHOL}, R.~Thiemann and A.~Yamada address polynomial interpolation,
including Lagrange interpolation and the case of integer polynomials,%
\footnote{\url{https://www.isa-afp.org/entries/Polynomial_Interpolation.html}}
and, using different {\HOL} algebraic structures, E.~Karayel also treats the
case of finite fields.%
\footnote{\url{https://www.isa-afp.org/entries/Interpolation_Polynomials_HOL_Algebra.html}}
In {\Lean}, the {\mathlib} library offers a thorough polished study,%
\footnote{\url{https://leanprover-community.github.io/mathlib4_docs/Mathlib/LinearAlgebra/Lagrange.html}}
by emphasizing the definition of the Lagrange polynomial basis, along with the
Lagrange nodes for the interpolation process, supported by various theorems to
validate that the Lagrange polynomial exactly matches the provided value at
each specified point.
These three developments are considering univariate polynomials only.

In our {\Rocq} work, we also define the Lagrange polynomial bases and show
various theorems.
The construction is different (we define them from reference Lagrange
polynomials, see Section~\ref{sec:LagPd1-Tgeom}), but the main difference is
that we are in a {\multivariate} setting.
We define Lagrange polynomial bases for~$d=1$ (with any degree), and for~$k=1$
(and any dimension).
This suffices for our needs.

\myskip

As stated in Section~\ref{sec:Intro:gen}, the~FEM is essentially used to
approximate~PDE solutions.
The formalization of ordinary differential equations~(ODE) and~PDE is therefore
also of interest.
There have been various formal developments in these domains.
In {\IsabelleHOL}, F.~Immler and J.~Hölzl~\cite{ImmHol12} present a
formalization of~ODE and the verification of rigorous numerical algorithms.
In {\Lean}, ODE~developments are also in progress%
\footnote{\url{\SciLeanODEURL}}.
In {\Mizar}, S.~Otsuki et al.~\cite{Ots:19:PDE} formalize a simple example for
linear~PDE and its solution using the method of separation of variables.
In {\HOL}, E.~Deniz et al.~\cite{Den:22:PDE} also use the method of separation
of variables to formally verify the solution of the~PDE, which allows modeling
the heat transfer in the slab under various initial and boundary conditions.
In {\Rocq}, M.~Tekriwal et al.~\cite{Tek21} formally prove the Lax equivalence
theorem for the finite difference method, which generalizes a part of the
mathematical proof of~\cite{BCF13}.

\section{Mathematical Presentation of Finite Elements}
\label{sec:FE:Maths}

For interested readers, we provide in Appendix~\ref{sec:Maths:FEM} a quick
mathematical introduction to the finite element \emph{method}, whose
formalization is our ultimate goal.
In this article, we formalize a brick of the method: the finite element that is
presented mathematically here.

In short, the~FEM consists in computing a discrete function~$\uh$ (``discrete''
meaning that it can be entirely represented by a finite set of values), that
approximates a continuous function~$u$ (``continuous'', by opposition to
``discrete'', meaning that it lives in an infinite dimensional functional
space), that is solution to the~PDE.

After some mathematical references with some of our formalization choices given
in Section~\ref{sec:math:ref:FEM}, the reference and current geometric
elements, as well as the transformation between the two, are introduced in
Section~\ref{sec:ref-cur-fe-geomap}.
Section~\ref{sec:fe-triple} is devoted to the mathematical definition of a~FE
as a triple.
This presentation ends in Section~\ref{sec:conform:FE} with the introduction of
the most common~FE, the Lagrange~FE over a simplex, that are fully formalized
in this paper (see Section~\ref{sec:FELagP}).

\subsection{A Few Mathematical References
  with Some of Our Formalization Choices}
\label{sec:math:ref:FEM}

Among the vast literature about the~FEM, we can cite a few references:
P.G.~Ciarlet~\cite{cia:fem:02},
I.~Babu{\v{s}}ka and T.~Strouboulis~\cite{bs:fem:01},
A.~Quarteroni and A.~Valli~\cite{qv:nap:94},
A.~Ern and J.-L.~Guermond~\cite{eg:tpf:04,eg:FE1:21,eg:FE2:21,eg:FE3:21},
S.C.~Brenner and L.R.~Scott~\cite{bs:mtf:08}, and
O.C.~Zienkiewicz, R.L.~Taylor and J.Z.~Zhu~\cite{ztz:fem:13}.
The first part of the Handbook of Numerical Analysis is dedicated to the
FEM~\cite{hna:cia:91,hna:wah:91,hna:rt:91,hna:bo:91,hna:fs:91}.

The present paper is mostly based on detailed pen-and-paper proofs
of~\cite{CM24}, that are mainly derived from~\cite{eg:FE1:21}.
The parts on algebra and affine geometry are mainly inspired by
B.~Gostiaux~\cite{gos:cms1:93,gos:cms4:95}.

\myskip

In reference works on~FE, such as~\cite{eg:FE1:21}, the building of simplicial
Lagrange~FE can be based on more powerful tools than strictly necessary, such
as differential calculus, {\eg} to characterize affine functions as those with
zero second order derivative.
This is mainly because those tools are put to good use afterwards to build
other~FE in more complex settings (such as Raviart–Thomas~FE), or for the
computation of integrals on the~FE.
But in fact, simplicial Lagrange~FE are essentially a matter of affine
properties, and it is possible to work with almost no tools of analysis, and to
employ purely affine arguments, only based on barycenters, for almost all steps
of their construction.
And this is done without having to deal with the linear part of affine objects,
or explicitly going down to the affine coordinates, except for a few very
simple cases.
Hence the consequent corpus on affine algebra developed in
Section~\ref{sec:AffineAlgebra}.
However, derivation remains a convenient way to establish the linear
independence of monomials in the space of {\multivariate} polynomials of given
maximum total degree, but this is no affine property.
And obviously, in the next steps, when using these~FE or extending to other~FE,
notions such as integration, differentiation, and normed spaces will become
necessary.

\subsection{Reference and Current Geometric Elements, Geometric Transformation}
\label{sec:ref-cur-fe-geomap}

In this paper, the vector quantities in mathematical expressions are written
with {\bf bold type}.
For instance, this is the case of points
\Notation{$\xx$}$\eqdef(x_0,\dots,x_{d-1})$ in~$\Rd$.

To solve a partial differential equation on an open bounded connected domain
$\Omega\subset\Rd$ by the~FEM, one needs a \emph{mesh} of~$\Omega$, see an
example in Figure~\ref{fig:aneurysm}.
The mesh is a collection of closed subsets of~$\Omega$ called \emph{mesh cells}
or \emph{geometric elements}~$K$, having nonempty interior, disjoint interiors,
and covering~$\Omegab$.
Other requirements are often made on meshes, see \AppSection~\ref{sec:mesh}.
In most cases, the mesh is made up of triangles or quadrangles ($d=2$), or
tetrahedra or hexahedra ($d=3$).

In the sequel, we only consider a single geometric element $K$ where the~FE is
defined.
We assume that this mesh cell $K$, often called the \emph{current} cell, is the
image of a \emph{reference} polyhedron~$\hK$ (such as the standard simplex,
also called here reference simplex) by a smooth invertible function, which we
call geometric mapping.
Let~$(\hvv_i)_{i\in[0..\nvtx)}$ be the~$\nvtx$ vertices of~$\hK$,
and~$(\vv_i)_{i\in[0..\nvtx)}$ be the ones of~$K$, where for all~$m,n\in\N$,
the notation~\Notation{$[m..n)$} represents the interval of integers
$[m,n)\cap\N$.
Then, the geometric mapping is often set as a combination using~$\nvtx$ shape
functions,
\begin{equation}
  \label{eq:def-geomap}
  \Tgeomv (\hxx) \eqdef \sum_{i \in [0..\nvtx)} \refLdi_i (\hxx) \vv_i,
\end{equation}
where~$\refLdi_i$ is a polynomial in~$\Rd$ that satisfies for all
$i,j\in[0..\nvtx)$, $\refLdi_i(\hvv_j)=\delta_{ij}$, with~$\delta$ being the
Kronecker delta function (see Section~\ref{sec:LagPd1-ref} in the simplicial
case).
Thus, for all $i\in[0..\nvtx)$, $\Tgeomv(\hvv_i)=\vv_i$.

When the geometric mapping is affine, the current cell~$K$ is a polyhedron.
See an example of an affine geometric mapping in Figure~\ref{fig:Tgeom}
(p.~\pageref{fig:Tgeom}).

\subsection{The Finite Element as a Triple}
\label{sec:fe-triple}

A~FE depends on three components: a geometric element~$K$, a discrete
approximation space~$P$ where the approximated solution~$\restr{\uh}{K}$ is
sought (typically a finite dimensional polynomial space), and a finite set of
linear forms~$\SSigma$ that specify the discrete quantities that are primarily
computed, and that are used to reconstruct~$\restr{\uh}{K}$.
The main property of a~FE is that the values of these linear forms uniquely
determine an approximation function in~$P$.

The construction of a~FE relies on several values.
The integer~\Notation{$d$} is the spatial dimension of the domain.
The unknown function~$\uh$ can be either scalar or vector, {\ie} belonging
to~\Notation{$\FRdRq$}, the space of functions from~$\Rd$ to~$\R^q$, where~$q$
is generally~1 or~$d$.
The approximation space~$P$ is often characterized by a degree~\Notation{$k$},
representing the accuracy of the approximation space, {\eg} the total degree of
polynomial functions in~\eqref{eq:pol-space-Pkd}.
And finally, \Notation{$\ndof$} is the number of linear forms in~$\SSigma$.

Following~\cite{cia:fem:02} and~\cite[Sec.~5.2, pp.~50--52]{eg:FE1:21}, a
finite element is defined as a triple \Notation{$(K,P,\SSigma)$}, where:
\begin{itemize}
\item
  $K$~is the {\em geometric element}, {\ie} a closed bounded subset of~$\Rd$
  with {\nonempty} interior.
  Typically, $K$~is a {\nondegenerate} $d$-polytope (the generalization of
  points, segments, polygons or polyhedra to dimension $d$), or the image of
  such a subset by a regular function.

\item
  $P$~is the {\em approximation space}, {\ie} a {\nonzero}, finite-dimensional
  vector space of functions from~$\Rd$ to~$\R^q$.
  It is meant to be a space of polynomial functions, or of polynomial
  functions composed with a given regular function (see pullback in
  Section~\ref{sec:TFE}).

\item
  $\SSigma\eqdef(\sigma_i)_{i\in[0..\ndof)}$ is the collection of~$\ndof$
  {\em local degrees of freedom}, {\ie}~$\ndof$ linear forms on~$P$, such that
  the application \Notation{$\PhiSSigma$}$%
  \eqdef(p\in P\mapsto(\sigma_i(p))_{i\in[0..\ndof)}\in\R^\ndof)$ is an
  isomorphism (a linear and bijective application between~$P$ and~$\R^\ndof$).
\end{itemize}

The bijectivity of~$\PhiSSigma$ is called {\em unisolvence}.
It is easy to see that unisolvence requires that the dimension of~$P$ is equal
to~$\ndof$.
The unisolvence property means that an approximation function of~$P$ is
uniquely characterized by the~$\ndof$ discrete values~$\UU$ in~$\R^\ndof$ such
that for all $i\in[0..\ndof)$, $\sigma_i(p)=U_i$.
For instance, in the case of nodal~FE, see Section~\ref{sec:conform:FE}, the
values at the nodes defines the approximation polynomial.

\myskip

Depending on the hypotheses on the geometric transformation~$\Tgeomv$ from the
reference to the current element ({\eg} see~\eqref{eq:def-geomap}), $K$~can be
a $d$-polytope, {\ie} with planar hyperfaces, or not, and can be convex, or
not.
Not all geometric elements are possible, though.
In this document, we assume that~$\Tgeomv$ is invertible and affine.
The {\nonaffine} case is outside the scope of the present study because it has
to be handled differently.
For instance, for a function $p\in P$, the function $p\circ\Tgeomv$ is not
in~$P$ in general, and this creates a discrepancy between the reference and
current elements.
Note that the geometric mapping is affine if and only if each of its components
is in~$\matPdi$.

In this work, the~FE triple is specified in Section~\ref{sec:FE:Coq}, using the
hypothesis of an affine geometric transformation, thus~$K$ is assumed to be a
$d$-polytope.

\subsection{A Finite Element, {$\FElagPdk$}}
\label{sec:conform:FE}

Let us focus on an important example, both for its simplicity and usage, the
simplicial Lagrange FE~\Notation{$\FElagPdk$}.
This~FE is one of the most commonly used for computations, as one can mesh
complex geometries with simplices, and the representation of polynomials by
nodal values is easy to use (for visualization or interpolation for instance).
The formal definition of~$\FElagPdk$ is the object of
Sections~\ref{sec:geometry}, \ref{sec:poly-space-Pdk}, and~\ref{sec:FELagP}.
Recall that the vector quantities are written with {\bf bold type}, such as the
{\multiindex}~$\aalpha$, the monomial function~$\XX^\aalpha$, and the family of
linear forms~$\SSigma^{\FElagPdk}$ that is defined below.

Let~$K$ be a simplex of~$\Rd$.
Assume that~$K$ is {\nondegenerate}, {\ie} its vertices are affinely
independent.
Let~$k\in\N$.
Let the {\multiindex} set be
\begin{equation}
  \label{eq:multi-ind-Adk}
  \Notation{$\calAdk$} \eqdef
  \left\{ \aalpha \in \N^d \st \len{\aalpha} \leq k \right\},
\end{equation}
where \Notation{$\len{\aalpha}$}$\eqdef\sum_{i\in[0..d)}\alpha_i$.
For~$\aalpha\in\calAdk$, let the monomial function of multi-degree~$\aalpha$ be
\begin{equation}
  \label{eq:X:alpha}
  \textstyle
  \Notation{$\XX^\aalpha$} \eqdef
  \left( \xx \in \Rd \longmapsto
  \prod_{i \in [0..d)} x_i^{\alpha_i} \in \R \right).
\end{equation}
The space of polynomials of degree at most~$k$ with~$d$ variables is defined by
\begin{align}
  \label{eq:pol-space-Pkd}
  \Notation{$\matPdk$}
  &\eqdef \Span{\XX^\aalpha}_{\aalpha \in \calAdk}\\
  \nonumber
  &=\textstyle \left\{
      \left( \xx \longmapsto
        \sum_{\aalpha \in \calAdk} c_\aalpha x_1^{\alpha_1} \ldots x_d^{\alpha_d}
      \right) : \ArRdR
    \ \right|
    \left.
      \vphantom{\left( \sum_{\aalpha \in \calAdk} \right)}
      (c_\aalpha)_{\aalpha \in \calAdk} \in \R
    \right\}.
\end{align}
Let~$(\aa_{\aalpha})_{\aalpha\in\calAdk}$ be pairwise distinct points in~$K$,
called the \emph{nodes} of~$K$ (fully defined in Section~\ref{sec:LagP-nodes}).
Let \Notation{$\SSigma^{\FElagPdk}$}%
$\eqdef\left(\sigma_\aalpha\right)_{\aalpha\in\calAdk}$ be the family of linear
forms on~$\FRd$ such that
\begin{equation*}
  \forall \aalpha \in \calAdk,\;
  \forall p \in \FRd,\quad
  \Notation{$\sigma_\aalpha$} (p) \eqdef p (\aa_{\aalpha}) \in \R.
\end{equation*}
The linear forms represent somewhat the type of unknowns one wants to compute.
In this case of \emph{nodal}~FE, a function is approximated by a discrete
function that is represented by its values at the nodes.
See other~FE linear forms in \AppSection~\ref{sec:FE:zoo}.

The triple $(K,\matPdk,\SSigma^{\FElagPdk})$ is the FE~$\FElagPdk$.
In the~FEM with~$\FElagPdk$, the approximated solution~$\uh$ is such that its
restriction to~$K$, $\restr{\uh}{K}$, lies in~$\matPdk$, and $\restr{\uh}{K}$
is entirely defined by its nodal values
$(\uh(\aa_{\aalpha}))_{\aalpha\in\calAdk}$, see unisolvence below.

\myskip

Note that the term ``Lagrange'' can be used for several distinct notions.
First, given~$n+1$ distinct points~$(\aa_i)_{i\in[0..n]}$ in~$\Rd$, the
\emph{Lagrange polynomials}~$(\calL_i)_{i\in[0..n]}$ denote the family of
polynomials that satisfies $\calL_i(\aa_j)=\delta_{ij}$.
Under some hypotheses, they constitute a basis of some finite dimensional
polynomial space, called the \emph{Lagrange basis}, see Section~\ref{sec:P1k}.
Then, in the context of~FE, the points~$(\aa_i)_{i\in[0..n]}$ are called
\emph{Lagrange nodes}.
Finally, the \emph{Lagrange finite element} is~$\FElagPdk$ or~$\FElagQ{d}{k}$,
according to the shape of~$K$ (either a $d$-simplex or a $d$-cuboid, see
\AppSection~\ref{sec:FE:zoo}).

Note also that in the literature, the term ``finite element'' often refers
to different notions.
The genuine meaning refers to a triple~$(K,P,\SSigma)$ that specifies how
functions are approximated (in~$P$) and represented (with~$\SSigma$) on a
cell~$K$.
The mesh cell~$K$ can be called \emph{(geometric) finite element}, or simply
\emph{element}.
When~$K$ and~$\SSigma$ are unambiguous, the approximation space~$P$ is often
assimilated with the~FE.
For instance, the Lagrange FE~$\FElagPdk$ is often denoted in the literature
as the polynomial space~$\matPdk$.
Finally, the \emph{finite element method} refers to a methodology to
solve some partial differential equations, as described in
\AppSection~\ref{sec:FEM:Poisson}.

In this article, we try and be as clear as possible, and distinguish the
different notions.
Thus, the~FE always refers to the triple.
In this respect, $\FElagPdk$ denotes the $d$-simplicial Lagrange~FE of
degree~$k$, whereas~$\matPdk$ denotes the space of polynomials of total degree
at most~$k$.

\myskip

\noindent
{\scriptsize\em
  [We recall that Sections~\ref{sec:Libs}, \ref{sec:Algebra},
  and~\ref{sec:AffineAlgebra} may be skipped if one is familiar with {\Rocq}
  and trusts the naming of the {\Rocq} definitions and lemmas.]}

\section{{\Rocq} Support Libraries}
\label{sec:Libs}

As stated in Section~\ref{sec:Intro:gen}, this work fits in a larger framework
of formally verified real analysis results and programs, mainly with {\Rocq}.
A full description of the {\Rocq} system is beyond the scope of this paper.
We just remind that it is a proof assistant, based on the calculus of inductive
constructions~\cite{Paulin-hdr,Werner-phd}, and therefore on type theory and
intuitionistic logic by default.

Without going into details, we just describe in Section~\ref{sec:Rocq-syntax} a
few {\Rocq}-specific keywords that are often used.
Moreover, for readers unfamiliar with the concrete syntax of \Rocq, we explain
unusual notations as and when required.
Then, Sections~\ref{sec:real_numbers} and~\ref{sec:Mathcomp} present the
{\Rocq} libraries that we use, and Section~\ref{sec:Classical} lists the axioms
we rely on.

\subsection{Focus on Some Specific Syntax of \Rocq}
\label{sec:Rocq-syntax}

A salient point is that we make extensive use of implicit arguments.
As mentioned in the {\Rocq} reference manual, ``An implicit argument of a
function is an argument which can be inferred from contextual knowledge''.%
\footnote{\url{https://rocq-prover.org/doc/V8.20.0/refman/language/extensions/implicit-arguments.html}}
For example, if a function input has type~$\Rd$, then the integer~$d$ can be
implicit.
In \Rocq, implicit arguments are enclosed in curly braces, as we can see for
example in the definition of \coqelkConcat{concatnF} and described at the end
of Section~\ref{sec:Monoid}.

Curly braces are not just for implicit types.
They are also used to distinguish between two \emph{or} and two \emph{exists}.
For example, \coqe!({ A } + { B }) : Set! and \coqe{(A \/ B) : Prop} both mean
``\coqe{A} or~\coqe{B}''.
The difference lies in the computational content of the first one.
Similarly, \coqe!{ x | A x } : Set! and \coqe{(\exists x, A x) : Prop} both
mean that ``there exists an~\coqe{x} such that~\coqe{A x} holds''.
Most of the time, we do not need any computational content, except to decide on
an equality or to use a witness in a statement.
See, for example, Sections~\ref{sec:real_numbers}, \ref{sec:Classical}
and~\ref{sec:FunSub}.

In {\Rocq}, there are several ways of declaring types and functions, depending
on their intended use.
In the following, we briefly outline the differences between the main keywords
we are using.
\begin{itemize}
\item \coqe{Inductive} allows defining types inductively, with possible
  recursive calls and the generation of an induction scheme.

\item \coqe{Variant} is similar to {\tt Inductive} but without recursive calls,
  and no induction scheme is generated.

\item \coqe{Definition} defines a term in the global environment.

\item \coqe{Fixpoint} allows defining functions by pattern matching over
  inductive objects using a fixed point construction.

\item \coqe{Let} is similar to {\tt Definition} but is for local definitions.
  When the section is closed, all persistent definitions and theorems within it
  that depend on the constant are wrapped with a let with the same declaration.

\item \coqe{Hypothesis} and \coqe{Variable} are synonymous, they postulate in
  the environment of a section that a type is inhabited (it can be a property).

\item \coqe{Context} declares variables in the context of the current section,
  like {\tt Variable}, but also allows implicit variables.
\end{itemize}

\subsection{{\Reals} and {\Coquelicot} Libraries}
\label{sec:real_numbers}

Since we are formalizing real analysis dedicated to problems in the field of
numerical analysis for and by classical numerical analysts, all our statements,
properties and theorems are classical (in the sense of classical logic, and not
intuitionistic logic).

For real numbers, we use the \Coquelicot~\cite{BLM15} library.
This library is an extension of the {\Reals}~\cite{May01} standard library.
The {\Reals} library was developed in the early 2000s.
It uses a classical axiomatization.
In particular, it contains a decidability property of the total order and a
multiplicative inverse operation represented as a total function,
\begin{lstlisting}
total_order_T : \forall r1 r2 : \R, { r1 < r2 } + { r1 = r2 } + { r2 < r1 }.
Rinv : \R\sp -> \R.
Rinv_r : \forall r, r <> 0 -> r * Rinv r = 1.
\end{lstlisting}
This part of the standard library provides a number of definitions and
properties of analysis such as limit, univariate derivative, and Riemann
integral.
In the early 2020s, {\Reals} was modified~\cite{Sem20} to include a
constructive common part, inspired by \CoRN~\cite{GeuNiq00}, but the backwards
compatibility of the classical part was retained and is what we and
{\Coquelicot} rely on.

\myskip

{\Coquelicot} was developed in the mid-2010s and provides a hierarchy of nested
algebraic structures, see~\cite{BLM15} and Section~\ref{sec:AlgStruct} for more
details, and~\cite{Mahboubi13} about {\Rocq} canonical structures.
By assuming functional extensionality, we provided canonical structures so that
functions fit into these algebraic structures.
This was done in our development for the proof of {\theLMt}, see~\cite{BCF17}.
We rely on the hierarchy of algebraic structures of the {\Coquelicot} library
at the base of which we contributed to add the commutative monoid structure
\coqe{AbelianMonoid} (available from version~3.4.0).
We also worked on {\substructure}s, addressed in
Section~\ref{sec:AlgSubstruct}.

\myskip

In the future (see Section~\ref{sec:Persp}), we will need floating-point
numbers and we will use the {\Flocq} library~\cite{BolMel11}, which is also
built on top of the {\Reals} standard library.

\subsection{{\MathComp} Library}
\label{sec:Mathcomp}

We also use parts of the Mathematical Components
(\MathComp)~\cite{MathComp_Ref} library and a small part of its analysis
companion (\MathCompAnalysis)~\cite{ACMRS18}, and some
\SSReflect~\cite{SSReflect_Ref} syntax, although most of our formalization is
``vanilla \Rocq''.
For example from {\MathComp}, we use the injective, bijective and involutive
properties from {\SSRfun}.
We use the {\MathComp} notations \Notation{\coqe{n.+1}} for the Peano
successor~\coqe{S n} and \Notation{\coqe{n.-1}} for the predecessor.
We also use the finite dependent type \coqe{ordinal}, ``bigops'', and binomial
coefficients (see Section~\ref{sec:Algebra}).
From {\MathCompAnalysis}, the only part we use is \soft{boolp}, for logical
results and axioms related to classical logic (see Section~\ref{sec:Classical}
below).
We explain other notations specific to {\SSReflect} and {\MathComp} when
required.

Among the {\MathComp} developments, there is a formalization of
multinomials.%
\footnote{\url{https://github.com/math-comp/multinomials/tree/2.4.0}}
Regrettably, there are not enough results on {\multivariate} polynomials for
our purpose. In particular the freedom of the multinomials of bounded total
degree is missing.
We found easier to rely on the {\multiindices} of Section~\ref{sec:MultiIndex},
which are also used for the Lagrange nodes in Section~\ref{sec:LagP-nodes},
that allow us to explicit a basis of~$\matPdk$.

As explained in Section~\ref{sec:real_numbers} and in more details in
Section~\ref{sec:AlgStruct}, we rely on canonical structures for nested
algebraic structures.
For instance, lemmas or notations such as~$+$ can be applied at the same time
to monoid elements, vector space elements, and functions from~$\R^d$ to~$\R^q$,
with {\Rocq} being able to instantiate the needed algebraic structure.
Unfortunately, the hierarchies of {\MathComp}/{\MathCompAnalysis} and
{\Coquelicot}, although similar, are incompatible: they cannot both be used in
the same development.
We therefore cannot embrace all of {\MathComp}, nor all of {\MathCompAnalysis}.

\subsection{Classical Aspects}
\label{sec:Classical}

Let us explicit the axioms we rely on, especially classical logic.
Indeed, the formalization and proof of {\theLMt}~\cite{BCF17} has shown us that
avoiding the classical axioms can lead to unnecessarily complicated statements
for this kind of mathematics.
As one objective is to convince mathematicians, that are mainly using classical
logic, we assume classical logic and other useful axioms.
We have also developed in {\RocqNumAnalysis} many easy lemmas dedicated to
logical facts in this framework in the file {\flkLogic}.

We require the {\Reals} standard library~\cite{May01}, which is based on
classical logic.
We require \soft{boolp} from the {\MathCompAnalysis}~\cite{ACMRS18} library,
that provides propositional and functional extensionality axioms, proof
irrelevance, and constructive indefinite description.
We rename the latter \coqelkLogic{ex_EX},
\begin{lstlisting}
Check ex_EX.
  : \forall (A : Type) (P : A -> Prop), (\exists\nsp x : A, P x) -> { x : A | P x }.
\end{lstlisting}
Finally, we require \soft{ClassicalChoice} from the standard library, that
provides (decidable) excluded middle (\coqelkLogic{classic} and
\coqelkLogic{classic_dec}), and a choice axiom,
\begin{lstlisting}
Check choice.
  : \forall (A B : Type) (R : A -> B -> Prop),
      (\forall\nsp x : A, \exists y : B, R x y) -> \exists f : A -> B, (\forall\nsp x : A, R x (f x)).
\end{lstlisting}

\section{Subsets and Algebra}
\label{sec:Algebra}

This section is dedicated to ``algebraic'' developments needed or useful for
the formalization of~FE.
These complements are mainly based on the {\Coquelicot} and {\MathComp}
libraries (see Section~\ref{sec:Libs}), and on our previous {\Opam} package
{\CoqNumAnalysis}.%
\footnote{\url{https://rocq-prover.org/p/coq-num-analysis/1.0.0}}
{\Coquelicot} for its hierarchy of canonical structures representing nested
mathematical algebraic structures, and {\MathComp} mainly for finite types
(``ordinals''), iterated operators (``bigops''), and
binomials.

For the formalization of Lebesgue integration for nonnegative measurable
functions in~\cite{BCF22,BCM23}, we provided features about subsets of a given
type, including support for possibly infinite families of subsets.
In the present work, we need complements about functions, including support for
restrictions of functions to subsets (Section~\ref{sec:FunSub}), and support
for finite families of any type (Section~\ref{sec:FF}).
All related files are now provided in the dedicated {\Opam} package
{\RNASubset}.%
\footnote{\url{https://rocq-prover.org/p/rocq-num-analysis-subset/2.2.0}}

For the formal proof of {\theLMt} in~\cite{BCF17}, we complemented the
{\Coquelicot} library with proofs that the type of functions to some algebraic
structure is an instance of the same canonical structure, and with minimal
support for linear maps.
Similarly, in the present work, we need to further add support for finitely
iterated operations such as sums and linear combinations
(Section~\ref{sec:AlgStruct}), and for the specific case of finite dimensional
linear algebra (Section~\ref{sec:FinDimMS}).
Just as for subsets, all related files are now provided in the dedicated
{\Opam} package {\RNAAlgebra}.%
\footnote{\url{https://rocq-prover.org/p/rocq-num-analysis-algebra/2.2.0}}

\subsection{Functions and Restrictions}
\label{sec:FunSub}

Functions are basic objects we have to deal with, be they exact solutions, or
piecewise polynomial approximations.
We need to deal with the cases where the input/output sets are empty or not.
We also need to handle subsets of function spaces, typically polynomials, and
bijections from/to theses subsets.
Unisolvence (see Section~\ref{sec:fe-triple}) is precisely a bijection between
a vector subspace of functions and a vector space.

The file {\flkFun} provides additional support for functions.
First, depending on whether the input and output types are inhabited or not.
For instance, given a proof~\coqe{H1} that some type~\coqe{T1} is empty,
\coqelkFun{fun_from_empty}~\coqe{H1} provides a witness that the type
\coqe{T1 -> T2} is inhabited, and \coqelkFun{fun_from_empty_is_unit} states
that this functional type is actually a single-element type ({\aka} unit type),
making the previous witness unique.

Then, based on \soft{Rocq.ssrfun} (where \coqe{bijective f} is an inductive type
taking the proofs that some function is the left and right inverse
of~\coqe{f}), support for surjective functions is added, and additional support
for injective and bijective functions is provided.

Finally, we provide support for the inverse of bijective functions.
For instance, the definition of the inverse \coqelkFun{f_inv} takes as argument
a proof of bijectivity~\coqe{Hf} of some function~$f$ and extracts the inverse
from \coqelkFun{bij_EX}~\coqe{Hf} that states the strong existence of a
left-and-right inverse for~$f$ (unsurprisingly, the proof of the latter needs
\coqelkLogic{ex_EX} to reify the inductive weak existential into a strong one
in~\coqe{Set}).

\myskip

{\flkFunS} provides support for restrictions of functions to subsets,
formalized as predicates (see~\cite{BCF22}), that is similar to the support for
regular functions.
All definitions, such as~\coqelkFunS{injS} and~\coqelkFunS{surjS}, have a name
with suffix~``\coqe{S}'' for ``subset''.
For instance, given a function \coqe{f : T1 -> T2} and subsets
\coqe{P1 : T1 -> Prop} and \coqe{P2 : T2 -> Prop}, the predicate
\coqelkFunS{funS}~\coqe{P1 P2 f := incl (}\coqelkFun{image}~\coqe{f P1) P2}
states that the image of~\coqe{P1} by~\coqe{f} is included in~\coqe{P2}, and
thus that the restriction of~\coqe{f} to both~\coqe{P1} (input) and~\coqe{P2}
(output) actually represents a function.
Similarly to the predicate \coqe{cancel} defined in \soft{Rocq.ssrfun}, given
an additional function \coqe{g : T2 -> T1}, the predicate
\coqelkFunS{canS}~\coqe{P1 f g} states that~\coqe{g} is the left inverse
of the restriction of~\coqe{f} to~\coqe{P1}, {\ie} ``cancels''~\coqe{f}
on~\coqe{P1}.

Note that the definition of bijectivity for a restricted function is slightly
more tricky than for a regular function, as we must ensure that the
restrictions of both the function and its inverse actually represent functions,
and thus satisfy the predicate \coqe{funS}, as shown in the following code
snippet.
\begin{lstlisting}
Definition bijS_spec?\lkFunS{bijS_spec}? : (T2 -> T1) -> Prop :=
  fun g => funS P1 P2 f /\ funS P2 P1 g /\ canS P1 f g /\ canS P2 g f.
Variant bijS?\lkFunS{bijS}? : Prop := BijS_ : \forall (g : T2 -> T1), bijS_spec g -> bijS.
\end{lstlisting}
See Section~\ref{sec:Rocq-syntax} for more details on the inductive-like
\code{Variant} definition.

The lemma \coqelkFunS{bijS_ex_uniq_rev} states that a proper restriction, {\ie}
satisfying the predicate \coqelkFunS{funS}, and such that all elements
in~\coqe{P2} have a unique preimage in~\coqe{P1}, also satisfies the
\coqelkFunS{bijS} predicate.
Its proof needs the use of \coqe{choice} to build the inverse of the
restriction.
The definition of the inverse of a restriction is similar to that of a regular
function, with the same reification of the inductive weak existential into a
strong existential in~\coqe{Set} using \coqelkLogic{ex_EX}.

\subsection{Finite Families}
\label{sec:FF}

Subsets of a given type~\coqe{T} are often represented in an intensional way,
either as a predicate~\coqe{P}\oftypeRR\coqe{T -> Prop}, or as the
corresponding sigma-type \coqe!{ x | P x }! (see
Section~\ref{sec:Rocq-syntax}).
However, it can be convenient to represent subsets in an extensional way as the
range of an enumerating function~\coqe{f : Idx -> T} for some type~\coqe{Idx}
of indices.
It is then advisable to choose an injective function.
For instance, this is the choice made with a finite type~\coqe{Idx} in
Section~\ref{sec:MultiIndex} for the collection~$\calAdk$ of {\multiindices}
see~\eqref{eq:multi-ind-Adk} in Section~\ref{sec:conform:FE}.

We also need support for arrays of items of the same type, where repetition of
the same value is not an issue.
We call them \emph{finite families}, or \emph{$n$-families} to emphasize their
length, be they actual vectors or points in~$\R^n$, or {\multiindices}
in~$\N^n$.
Rather than tuples, we also choose to use enumerating functions from a finite
type as above and share the same developments.

This can seem (and sometimes is) a burden to handle finite families of a given
size, with possibly casts between equal size values.
Nevertheless, it is related to safety proofs and preventing out-of-array
accesses.
We ensure the size of each family/vector, even if handling this size means
additional proofs.
Moreover, we sometimes explicitly need the size of the family, as in
Section~\ref{sec:Adk}.

Given a type~\coqe{E} and some length \coqe{n : \nat}, the type of
\coqe{n}-families on~\coqe{E} is \coqe{[0..n) -> E},
denoted~\Notation{\coqe{'E^n}}, where the type \coqe{ordinal n} (from
\soft{fintype} in the {\MathComp} library), denoted here
\Notation{\coqe{[0..n)}}, represents the interval $[0..n)\subset\N$.
The type~\coqe{[0..0)}, denoted~\Notation{$\emptyset$}, being empty, \coqe{'E^0}
is a unit type (\coqelkFF{hat0F_unit}).
For {\nonzero} length, \coqe{[0..n.+1)} is denoted \Notation{\coqe{[0..n]}},
and then, \coqe{ord0} and~\coqe{ord_max} from {\MathComp} respectively
represent the bounds~0 and~$n$.

{\flkFF} provides support for structural operations on finite families where
most operators are based on transformations on ordinals provided in {\flkOrd}.
All definitions about finite families have a name containing~``\coqe{F}'' for
``family'', usually as suffix.
Let us describe some of the most useful operators.
The operator \coqelkFF{constF} builds a constant family, the binary predicate
\coqelkFF{eqAF} states equality of items for all indices, and \coqelkFF{inclF}
states that all items belong to a given subset.

Moreover, we are interested in \subfamilies: \coqelkFF{liftF_S} removes the
first item,
\coqelkFF{skipF}\oftypeRR\coqe{[0..n] ->}\linebreak[0]\coqe{'E^n.+1 ->'E^n} removes
the specified item, it serves as a projector from~\coqe{'E^n.+1}
onto~\coqe{'E^n}, and given a proof \coqe{H : n1 <== n2},
\coqelkFF{widenF}~\coqe{H : 'E^n2 -> 'E^n1} only keeps the~\coqe{n1} first
items from \coqe{n2}-families.

Finally, \coqelkFF{insertF}~\coqe{: [0..n] -> E -> 'E^n -> 'E^n.+1} inserts the
specified value at the specified location, it serves as an injection
from~\coqe{'E^n} to~\coqe{'E^n.+1}.
Given a function \coqe{f : E -> F}, \coqelkFF{mapF}~\coqe{f : 'E^n -> 'F^n}
applies~\coqe{f} to all items of the argument \coqe{n}-family on~\coqe{E},
\coqelkFF{to_listF} and \coqelkFF{of_listF} provide bijections between types
\coqe{\exists n, 'E^n} and \coqe{list E}, and
\coqelkFF{gather}~\coqe{: '(E -> F)^n -> E -> 'F^n} transforms a family of
functions into the function that gathers all the results.

Now, let us discuss casts.
Given a proof \coqe{H : n1 = n2}, \coqelkFF{castF}~\coqe{H} casts an
\coqe{n1}-family into an \coqe{n2}-family.
This may seem silly from a mathematical viewpoint but it has often proved
necessary.
Many operators on finite families operate through casts.
For instance, given a proof \coqe{H : n = n1 + n2} and an
\coqe{n}-family~\coqe{A}, \coqelkFF{firstF}~\coqe{(castF H A)} extracts the
first items of~\coqe{A} into an \coqe{n1}-family.
And given \coqe{A1 : 'E^n1}, \coqe{A2 : 'E^n2} and a proof
\coqe{H : n1 + n2 = n}, \coqe{castF H (}\coqelkFF{concatF}~\coqe{A1 A2)}
concatenates the two families into an \coqe{n}-family.

More operators on finite families are presented in the next section, in the
context of some algebraic structure.

\subsection{Operators in Algebraic Structures}
\label{sec:AlgStruct}

\newcommand{\newC}{darkblue}
\newcommand{\txtnewC}[1]{\textcolor{\newC}{#1}}

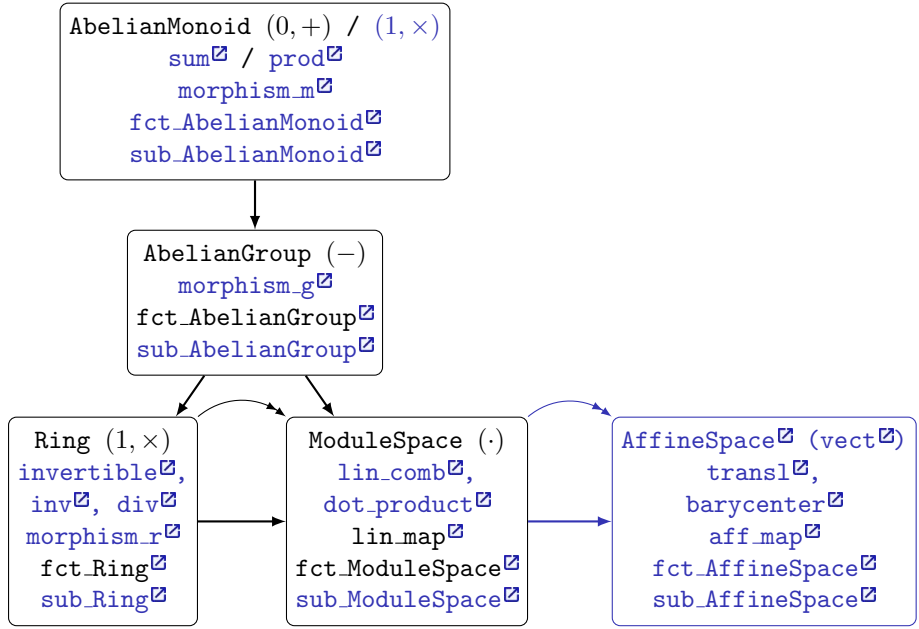
\begin{figure}[ht]
  \centering
  \begin{tikzpicture}
    \node[draw, rounded corners=3pt] (AM) at (4,7.2)
      {\minitab[c]{%
          \tt AbelianMonoid $(0,+)$ / \txtnewC{$(1,\times)$}\\
          \tt \txtnewC{\LLKM{Monoid_sum}{sum}} /
          \txtnewC{prod\lkProd{prod\_nat}}\\
          \tt \txtnewC{\LLKM{Monoid_morphism}{morphism\_m}}\\
          \tt \txtnewC{\LLKH{fct\_AbelianMonoid}}\\
          \tt \txtnewC{\LLKM{Monoid_sub}{sub\_AbelianMonoid}}}};
    \node[draw, rounded corners=3pt] (AG) at (4,4.4)
      {\minitab[c]{%
          \tt AbelianGroup $(-)$\\
          \tt \txtnewC{\LLKG{Group_morphism}{morphism\_g}}\\
          \tt \LLKH{fct\_AbelianGroup}\\
          \tt \txtnewC{\LLKG{Group_sub}{sub\_AbelianGroup}}}};
    \node[draw, rounded corners=3pt] (R) at (2,1.5)
      {\minitab[c]{%
          \tt Ring $(1,\times)$\\
          \tt \txtnewC{\LLKR{Ring_compl}{invertible},}\\
          \tt \txtnewC{\LLKR{Ring_compl}{inv}, \LLKR{Ring_compl}{div}}\\
          \tt \txtnewC{\LLKR{Ring_morphism}{morphism\_r}}\\
          \tt \LLKH{fct\_Ring}\\
          \tt \txtnewC{\LLKR{Ring_sub}{sub\_Ring}}}};
    \node[draw, rounded corners=3pt] (MS) at (6,1.5)
      {\minitab[c]{%
          \tt ModuleSpace $(\cdot)$\\
          \tt \txtnewC{\LLKMS{ModuleSpace_lin_comb}{lin\_comb},}\\
          \tt \txtnewC{\LLKMS{ModuleSpace_lin_comb}{dot\_product}}\\
          \tt \LLKMS{ModuleSpace_lin_map}{lin\_map}\\
          \tt \LLKH{fct\_ModuleSpace}\\
          \tt \txtnewC{\LLKMS{ModuleSpace_sub}{sub\_ModuleSpace}}}};
    \node[draw=\newC, rounded corners=3pt] (AS) at (10.7,1.5)
      {\minitab[c]{%
          \tt \txtnewC{\LLKAS{AffineSpace_def}{AffineSpace}
            (\LLKAS{AffineSpace_def}{vect})}\\
          \tt \txtnewC{\LLKAS{AffineSpace_def}{transl},}\\
          \tt \txtnewC{\LLKAS{AffineSpace_baryc}{barycenter}}\\
          \tt \txtnewC{\LLKAS{AffineSpace_aff_map}{aff\_map}}\\
          \tt \txtnewC{\LLKAS{AffineSpace_def}{fct\_AffineSpace}}\\
          \tt \txtnewC{\LLKAS{AffineSpace_sub}{sub\_AffineSpace}}}};
    \draw[->, >=latex, thick] (AM) to (AG);
    \draw[->, >=latex, thick] (AG) to (R);
    \draw[->, >=latex, thick] (AG) to (MS);
    \draw[->, >=latex, thick] (R) to (MS);
    \draw[->>, >=latex, thin]
      (R.north east) to [bend left=40] (MS.north west);
    \draw[->, >=latex, thick, color=\newC] (MS) to (AS);
    \draw[->>, >=latex, thin, color=\newC]
      (MS.north east) to [bend left=40] (AS.north west);
  \end{tikzpicture}
  \caption[Hierarchy of algebraic structures.]{%
    Hierarchy of algebraic structures used in the present development.\\
    Additions to the {\Coquelicot} library (version~3.4.0) and to previous
    work~\cite{BCF17} are in blue.
    The main points are: finite iterations of operators using ``bigops'' from
    \soft{MathComp} (\coqelkSum{sum}/\coqe{prod}\lkProd{prod_R} and
    \coqelkLC{lin_comb}), most morphisms, algebraic {\substructure}s, and the
    affine space structure have been added.
    Downward arrows specify inheritance, horizontal arrows specify parameters,
    and the bent double-headed arrows mention that the input structure is
    shown to be an instance of the output one.
    For instance, any \coqe{ModuleSpace} has the \coqe{AffineSpace} structure
    over itself.
    Specifics of affine algebra are presented in
    Section~\ref{sec:AffineAlgebra}.}
  \label{fig:AlgStruct}
\end{figure}

For mathematical algebraic structures, the present work relies on the
{\Coquelicot}~\cite{BLM15} hierarchy of canonical structures~\cite{Mahboubi13},
including the \coqe{AbelianMonoid} structure that we recently added to the
library.
Figure~\ref{fig:AlgStruct} depicts the part of the hierarchy needed for~FE,
where new features are in blue.
Support for functions to algebraic structures, denoted \coqe{fct_<Alg_Struct>},
such as \coqelkH{fct_ModuleSpace}, is provided in {\flkH}, and almost all other
features have their own file such as {\flkSum} and {\flkLM}.
Our inspiration mainly comes from the mathematical book~\cite{gos:cms1:93}.

\subsubsection{Operators in Monoids}
\label{sec:Monoid}

We first describe some complements about the \coqe{AbelianMonoid} structure,
taking advantage of the fact that the structure contains a zero.
For instance, the notation
\coqelkMonF{insert0F}~\coqe{:= fun i => insertF i 0} serves as an injection
from~\coqe{'G^n} to~\coqe{'G^n.+1}.
We define the new operator on families
\coqelkMonF{itemF}\oftypeRR\coqe{\forall (n : \nat), G -> [0..n) -> 'G^n} that
outputs the family made of zeros except a given value in a given position.
We also prove \coqelkMonF{hat0F_eq}, that states that~\coqe{'G^0} is the unit
type with the single value~$0$ (as it is a monoid).

\myskip

Let us now take advantage of the monoid law.
The introduction of the commutative monoid structure opens the way to define
finite sums for both~\coqe{nat} and~\coqe{R}, but also finite products, and
even finite composition of functions (dropping commutativity).
Technically, this is implemented using ``bigops'' from {\MathComp} that
finitely iterate an operator that is assumed to be associative, with an
identity element, and possibly commutative, {\ie} endowing the underlying type
with a, possibly commutative, monoid structure.
Note that this is not fully automatic since {\Coquelicot} and {\MathComp}
hierarchies are not compatible, although they both represent the same
mathematical objects.
A vanilla-{\Rocq} interface is provided through wrappers around the main
results from \soft{bigop}, and the interaction with most operators on families
is provided.
For instance, \coqelkSum{sum_ind_l} states that the sum of a family is equal to
the sum of its first element and the sum of the rest of the family, it is
proved using \coqe{big_ord_recl}, and alike, \coqelkSum{sum_skipF}, that
decomposes the sum into any item and the rest, is similar to \coqe{bigD1}.

In practice, we need both~$(\N,+)$ and~$(\N,\times)$ to be monoids.
The {\Coquelicot} structure \coqe{Abelian}\linebreak[0]\coqe{Monoid} is fully
abstract, but the identity element is called \coqe{zero}, the operator
\coqe{plus}, and in everybody's mind, this has an additive connotation.
Moreover, only one instance of a given canonical structure can be assigned to a
given type.
Nonetheless, we define aliases \coqelkProd{nat_mul}~\coqe{:= \nat} and
\coqelkProd{R_mul}~\coqe{:= \R} in {\flkProd}, provide two-way coercions, and
endow these ``new types'' with a multiplicative \coqe{AbelianMonoid} structure.
Then, the specific products \coqelkProd{prod_nat} and \coqelkProd{prod_R} are
simply defined as the generic \coqelkSum{sum}, and all results are available.
Only a few are supplied with a dedicated name.

The iterated composition of functions is similar, with the identity function as
the identity element, but the underlying monoid structure is not commutative.
Hence, it cannot be represented by an \coqe{AbelianMonoid} structure,
and~\coqe{sum} cannot be recycled a second time.
Thus, we provide in {\flkComp} the definition of \coqelkComp{comp_n} as another
``bigop'' from {\MathComp} and with a similar vanilla-{\Rocq}-like interface.
Results about \coqe{comp_n} are mainly about the preservation of morphisms
(that transports algebraic laws).
For instance, \coqelkComp{comp_n_mm} states that the iterated composition of
monoid morphisms is a monoid morphism (where~``\coqe{mm}'' stands for
``monoid morphism'').
It is also deployed for morphisms of other algebraic structures.

\myskip

Finally, using the sum (denoted here~\Notation{\coqe{\\sum}}) for natural
numbers, support for the concatenation of (not necessarily rectangular)
families of families is provided in {\flkConcat}.
Given an $n$-family~\coqe{b} of integers and given an $n$-family of
families~\coqe{A} such that~\coqe{A i} has size~\coqe{b i}, we construct the
flattened family of the elements in the expected order, that is of
size~\coqe{\sum\sp b}.
This is obtained through the concatenation of lists (from the standard {\Rocq}
library) and family-to-list transformations (see Section~\ref{sec:FF}),
\begin{lstlisting}
Definition concatnF?\lkConcat{concatnF}? {E : Type} {n} {b : '\nat^n} : (\forall\nsp i, 'E^(b i)) -> 'E^(\sum b) :=
  fun A => castF _ (of_listF (List.concat (to_listF (fun i => to_listF (A i))))).
\end{lstlisting}
where the proof that the concatenated family is actually of length
\coqe{\sum\sp b} is omitted here for legibility, and simply represented by the
unspecified argument~``\coqe{_}'', and the expressions between curly
braces~\verb+{ }+ represents implicit arguments (see
Section~\ref{sec:Rocq-syntax}).
Intuitively, this corresponds to block concatenation, that
will also be useful for block matrices (used in the future for the~FEM,
more in \AppSection~\ref{sec:lin:system}).
Main results about \coqe{concatnF} include \coqelkConcat{concatnF_sortedF}
which states the conditions so that the concatenation of sorted families is
also sorted (that are easier to state and prove when the families are
{\nonempty}), and \coqelkConcat{sum_assoc} stating that the sum of a
concatenation is the double sum of the family of families.

\subsubsection{Operators in Rings}
\label{sec:Ring}

{\flkRing} is for additional definitions and results specific to the
\coqe{Ring} structure, having also a one and a multiplication.
For instance, the predicate \coqelkRing{invertible} states that an element of a
ring has a left-and-right inverse for the multiplicative law, and the unary
operator~\coqelkRing{inv} is the total function extracting this inverse when
the argument is indeed invertible, and outputing~0 otherwise, similar to what
is done in the {\Reals} with \coqe{Rinv}, see Section~\ref{sec:real_numbers}.

For barycenters (see Section~\ref{sec:AffineSpace}), we use the family operator
\coqelkRingF{part1F}\oftypeJ\coqe{[0..n] -> 'K^n -> 'K^n.+1} that
inserts ``one minus the sum'' at the specified location, such that the output
is of sum~1 (thus creating somehow a partition of unity, hence its name).
The notation \coqelkRingF{part1F0}~\coqe{:= part1F ord0} serves as an injection
from~\coqe{'K^n} onto the subset of~\coqe{'K^n.+1} of elements of sum~1.

Note that a singleton, usually denoted~\Notation{$\{0\}$}, can be endowed with
a ring structure where identity elements for both additive and multiplicative
laws are the same.
Some results only hold on specific rings.
Thus, we also provide predicates such as \coqelkRing{is_field} for {\nonzero}
commutative rings in which all {\nonzero} elements have an inverse, {\aka}
fields, and \coqelkRChar{is_field_not_charac_2} for fields in which~$1+1$ is
distinct from~0 ({\eg} to be able to define the middle of two points).

The Kronecker delta function needs (distinct)~0 and~1, hence it is defined in
rings, in {\flkKron}, and some results only hold on {\nonzero} rings.
This should be generalized to \emph{semirings} (in which the existence of an
additive inverse for all elements is dropped), and thus would also apply
to~\coqe{\nat}.
But such an algebraic structure is not yet present in {\Coquelicot}'s
hierarchy.

\subsubsection{Operators in Module Spaces}
\label{sec:ModuleSpace}

{\flkMS} collects additional definitions and results about module spaces, {\ie}
vector spaces with a ring of scalars (not necessarily a field).
The scalar multiplication of a vector is called~\coqe{scal}.
Since there is no canonical structure for either commutative rings or fields in
{\Coquelicot}'s hierarchy yet, we choose to state results needing that
assumptions only for vector spaces over the field we use here, namely the real
numbers.

Given a module space~\coqe{E} on some ring~\coqe{K}, given an $n$-family of
scalars \coqe{a : 'K^n} and an $n$-family of vectors \coqe{u : 'E^n},
\coqelkMSF{scalF}~\coqe{a u} is the $n$-family with items
\coqe{scal (a i) (u i)}, and its sum is the linear combination
\coqelkLC{lin_comb}~\coqe{a u}.
For instance, results about \coqe{lin_comb} include \coqelkLC{lc_nil} stating
that the linear combination of the empty family is~0, and
\coqelkLC{lc_filter_n0F_l} stating that the linear combination remains the same
when filtering out zero scalars, or vectors (with suffix~\coqe{_r}).
Note the common prefix~``\coqe{lc_}'', that stands for ``linear combination''.

More results on finite-dimensional module spaces are given in
Section~\ref{sec:FinDimMS}.

\subsubsection{Predicates for Algebraic {\Substructure}s}
\label{sec:AlgSubstruct}

Given a type~\coqe{E} equipped with some algebraic structure, subsets
of~\coqe{E} that are closed under the laws associated with the structure are of
particular interest, such as polynomials being a vector subspace of the vector
space of functions.
This is declined for each algebraic structure in the dedicated file
\soft{$<$Alg\_Struct$>$\_sub}.
As a general naming rule, \coqe{<op>_closed} is a predicate stating closedness
under some operator~\coqe{<op>}, such as \coqelkMonS{plus_closed} and
\coqelkMonS{sum_closed} in abelian monoids, and \coqelkMSS{lc_closed} in module
spaces.
Then, \coqe{compatible_<s>} expresses closedness under all operations
constitutive to the algebraic structure corresponding to the
shorthand~``\coqe{<s>}'': ``\coqe{m}'', ``\coqe{g}'', ``\coqe{r}'',
and~``\coqe{ms}'', respectively for \coqe{AbelianMonoid}, \coqe{AbelianGroup},
\coqe{Ring}, and \coqe{ModuleSpace}.
Associated statements have~``\coqe{c<s>}'' in their name (where~``\coqe{c}''
stands for ``compatible'').
For instance, \coqelkMonS{compatible_m} states both \coqe{plus_closed} and
\coqelkMonS{zero_closed} ({\ie} contains~0), and \coqelkMSS{cms_lc} states that
\coqelkMSS{compatible_ms} implies \coqe{lc_closed}.

Compatible subsets can also be equipped with the same algebraic structure.
This is formalized following the representation of subsets by dependent types,
but this ultimately proved unnecessary for the present formalization work.

\subsubsection{Kernel}
\label{sec:Ker}

A subspace of interest is the kernel of a function.
The kernel (or nullspace) of a function can be defined in commutative monoids,
as it only needs a zero (with an additive interpretation of it).
But of course, most interesting properties about injectivity and bijectivity
are only valid in groups, and much more in module spaces.
The kernel is the preimage of the zero subspace~$\{0\}$.
Note that the kernel of the restriction of a function to some subset is the
intersection of the kernel and the subset.
We also provide the predicate \coqe{KerS0} stating that the kernel of a
restriction is the zero subspace, corresponding to the injectivity of the
restricted function (when it is a morphism).
Given \coqe{G1 G2 : AbelianMonoid}, a subset \coqe{PG1 : G1 -> Prop} and a
function \coqe{f : G1 -> G2},
\begin{lstlisting}
Definition Ker?\lkMonS{Ker}? : G1 -> Prop := preimage f (singleton 0).
Definition KerS?\lkMonS{KerS}? : G1 -> Prop := inter PG1 (Ker f).
Definition KerS0?\lkMonS{KerS0}? : Prop := \forall x1, PG1 x1 -> f x1 = 0 -> x1 = 0.
\end{lstlisting}
For a family of functions \coqe{f : '(G1 -> G2)^n}, we provide
\begin{lstlisting}
Lemma KerS0_gather_equiv?\lkMonS{KerS0_gather_equiv}? :
  KerS0 PG1 (gather f) <-> (\forall\nsp x1 : G1, PG1 x1 -> (\forall\nsp i, f i x1 = 0) -> x1 = 0).
\end{lstlisting}
where \coqelkFF{gather} is defined in Section~\ref{sec:FF}, that is used for
the unisolvence property (see Section~\ref{sec:FE-general}).
When~\coqe{f} is a morphism and~\coqe{PG1} is compatible with the algebraic
structure, \coqe{Ker f} and \coqe{KerS PG1 f} are also compatible with the same
algebraic structure.
Moreover, we provide the equivalence between~\coqe{KerS0} and the bijectivity
of the restriction of a linear mapping to module subspaces of the same
dimension.
Given \coqe{E1 E2 : ModuleSpace R_Ring}, \coqe{PE1 : E1 -> Prop},
\coqe{PE2 : E2 -> Prop} and \coqe{f : E1 -> E2},
\begin{lstlisting}
Lemma lmS_bijS_KerS0?\lkFDLMR{lmS_bijS_KerS0}? :
  have_same_dim PE1 PE2 -> lin_map f -> funS PE1 PE2 f ->
  KerS0 PE1 f <-> bijS PE1 PE2 f.
\end{lstlisting}
where \coqelkFunS{funS} and \coqelkFunS{bijS} are defined in
Section~\ref{sec:FunSub} and \coqelkFDMS{have_same_dim} in
Section~\ref{sec:FinDimMS} below.
It is a consequence of the rank--nullity theorem (see also
Section~\ref{sec:FinDimMS}).
Note that this result needs scalars in a field, and is here only stated for the
real numbers.

\subsection{Finite-dimensional Linear Algebra}
\label{sec:FinDimMS}

Although~FEM requires infinite-dimensional spaces for functional spaces, it is
also based on a variety of finite-dimensional vector subspaces,
such as the approximation spaces with their polynomial bases and dual bases.
Files in folder \texttt{Finite\_dim} provide support for module subspaces of
finite dimension.
Again, definitions and results needing a commutative ring of scalars, or a
field, are stated for the field of real numbers only.

Given a module space~\coqe{E} on some ring~\coqe{K}, a subset
\coqe{PE : E -> Prop} and a finite family of vectors \coqe{B : 'E^n},
\coqelkFDMS{lin_span}~\coqe{B} is the linear span of~\coqe{B} ({\ie} the subset
of all linear combinations of~\coqe{B}),
the predicate \coqelkFDMS{lin_gen}~\coqe{PE B} states that~\coqe{B}
spans~\coqe{PE} ({\ie} is a linear generator of~\coqe{PE}),
\coqelkFDMS{lin_indep}~\coqe{B} that~\coqe{B} is linearly independent, or free,
\coqelkFDMS{basis}~\coqe{PE B} that~\coqe{B} is a basis of~\coqe{PE} ({\ie} is
generator and linearly independent), \coqelkFDMS{has_dim}~\coqe{PE n}
that~\coqe{PE} is of dimension~\coqe{n} ({\ie} admits a basis of
length~\coqe{n}), and given another module space~\coqe{F} on the same
ring~\coqe{K} and a subset \coqe{PF : F -> Prop},
\coqelkFDMS{have_same_dim}~\coqe{PE PF} states that~\coqe{PE} and~\coqe{PF}
have the same dimension.
Note that a total function for the dimension with values in~\coqe{\nat} would
not be satisfactory since all values are meaningful (including~0), and hence
there would be no way to discriminate infinite-dimensional subspaces, and
subsets that are not subspaces.
Thus, we opted for a predicate version, which we preferred to an option type.

We provide popular results such as
the dimension theorem (\coqelkFDBR{dim_uniq}),
the rank--nullity theorem (\coqelkFDLMR{rank_nullity_thm}), and
the incomplete basis theorem (\coqelkFDBR{incomplete_basis_thm}, close to the
Steinitz exchange lemma).
Their formalization from the math book~\cite{gos:cms1:93} is quite
straightforward, thanks to the various operators on finite families defined in
Sections~\ref{sec:FF} and~\ref{sec:AlgStruct}.
These results need a field for the scalars, they are stated for vector spaces
on~\coqe{\R}.

\subsection{Binomials}
\label{sec:Binom}

We also need additional results about the type~\coqe{\nat} of Peano natural
numbers.
For instance, we provide several variants of double induction schemes,
including the following well-founded double induction,
\begin{lstlisting}
Lemma nat_ind2_strong?\lkNat{nat_ind2_strong}? : \forall (P : \nat\sp -> \nat\sp-> Prop),
  (\forall\nsp m n, (\forall\nsp m1 n1, m1 <== m -> n1 <== n -> (m1, n1) <> (m, n) -> P m1 n1) -> P m n) ->
  \forall m n, P m n.
\end{lstlisting}

\myskip

In Section~\ref{sec:Adk}, we construct {\multiindices}, that are families whose
size is a binomial.
We therefore need additional definitions and results about binomial
coefficients, based on \soft{binomial} from {\MathComp} where~$\binom{m}{n}$ is
denoted~\Notation{\coqe{'C (m, n)}}.
A useful result is \coqelkBinom{binom_rising_sum_l} stating that
\begin{equation}
  \label{eq:binom-rising-sum}
  \forall m, n \in \N,\quad
  \sum_{i = 0}^n \binom{m + i}{m} = \binom{m + 1 + n}{m + 1}.
\end{equation}

Moreover, binomials are at least~1 when the second (or lower) argument is at
most equal to the other ({\eg} see~\cite{BCH24}), and it may be convenient to
structurally grant that the value is {\nonzero}.
Hence the
\begin{lstlisting}
Definition pbinom?\lkBinom{pbinom}? (m n : \nat) : \nat\sp:= ('C (m + n, m)).-1.
\end{lstlisting}
that returns the natural number $\binom{m+n}{n}-1$.
This is precious in the sequel where the subspace of $d$-variable polynomials
of total degree at most~$k$ is shown to be of dimension~$\binom{d+k}{k}$,
{\ie}~\coqe{(pbinom d k).+1}, thus avoiding technical casts of finite families
when a {\nonzero} size is mandatory.

\section{Affine Algebra}
\label{sec:AffineAlgebra}

As claimed in Section~\ref{sec:math:ref:FEM}, the construction of simplicial
Lagrange~FE is largely a matter of  affine objects such as affine
mappings and affine subspaces ({\eg} hyperplanes), in particular for the
geometric aspects (see Section~\ref{sec:geometry}).
Actually, geometrical transformations are based on the Lagrange polynomials of
degree at most~1 (in~$\matPdi$), that happen to be equal to the barycentric
coordinates, the key ingredient in affine algebra, and this explains the
developments made for~$\matPdi$, see Sections~\ref{sec:LagPd1}
and~\ref{sec:P1k}.

In practice, affine spaces are often already equipped with a vector space
structure, such as the Euclidean spaces~$\Rd$, or more generally the module
spaces~$K^d$ for any ring~$K$.
They behave as if the zero vector had lost its special status, and that any
other vector (then called \emph{point}) could play a similar role.
In this context, affine subspaces are known to be vector subspaces
{\em translated by some vector}, and affine mappings are known to be linear
mappings {\em shifted by a constant}.
Thus, affine algebra can reduce to ``shifted linear algebra'', but the roles of
points and vectors are then confused.

However, the notion of barycenter brings symmetry in formulas, and more
importantly, it is the central concept around which all affine aspects revolve.
For instance, B.~Gostiaux writes in~\cite[Chap.~1, p.~5]{gos:cms4:95}:
``une propriété vraiment affine se traduit par des barycentres''
[a truly affine property translates into barycenters].
Thus, rather than dealing with the linear part of affine objects, we decided to
focus on their true affine nature
following~\cite{gos:cms4:95}, and provide the bases of a full-flavored
affine algebra library.
This vision has the great advantage of separating linear aspects from affine
ones.

Section~\ref{sec:AffineSpace} describes the abstract construction of affine
spaces, the definition and first properties of barycenters and affine mappings,
and the specific case of finite dimension is discussed in
Section~\ref{sec:FinDimAS}.

\subsection{Affine Space and Barycenter}
\label{sec:AffineSpace}

An affine space over a module space~\coqe{V} (the type of vectors) is any
nonempty type~\coqe{E} (the type of points) equipped with an external law
\coqelkAS{vect}~\coqe{: E -> E -> V}, denoted with the infix
symbol~\Notation{\coqe{\\vect}}, so that \coqe{A \vect\sp B} represents the
usual notation~$\vect{AB}$ in~\coqe{V}.
It satisfies the Chasles relation,
\begin{lstlisting}
Lemma vect_chasles?\lkAS{vect_chasles}? : \forall (A B C : E), (A \vect\sp B) + (B \vect\sp C) = A \vect\sp C.
\end{lstlisting}
and is such that \coqe{vect O} is bijective for all point~\coqe{O} (where
\coqe{O} stands for a possible ``origin'' of the affine space).
Equivalently, the type~\coqe{E} can also be equipped with an external law
\coqelkAS{transl}~\coqe{: E -> V -> E}, denoted with the infix
symbol~\Notation{\coqe{\\transl}}, that is ``associative'',
\begin{lstlisting}
Lemma transl_assoc?\lkAS{transl_assoc}? : \forall (A : E) (u v : V), (A \transl\sp u) \transl\sp v = A \transl\sp (u + v).
\end{lstlisting}
and is such that \coqe{transl O} is bijective for all
point~\coqe{O} (the inverse of \coqe{vect O}).
Either \coqe{vect} or \coqe{transl} can be built from the other using the
bijectivity property.

The \coqe{AffineSpace} structure is another entry point in the hierarchy of
algebraic structures.
Indeed, the type~\coqe{E} of points does not need any prior structure.
But it is parameterized by a \coqe{ModuleSpace}, which is itself parameterized
by a \coqe{Ring} (see Figure~\ref{fig:AlgStruct}).
Any module space~\coqe{V} can be equipped with an \coqe{AffineSpace} structure
on itself using subtraction, \coqelkAS{ms_vect}~\coqe{(A B : V) := B - A}, and
then the translation is simply the addition,
\coqelkAS{ms_transl_eq}~\coqe{(A u : V) : A \transl\sp u = A + u} (the
\coqe{ms_} prefix stands for ``module space'').

Symmetrically, any affine space~\coqe{E} on some vector space~\coqe{V} can be
equipped with a \coqe{ModuleSpace} structure by choosing any
origin~\coqe{O : E}, defined as the zero.
Then, addition of points would be
\coqe{fun (A B : E) => O \transl\sp ((O \vect\sp A) + (O \vect\sp B)) : E},
and the scalar multiplication of a point would be
\coqe{fun (x : K) (A : E) => O \transl\sp (scal x (O \vect\sp A)) : E}.
Such laws yield a valid canonical structure, and the resulting module space is
obviously isomorphic to the initial~\coqe{V}.
But this generates a cycle in the instantiation rules of the canonical
structures, that further confuses the typing system in the presence of module
spaces used as affine spaces, and thus should be used with caution.

\myskip

We now describe new operators on finite families specific to affine spaces.
Let~\coqe{V} be a module space on some ring~\coqe{K} and~\coqe{E} be an affine
space on~\coqe{V}.
Given points~\coqe{O : E} and~\coqe{A : 'E^n}, \coqelkASF{vectF}~\coqe{O A} is
the \coqe{n}-family of vectors \coqe{O \vect\sp A i}.
The law~\coqe{transl} can be lifted similarly into \coqelkASF{translF}.
The operators~\coqe{vectF O} and~\coqe{translF O} are inverses of each other,
and they both commute with most operations on finite families.
Given~\coqe{B : 'E^n.+1} and some index~\coqe{i0},
\coqelkASF{frameF}~\coqe{i0 B} is the \coqe{n}-family of vectors
\coqe{vectF (B i0) (}\coqelkFF{skipF}~\coqe{i0 B)}, {\ie}
$(\vect{B_{i_0}B_j})_{j\in[0..n]\setminus\{i_0\}}$.

\myskip

Given scalars~\coqe{L : 'K^n}, points~\coqe{A : 'E^n} and~\coqe{G : E}, the
predicate \coqelkASB{aff_comb}~\coqe{L A G} stipulates that
\coqelkLC{lin_comb}~\coqe{L (vectF G A)} vanishes, which is equivalent to
\begin{equation*}
  \forall O \in E,\quad
  \left( \sum_{i = 0}^{n - 1} L_i \right) \vect{OG}
  = \sum_{i = 0}^{n - 1} L_i \vect{OA}_i.
\end{equation*}
When the scalar~\coqe{\sum\sp L} is invertible ({\ie} {\nonzero} in a field,
see Section~\ref{sec:Ring}), lemma \coqelkASB{ac_uniq_r} (where~``\coqe{ac}''
stands for ``affine combination'') states the uniqueness of such a
point~\coqe{G}, and \coqelkASB{baryc_EX} states its constructive existence.
On the field of real numbers, this reads
\begin{equation*}
  \sum_{i = 0}^{n - 1} L_i \neq 0 \IMPLIES
  \exists ! G \in E,\quad \sum_{i = 0}^{n - 1} L_i \vect{GA}_i = 0.
\end{equation*}

Then, \coqelkASB{barycenter}~\coqe{L A} is a total function extracting that
point~\coqe{G} when \coqe{invertible (\sum\sp L)} holds, and outputing some
point in the inhabited type~\coqe{E} otherwise.
Note that when \coqe{\sum\sp L = 1} and the real scalars~$L_i$'s are
{\nonnegative} (and thus lower than~1), the combination is called
{\em convex combination}.
The coefficients can be called \emph{weights}, with the analogy between the
barycenter and the center of mass of the masses~$L_i$'s located at the
points~$A_i$'s.
When~\coqe{L} is a constant scalar family, the barycenter is called
\emph{isobarycenter}, moreover when there are only two points, it is called the
\emph{middle}.

For instance in~$\Rd$, the middle~$M$ of two points~$A$ and~$B$ can be computed
through $M$\equalsRR$A+\frac{B-A}{2}=\frac{A+B}{2}$.
The first formula uses the coordinate frame $(A,\vect{AB})$ with the single
coordinate~$\frac{1}{2}$.
The second formula is the barycentric view, with two coordinates
$(\frac{1}{2},\frac{1}{2})$ summing up to~1, in which~$A$ and~$B$ play a
symmetric role.

Interesting results include the case of Kronecker weights, the possibility to
skip zero weights, insensitivity to permutations, and commutativity of
barycenters (for a table/matrix of points),
\begin{lstlisting}
Lemma baryc_l_kron_r?\lkASB{baryc_l_kron_r}? : \forall {n} (A : 'E^n) (i : [0..n)), barycenter (kron i) A = A i.
Lemma baryc_skip_zero?\lkASB{baryc_skip_zero}? : \forall {n} i0 L (A : 'E^n.+1), invertible (\sum L) -> L i0 = 0 ->
  barycenter L A = barycenter (skipF i0 L) (skipF i0 A).
Lemma baryc_permutF?\lkASB{baryc_permutF}? :
  \forall {n} {p : [0..n) -> [0..n)} {L} {A : 'E^n}, injective p -> invertible (\sum L) ->
  barycenter (permutF?\lkFF{permutF}? p L) (permutF p A) = barycenter L A.
Lemma baryc2_rr_R?\lkASB{baryc2_rr_R}? : \forall {n1 n2} L1 L2 (A12 : 'E^{n1,n2}), \sum\sp L1 <> 0 -> \sum\sp L2 <> 0 ->
  barycenter L2 (barycenter L1 A12) = barycenter L1 (mapF (barycenter L2) A12).
\end{lstlisting}
The latter needs a field of scalars (stated here for~$\R$);
its proof is quite long and cumbersome, it is based on a similar result that is
much easier to prove for linear combinations (\coqelkMSR{lc2_rr_R}).
In module spaces equipped with their canonical affine space structure over
themselves, when \coqe{\sum\sp L = 1}, \coqelkASB{baryc_ms_eq} states that
\coqe{barycenter L A = lin_comb L A}.
We also provide specialized helpers such as \coqelkASB{barycenter_ms} that
expects points in a module space and not in an affine space.
This is due to an incorrect canonical structures instantiation for functions to
affine spaces.
This is explained in more details in Figure~\ref{fig:diamond-pbs} and
Section~\ref{sec:Diff}.

\myskip

Affine mappings are functions between affine spaces that preserve barycenters
(``\coqe{am}'' in the names of statements stands for ``affine map'').
Equivalently, they are functions~$f:E_1\to E_2$ such that
$\vect{u}_1\in V_1\mapsto\vect{f(O_1)f(O_1+\vect{u}_1)}\in V_2$ is a linear map
for all~$O_1$.
In {\Rocq} (see Section~\ref{sec:Rocq-syntax} for more details on the
\code{Context} keyword),
\begin{lstlisting}
Context {K : Ring}.
Context {V1 V2 : ModuleSpace K}.
Context {E1 : AffineSpace V1}.
Context {E2 : AffineSpace V2}.
Variable f : E1 -> E2.
Variable O1 : E1.

Definition aff_map?\lkAM{aff_map}? : Prop := \forall n L (A1 : 'E1^n),
  invertible (\sum L) -> f (barycenter L A1) = barycenter L (mapF f A1).
Definition fct_lm?\lkAM{fct_lm}? : V1 -> V2 := fun u1 => f O1 \vect\sp f (O1 \transl\sp u1).
Lemma am_lm_equiv?\lkAM{am_lm_equiv}? : aff_map f <-> lin_map (fct_lm f O1).
\end{lstlisting}
Note that \coqe{fct_lm} is independent of~\coqe{O1} (see
\coqelkAM{am_flm_orig_indep}).
Interesting results include
\coqe{comp_am}\linebreak[0]\coqe{_ms}\lkAM{comp_am_ms}, stating that the
composition of affine mappings is an affine mapping, and
\coqelkAM{am_bij_compat}, stating that the inverse of a bijective affine
mapping is an affine mapping.
Moreover, in the case of module spaces equipped with their canonical affine
space structure, the gathering of a family of affine mappings is an affine
mapping,
\begin{lstlisting}
Context {K : Ring}.
Context {E1 E2 : ModuleSpace K}.

Lemma am_gather_equiv_ms?\lkAM{am_gather_equiv_ms}? :
  \forall {n} (f : '(E1 -> E2)^n), aff_map_ms (gather f) <-> (\forall\nsp i, aff_map_ms (f i)).
\end{lstlisting}

We also provide specialized helpers such as \coqelkAM{aff_map_ms} that
expect a function between module spaces and not affine spaces.
Again, this is due to an incorrect canonical structures instantiation for
functions to affine spaces.

\myskip

Affine subspaces are defined as the subsets that are closed under the
barycenter operation (similar to what is done in
Section~\ref{sec:AlgSubstruct}).
\begin{lstlisting}
Context {K : Ring}.
Context {V : ModuleSpace K}.
Context {E : AffineSpace V}.

Definition compatible_as?\lkASS{compatible_as}? (PE : E -> Prop) : Prop :=
  \forall n L (A : 'E^n), invertible (\sum L) -> inclF A PE -> PE (barycenter L A).
\end{lstlisting}
In module spaces equipped with the affine space structure, the kernel of an
affine mapping is shown to be an affine subspace (\coqelkASS{Ker_cas}).

\subsection{Finite-dimensional Affine Algebra and Barycentric Coordinates}
\label{sec:FinDimAS}

Definitions for finite-dimensional affine spaces are similar to those for
vector spaces in Section~\ref{sec:FinDimMS}, and are also in folder
\texttt{Finite\_dim}.
To start with, the affine span of a family of points is the subset made of all
barycenters of theses points (instead of the linear combinations of a family of
vectors for the linear span), and affine independence of a {\nonempty} family
of points states that the family of vectors from the first point to all the
others~$(\vect{A_0A_i})_{i\in[1..n]}$ is linearly independent.
Then, affine generator, affine basis and affine dimension are the same as in
linear algebra, except that the cardinality of the (affine) basis is one more
than the corresponding module space dimension.
\begin{lstlisting}
Context {K : Ring}.
Context {V : ModuleSpace K}.
Context {E : AffineSpace V}.

Variant aff_span?\lkFDAS{aff_span}? {n} (A : 'E^n) : E -> Prop :=
  Aff_span : \forall L, invertible (\sum L) -> aff_span A (barycenter L A).
Definition aff_indep?\lkFDAS{aff_indep}? {n} : 'E^n.+1 -> Prop := fun A => lin_indep (frameF ord0 A).

Variable PE : E -> Prop.

Definition aff_gen?\lkFDAS{aff_gen}? {n} (A : 'E^n) : Prop := PE = aff_span A.
Definition aff_basis?\lkFDAS{aff_basis}? {n} (A : 'E^n.+1) : Prop := aff_gen A /\ aff_indep A.
Variant has_aff_dim?\lkFDAS{has_aff_dim}? n : Prop :=
  Aff_dim : \forall (A : 'E^n.+1), aff_basis A -> has_aff_dim n.
\end{lstlisting}

As expected, \coqelkFDASp{aff_span_cas} states that affine spans are affine
subspaces, and
\coqe{aff_indep_all_}\linebreak[0]\coqe{equiv}\lkFDAI{aff_indep_all_equiv}
states that affine independence can be equivalently expressed with any point,
and not just the first.
Moreover, \coqelkFDAI{aff_indep_skipF} states that removing any point keeps the
affine independence property, and \coqelkFDAI{aff_indep_permutF} states that
this is the same when permuting the points.
Then, \coqelkFDAB{aff_dim_uniq} states that the affine dimension is unique, and
\coqelkFDAB{aff_indep_aff_basis} states that an affinely independent family is
a basis if it is of the correct length.
Finally, \coqelkFDAB{has_aff_dim_Rn} states that~$\R^n$ is of affine
dimension~$n$.

A \emph{simplex} of dimension~$d$ is defined by~$d+1$ affinely independent
points of~$\Rd$ ({\ie} an affine basis of dimension~$d$), that are its
vertices.
The simplices are tetrahedra in three dimensions, triangles in the plane, line
segments on the line, and the point itself in zero dimension.

\myskip

Consider now a family of points \coqe{A : 'E^n}.
Then, for any point in their affine span, \coqe{aff_span_EX} states the
(constructive) existence of weights of sum~1 such that the point is the
barycenter with these weights (with the help of axiom \coqelkLogic{ex_EX}, see
Section~\ref{sec:Classical}),
\begin{lstlisting}
Lemma aff_span_EX?\lkFDAS{aff_span_EX}? :
  \forall {n} {A : 'E^n} {B}, aff_span A B -> { L | \sum\sp L = 1 /\ B = barycenter L A }.
\end{lstlisting}
Then, the barycentric coordinates relative to~$n$ points~$\AA$,
denoted~\Notation{$\bc{\AA}$}, are the~$n$ total functions extracting those
weights when the point is in the affine span, and zero otherwise.
In {\Rocq},
\begin{lstlisting}
Definition baryc_coord?\lkFDAS{baryc_coord}? {n} (A : 'E^n) (B : E) : [0..n) -> K :=
  match in_dec?\lkSdec{in_dec}? (aff_span A) B with
  | left HB => proj1_sig (aff_span_EX HB)
  | right _ => 0
  end.
\end{lstlisting}
When the family of points is {\nonempty} and affinely independent, the
barycentric coordinates are the unique such weights of sum~1 (\coqe{bc_} stands
for ``barycentric coordinates''),
\begin{lstlisting}
Lemma bc_uniq?\lkFDAI{bc_uniq}? : \forall {n} {A : 'E^n.+1}, aff_indep A ->
  \forall {B L}, aff_span A B -> \sum\sp L = 1 -> barycenter L A = B -> L = baryc_coord A B.
\end{lstlisting}

Now, let~$E$ be a module space over a ring~$K$ equipped with the affine
structure over itself.
Let~$\AA$ be~$n+1$ affinely independent points and let~$B$ be in the affine
span of~$\AA$.
From \coqe{bc_uniq}, we have $B=\sum_{i\in[0..n]}\bci{\AA}{i}(B)A_i$, and thus
\begin{equation*}
  \vect{A_0B} = B - A_0
  = \sum_{i \in [1..n]} \bci{\AA}{i}(B) A_i + (\bci{\AA}{0}(B) - 1) A_0
  = \sum_{i \in [1..n]} \bci{\AA}{i}(B) \vect{A_0A_i},
\end{equation*}
since $\sum_{i\in[0..n]}\bci{\AA}{i}(B)=1$.
This means that~$(\bci{\AA}{i}(B))_{i \in [1..n]}$ are the~$n$ coordinates of~$B$
in the frame~$(A_0,(\vect{A_0A_i})_{i\in[1..n]})$, where~$A_0$ plays a specific
role.
To sum up, with an additional component (such that the sum becomes~1), the
barycentric coordinates allow to give all points in~$\AA$ a symmetric role.

\myskip

Again, we also provide specialized helpers such as \coqelkFDAS{aff_indep_ms}
for module spaces of functions equipped with the affine space structure over
themselves.
We also give results about the interaction with affine mappings such as
\coqelkFDAM{bc_am_ms}, stating that the barycentric coordinates in an affine
basis are affine mappings ({\ie} $(B\mapsto\bc{\AA}(B))$ is affine), and
\coqelkFDAM{am_ext_gen_full_ms}, stating that two affine mappings are equal as
soon as they are equal on an affine generator.
Moreover, \coqelkFDAM{am_aff_span_ms} states that the image of an affine span
by an injective affine mapping is the affine span of the images, and
\coqelkFDAM{am_aff_indep_ms} states that the images of affinely independent
points by an injective affine mapping are affinely independent.
Finally, \coqelkFDAM{am_has_aff_dim_ms} states that injective affine mappings
preserve the affine dimension.

\section{Formalization of Finite Elements}
\label{sec:FE:Coq}

In this Section we define the {\Rocq} type of a finite element.
In Section~\ref{sec:fe-triple}, a~FE is defined as a triple~$(K,P,\SSigma)$
together with some properties, that are included in a record type.
Furthermore, this record is designed to ask minimal effort to build instances.
Note also that an instance of this record is built in Section~\ref{sec:FELagP}.

In the mathematical literature, the degenerate cases~$\ndof=0$ and~$d=0$ are
usually prohibited.
Indeed, the absence of degree of freedom is of no help in resolving actual~PDE
problems, and~PDE usually do not live in the space~$\R^0$ reduced to a single
point (which is different from considering a specific equation at some point of
a space of {\nonzero} dimension).
However, they can be included in the mathematical theory and perfectly fit the
formalization.
Specifically avoiding them would lead to dispensable hypotheses in lemmas.
Therefore, we have chosen to address them.
Section~\ref{sec:FE-general} is first devoted to the general record type, and
then, Section~\ref{sec:FE-degenerate} presents the two degenerate cases
($\ndof=0$ and~$d=0$).

\subsection{{\Rocq} Definition as a Record}
\label{sec:FE-general}

Let us start by presenting some definitions useful for the geometric
element~$K$.
As specified in Section~\ref{sec:fe-triple}, $K$ is a {\nondegenerate}
polytope ({\ie} with a {\nonempty} interior).
In this article, we focus on the simplex shape only, but anticipating future
works, we also provide the possibility of using cuboids (there is a field
called \coqe{shape} in the record type below),
\begin{lstlisting}
Inductive shape_type?\lkFE{shape_type}? :=  Simplex | Cuboid.
\end{lstlisting}
The number of vertices (field \coqe{nvtx}) is given by
\begin{lstlisting}
Definition nvtx_of_shape?\lkFE{nvtx_of_shape}? (d : \nat) (shp : shape_type) : \nat\sp:=
  match shp with Simplex => d.+1 | Cuboid => 2^d end.
\end{lstlisting}
Indeed, a simplicial polytope in dimension~$d$ has~$d+1$ vertices, such as a
triangle for~$d=2$ or a tetrahedron for~$d=3$.

The vertices (field \coqe{K_vertices}) are a family of points in~$\Rd$ of the
appropriate size \coqe{nvtx_of_shape}.
Then, $K$~is specified as their convex hull (field \coqe{K_geom}), {\ie} the
subset of all convex combinations (see Section~\ref{sec:AffineSpace}),
\begin{lstlisting}
Variant convex_hull?\lkGeom{convex_hull}? {E : ModuleSpace R_Ring} {n} (vtx : 'E^n) : E -> Prop :=
  | Cvx : \forall L, (\forall\nsp i, 0 <== L i) -> \sum\sp L = 1 -> convex_hull vtx (barycenter L vtx).
\end{lstlisting}
Note that the set~$K$ is constructed from the vertices, and thus the geometry
is entirely defined by the vertices.
In the current development, the {\nondegeneracy} of the geometrical element is
not explicitly specified, see Section~\ref{sec:Design}, as we rely instead on
the affine independence of the vertices of the simplex.

Considering the case of scalar unknown functions from~$\Rd$ to~$\R$ only
($q=1$, see Section~\ref{sec:fe-triple}), the approximation space~$P$ is a
finite-dimensional vector subspace of~$\FRd$, the space of functions from~$\Rd$
to~$\R$ represented in {\Rocq} as \coqelkFF{FRd}~\coqe{d} (fields
\coqe{P_approx} for the subset and \coqe{P_approx_has_dim} for its dimension).
The degrees of freedom~$\SSigma$ are a collection of~$\ndof$ linear forms
in~$\FRdp$ (fields \coqe{ndof}, \coqe{S_dof}, and \coqe{S_dof_lm}).
We could equivalently have defined~$\Sigma$ as a linear map between~$P$
and~$\R^\ndof$ (the \coqe{gather S_dof} below), but having direct access to
each degree of freedom seemed  closest to the mathematical literature.

Then the record type~\coqe{FE} is:
\begin{lstlisting}
Record FE?\lkFE{FE}? (d : \nat) : Type := mk_FE {
  shape : shape_type;
  nvtx : \nat\sp := nvtx_of_shape d shape;
  K_vertices : [0..nvtx) -> '\R^d;
  K_geom : '\R^d -> Prop := convex_hull K_vertices;?\hfill?(* geometric element K *)
  ndof : \nat;
  P_approx : FRd d -> Prop;?\hfill?(* approximation space P *)
  P_approx_has_dim : has_dim P_approx ndof;
  S_dof : [0..ndof) -> FRd d -> \R;?\hfill?(* degrees of freedom \Sigma\sp *)
  S_dof_lm : \forall i, lin_map (S_dof i);
  unisolvence_inj : KerS0 P_approx (gather S_dof);
}.
\end{lstlisting}
where \coqe{mk_FE} is the type constructor of~\coqe{FE}.

The fields \coqe{ndof}, \coqe{P_approx}, and \coqe{S_dof} specify the data
parts of~$P$ and~$\SSigma$.
The proof parts reduce to three fields: first, \coqe{P_approx_has_dim} where
the dimension of \coqe{P_approx} is equal to the size of~$\SSigma$, then
\coqe{S_dof_lm} for the linearity of the degrees of freedom, and last
\coqe{unisolvence_inj} for the sole injectivity of~$\PhiSSigma$, expressed as
\coqelkMonS{KerS0}~\coqe{P_approx (gather S_dof)}.
Indeed, when~$\dim P=\ndof$ and~$\PhiSSigma$ is linear, its properties of
injectivity and bijectivity are equivalent (see Section~\ref{sec:Ker}), and we
provide bijectivity through lemma \coqelkFE{unisolvence} as specified in
Section~\ref{sec:fe-triple}.
As mentioned above, the record is thus designed to ask minimal effort to build
instances.

Note that, as total functions are usually easier to use, {\eg}
see~\cite{May01,BLM15}, we consider total functions instead of functions on
subsets: the approximation space~$P$ is a subset of the space of functions from
the whole~$\Rd$ to~$\R$, instead of $\Omega\rightarrow\R$, or even
$K\rightarrow\R$.

\myskip

The following extensionality result states that it is sufficient to check
equality of data fields to ensure the equality of two instances of the record
type~\coqe{FE}.
\begin{lstlisting}
Lemma FE_ext?\lkFE{FE_ext}? :
  \forall {d} {fe1 fe2 : FE d} (Hs : shape fe1 = shape fe2) (Hn : ndof fe1 = ndof fe2),
    K_vertices fe1 = castF _ (K_vertices fe2) -> P_approx fe1 = P_approx fe2 ->
    S_dof fe1 = castF _ (S_dof fe2) -> fe1 = fe2.
\end{lstlisting}

\subsection{Degenerate Cases}
\label{sec:FE-degenerate}

We now focus on the two degenerate cases, and build the corresponding
instances, as they are rarely discussed and may seem counterintuitive.

\myskip

When~$\ndof=0$, the approximation space~$P$ is of dimension~0, {\ie} it reduces
to the sole zero function (see \coqelkFDB{has_dim_0_rev}), and the empty family
of degrees of freedom~$\SSigma$ is zero  (see \coqelkMonF{hat0F_eq} in
Section~\ref{sec:Monoid}).
Of course, all degrees of freedom are linear, since there are none.
Unisolvence is also straightforward since the only approximation function is
the zero function.

Moreover, lemma \coqelkFE{FE_ndof_0_uniq} states that given any geometric
element~$K$ (shape and vertices), there is only one~FE on~$K$ with
no degree of freedom.

\myskip

When~$d=0$, there is only one spatial point ($\R^0=\{0\}$), and only one
geometric element~$K$, equal to the whole space (\coqelkMonT{hatm0T_eq}).%
\footnote{This really is ``the realm of Pointland'',
  from Edwin A. Abbott's {\em Flatland} (1884).}
The functional space~$\FRo$ is the one-dimensional vector space of constant
functions (\coqelkFDBR{FR0_has_dim}), since there is only one input value;
it is generated by the constant function~$\oone$ (\coqelkFDLG{FR0_lin_gen}).
And if not of dimension~0 ({\ie} not the previous case above), the
approximation space~$P$ is also of dimension~1 (\coqelkFDBR{FR0_sub_dim}),
{\ie} equal to the whole space~$\FRo$ (\coqelkFDBR{FR0_sub_eq}).
The dual space~$\FRop$ is generated by the linear form~$(f\mapsto f(0))$
evaluating constant functions at the single spatial point
(\coqelkFDLG{FR0_dual_lin_gen}), and two linear forms are equal if they
coincide on the constant function~$\oone$ (\coqelkMSR{FR0_dual_ext}).
Then, the unisolvence property is equivalent to the {\noncancellation} of the
degree of freedom on~$\oone$ (\coqelkFE{FE_d_0_S_dof_n0} and
\coqelkFE{unisolvence_inj_d_0}).

We can now construct a record for~$d=0$ and any shape~\coqe{shp},
\begin{lstlisting}
Definition FE_d_0?\lkFE{FE_d_0}? : FE 0 :=
  mk_FE shp 0 P_approx_has_dim_d_0?\lkFE{P_approx_has_dim_d_0}? S_dof_lm_d_0?\lkFE{S_dof_lm_d_0}? unisolvence_inj_d_0.
\end{lstlisting}

Again, lemma \coqelkFE{FE_d_0_uniq} states that given any shape (which is
meaningless here), there is only one~FE on the single zero-dimensional
geometric element, up to a scaling factor, that is~$\sigma_0(\oone)$, on the
{\nonzero} single degree of freedom.

\section{Affine Geometric Transformations}
\label{sec:geometry}

This section addresses geometrical issues such as how to map a geometrical
element onto another, and how to control the mapping of the hyperplanes
carrying the hyperfaces of a geometrical element.

One of the main ingredients of the~FEM is to bring computations on general mesh
cells back to the sole case of the standard reference element (see
Section~\ref{sec:ref-cur-fe-geomap}).
The function mapping the reference element to any given cell of the mesh is
called the {\em geometric transformation};
it is supposed to be a smooth invertible mapping, such as a family of
polynomial functions with invertible Jacobian.
The action of such a geometric transformation can be seen as a change in the
local representation of the spatial dimensions.
In this study, we limit ourselves to affine transformations on the whole
space~$\Rd$, which are sufficient for our case of interest: the simplices.

Section~\ref{sec:geom-ref} describes the reference element.
The geometric transformation mapping the reference element onto a current one
is built in Section~\ref{sec:LagPd1-Tgeom}, it is based on Lagrange polynomials
of degree at most~1.
Then, Section~\ref{sec:Tgeom-and-Hface} is about face hyperplanes and how they
are mapped by the geometric transformation.
Finally, Section~\ref{sec:TFE} focuses on the geometric transformation of a
full~FE.

\subsection{Reference Geometry}
\label{sec:geom-ref}

The reference element associated with the simplicial Lagrange~FE is the
reference simplex~$\hK$.
Its extremities are called the {\em reference vertices},
denoted~\Notation{$\hvv$}.
Quantities related to the reference element are all referred to with a
\Notation{``hat''} notation.

Let~$d$ be the spatial dimension.
The~$d+1$ reference vertices~$\hvv$ in~$\Rd$ are defined by
\begin{equation*}
  \hvv_0 \eqdef \zzero,
  \AND
  \forall i \in [1..d],\; \hvv_i \eqdef \ddelta_{i - 1},
\end{equation*}
where~$\zzero$ is the zero family, and~$(\ddelta_i)_{i\in[0..d-1]}$ is the
Kronecker family, such that~$\delta_{ij}=0$ if~$i\neq j$ and~$\delta_{ii}=1$,
here of length~$d$.
It is straightforwardly formalized as
\begin{lstlisting}
Context {d : nat}.

Definition vtx_ref?\lkLdiref{vtx_ref}? (i : [0..d]) : '\R^d :=
  match ord_eq_dec?\lkOrd{ord_eq_dec}? i ord0 with
  | left _ => 0
  | right Hi => kron (lower_S Hi)
  end.
\end{lstlisting}
where~\coqelkKron{kron} is the Kronecker delta function,
and~\coqelkOrd{lower_S} is the decrement of an ordinal from the proof it is
{\nonzero}.

The reference vertices are affinely independent (see
Section~\ref{sec:FinDimAS}),
\begin{lstlisting}
Lemma vtx_ref_aff_indep?\lkLdiref{vtx_ref_aff_indep}? : aff_indep_ms vtx_ref.
\end{lstlisting}
They also span the affine space~$\Rd$, and thus constitute an affine basis
of~$\Rd$.
As a consequence (see Section~\ref{sec:FinDimAS}), any point~$\hxx\in\Rd$ is
the barycenter of the reference vertices~$\hvv$ with the weights being its
\emph{reference} barycentric coordinates
\Notation{$\hbc$}$(\hxx)\eqdef\bc{\hvv}(\hxx)$,
\begin{lstlisting}
Lemma bc_vtx_ref_uniq?\lkLdiref{bc_vtx_ref_uniq}? : \forall (x_ref : '\R^d) (L : '\R^d.+1), \sum\sp L = 1 ->
  barycenter_ms L vtx_ref = x_ref -> L = baryc_coord_ms vtx_ref x_ref.
\end{lstlisting}
The equality of affine mappings can be checked on an affine basis only, thus we
have the following extensionality result
\begin{lstlisting}
Lemma am_ext_vtx_ref?\lkLdiref{am_ext_vtx_ref}? : \forall {f g : '\R^d -> '\R^d}, aff_map_ms f -> aff_map_ms g ->
  (\forall\nsp i, f (vtx_ref i) = g (vtx_ref i)) -> f = g.
\end{lstlisting}

\subsection{Lagrange Polynomials and Geometric Transformation}
\label{sec:LagPd1-Tgeom}

Lagrange polynomials are interpolation polynomials.
The expression of Lagrange polynomials is simple when~$d=1$, or when~$k=1$.
The former case is treated in Section~\ref{sec:P1k}.
In the affine case (degree at most~1), Lagrange polynomials have a very simple
expression when the interpolation points are the reference vertices
(Section~\ref{sec:LagPd1-ref}).
In the general case, the affine transformation that maps the reference vertices
to some~$d+1$ (current) vertices (Section~\ref{sec:Tgeom}) is the convenient
way to define the (current) Lagrange polynomials (Section~\ref{sec:LagPd1}).

The affine Lagrange polynomials are actually shown to be the barycentric
coordinates with respect to the chosen current vertices, and the geometric
transformation is at the core of the construction and use of the simplicial
Lagrange~FE.

\subsubsection{$\refLdi$ Reference Lagrange Polynomials}
\label{sec:LagPd1-ref}

Given some dimension~$d$, the reference Lagrange polynomials
\Notation{$\refLLdi$}$:(\ArRdR)^{d+1}$ of degree at most~1 are defined for
all~$\hxx\in\Rd$ as
\begin{equation*}
  \refLdi_0 (\hxx) \eqdef 1 - \sum_{j \in [0..d)} \hx_j,
  \AND
  \forall i \in [1..d],\;
  \refLdi_i (\hxx) \eqdef \hx_{i - 1}.
\end{equation*}
This family of functions is formalized as
\begin{lstlisting}
Context {d : nat}.

Definition LagPd1_ref?\lkLdiref{LagPd1_ref}? (i : [0..d]) : FRd d :=
  match ord_eq_dec i ord0 with
  | left _ => fun x_ref => 1 - \sum\sp x_ref
  | right Hi => fun x_ref => x_ref (lower_S Hi)
  end.
\end{lstlisting}

The reference Lagrange polynomials are of sum~1
(\coqelkLdiref{LagPd1_ref_sum}), and satisfy
\begin{equation}
  \label{eq:LagPd1-ref-baryc}
  \forall \hxx \in \Rd,\quad
  \sum_{i \in [0..d]} \refLdi_i (\hxx) \hvv_i = \hxx
\end{equation}
(\coqelkLdiref{LagPd1_ref_barycF}).
Thus, from \coqelkLdiref{bc_vtx_ref_uniq} (see Section~\ref{sec:geom-ref}), the
reference Lagrange polynomials are indeed the barycentric coordinates in the
system of reference vertices, $\refLLdi=\hbc$ (see
\coqelkLdiref{LagPd1_ref_bcF}).
As such, the reference Lagrange polynomials are affine mappings (see
Section~\ref{sec:FinDimAS}, \coqelkLdiref{LagPd1_ref_am}), they do interpolate
at the reference vertices, $\refLdi_j(\hvv_i)=\delta_{ij}$ (see
\coqelkLdiref{LagPd1_ref_kron_vtx}), and~$\refLLdi$ is injective, and
surjective onto the hyperplane of~$\Rdpi$ of equation~$\sum_{i\in[0..d]}x_i=1$
(see Section~\ref{sec:FunSub} for \coqe{surjS}),
\begin{lstlisting}
Lemma LagPd1_ref_inj?\lkLdiref{LagPd1_ref_inj}? : injective (gather LagPd1_ref).
Lemma LagPd1_ref_surjS?\lkLdiref{LagPd1_ref_surjS}? : surjS fullset (fun x => \sum\sp x = 1) (gather LagPd1_ref).
\end{lstlisting}

In Section~\ref{sec:P1k}, they are also shown to form a basis of the
space~$\matPdi$ of $d$-variate affine polynomials.

\subsubsection{Geometric Transformation}
\label{sec:Tgeom}

\begin{figure}[ht]
  \centering
  \resizebox{0.8\textwidth}{!}{\input{fig_Tgeom}}
  \caption[Affine geometric transformation in~2D]{%
    Affine geometric transformation~$\Tgeomv$ in the case~$d=2$.\\
    Given $\vv=(\vv_0,\vv_1,\vv_2)$, $\Tgeomv$~maps the reference
    triangle~$\hK$ to the triangle~$K$ of vertices~$\vv$.}
  \label{fig:Tgeom}
\end{figure}
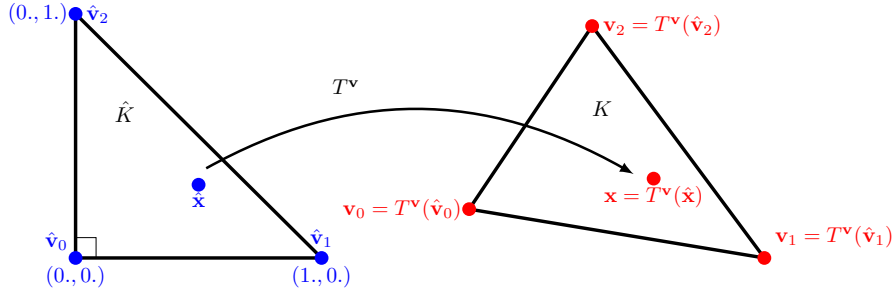

Given~$d+1$ current vertices~$\vv$ of a current geometric
element~$K\subset\Rd$, an affine mapping sending reference vertices~$\hvv$
onto~$\vv$ preserves barycenters (see the definition of \coqe{aff_map} in
Section~\ref{sec:AffineSpace}).
Thus, the affine geometric transformation~\Notation{$\Tgeomv$} outputs the
point with the same barycentric coordinates as the input point in the
system~$\hvv$, but in the system of vertices~$\vv$ (see Figure~\ref{fig:Tgeom}
for an example in the case~$d=2$).
It is defined by (see~\eqref{eq:def-geomap} in
Section~\ref{sec:ref-cur-fe-geomap})
\begin{equation*}
  \Tgeomv : \hxx \in \Rd
  \longmapsto \sum_{i \in [0..d]} \refLdi_i (\hxx) \vv_i \in \Rd,
\end{equation*}
and in {\Rocq} (emphasizing the barycenter, since the weights add up to~1),
\begin{lstlisting}
Context {d : nat}.

Definition T_geom?\lkTgeom{T_geom}? (vtx : [0..d] -> '\R^d) : '\R^d -> '\R^d :=
  fun x_ref => barycenter_ms (fun i => LagPd1_ref i x_ref) vtx.
\end{lstlisting}

The geometric transformation~$\Tgeomv$ is an affine mapping
(\coqelkTgeom{T_geom_am}) since
\begin{equation*}
  \Tgeomv - \Tgeomv \zzero =
  \left( \hxx \longmapsto \sum_{i \in [0..d]} \left(
      \refLdi_i (\hxx) - \refLdi_i (\zzero) \right) \vv_i \right)
\end{equation*}
is a linear map (because the~$\refLdi_i$'s are affine mappings).
We have~$\Tgeomv\hvv=\vv$ (\coqelkTgeom{T_geom_vtxF}), which actually means for
all $i\in[0..d]$, $\Tgeomv\hvv_i=\vv_i$, and this provides the formula for the
composition of geometric transformations,
\begin{equation}
  \label{eq:Tgeom_v_circ_Tgeom_w}
  \forall \vv, \ww \in (\Rd)^{d + 1},\qquad
  \Tgeomv \circ \Tgeomw = \Tgeom{\Tgeomv\ww},
\end{equation}
that is formalized as
\begin{lstlisting}
Lemma T_geom_comp?\lkTgeom{T_geom_comp}? : \forall (vtx1 vtx2 : [0..d] -> '\R^d),
  T_geom vtx1 \o T_geom vtx2 = T_geom (mapF (T_geom vtx1) vtx2).
\end{lstlisting}
Indeed, from \coqelkLdiref{am_ext_vtx_ref}, it suffices to check that both
functions agree on~$\hvv$, which is the case.

Moreover, $\Tgeomhv$~is the identity map (\coqelkTgeom{T_geom_ref}).
Introducing \Notation{$\Scal{s}$}$:\Rd\to\Rd$ the scaling by~$s$, defined for all
$i\in[0..d)$ by $(\Scal{s}\xx)_i\eqdef s*x_i$, we have
(\coqelkTgeom{T_geom_scal_ref})
\begin{equation}
  \label{eq:Tgeom-scal}
  \forall s \in\Rstar,\quad \Tgeom{\Scal{s} \hvv} = \Scal{s}.
\end{equation}
And the geometric elements~$\hK$ and~$K$ being the convex hull of their
vertices, we also have $\Tgeomv(\hK)$\equalsRR$K$ (see
\coqelkTgeom{T_geom_K_geom_eq} and Figure~\ref{fig:Tgeom}).

\myskip

Assume now that the vertices~$\vv$ are affinely independent.
Then, the barycentric coordinates in the system~$\vv$ are unique, and
injectivity and surjectivity of~$\Tgeomv$ follow that of the reference Lagrange
polynomials~$\refLLdi$ (see Section~\ref{sec:LagPd1-ref}).
Thus, $\Tgeomv$ is bijective onto the whole~$\Rd$, and we can build its
inverse~$\Tgeomvinv$ (see Section~\ref{sec:FunSub}).
In {\Rocq},
\begin{lstlisting}
Context {vtx : [0..d] -> '\R^d}.
Hypothesis Hvtx : aff_indep_ms vtx.

Lemma T_geom_bij?\lkTgeom{T_geom_bij}? : bijective (T_geom vtx).
Definition T_geom_inv?\lkTgeom{T_geom_inv}? : '\R^d -> '\R^d := f_inv T_geom_bij.
\end{lstlisting}
Note that~$\Tgeomv$ is also proved to be a bijection from~$\hK$ onto~$K$
(\coqelkTgeom{T_geom_bijS_K_geom}).
The inverse geometric transformation~$\Tgeomvinv$ is also an affine map
(\coqelkTgeom{T_geom_inv_am}), and we have $\Tgeomvinv\vv$\equalsRR$\hvv$ (see
\coqelkTgeom{T_geom_inv_vtx}) and $\Tgeomvinv(K)=\hK$ (see
\coqelkTgeom{T_geom_inv_K_geom_eq}).

\myskip

As explained in Section~\ref{sec:Libs}, we have made implicit some variables.
Here, the function \coqe{T_geom} requires only the vertices (and not the
dimension), while \coqe{T_geom_inv} requires only the proof of affine
independence of the vertices (and neither the vertices, nor the dimension).
It reduces the number of parameters to be given, hence it improves the
readability.

\subsubsection{$\curLdiv$ Lagrange Polynomials}
\label{sec:LagPd1}

The geometric transformation, or rather its inverse, is a convenient way to
bring computations back to the simpler case of the reference simplex.
Given~$d+1$ affinely independent vertices~$\vv$, the associated affine Lagrange
polynomials \Notation{$\curLLdiv$}$:\ArRdRdpi$ are defined as
\begin{equation*}
  \forall i \in [0..d],\quad \curLdiv_i \eqdef \refLdi_i \circ \Tgeomvinv.
\end{equation*}
This is formalized as
\begin{lstlisting}
Context {d : nat}.
Context {vtx : [0..d] -> '\R^d}.
Hypothesis Hvtx : aff_indep_ms vtx.

Definition LagPd1?\lkLdi{LagPd1}? (i : [0..d]) : FRd d := fun x => LagPd1_ref i (T_geom_inv Hvtx x).
\end{lstlisting}

Unsurprisingly, all affine properties of the reference affine Lagrange
polynomials~$\refLLdi$ are transported to~$\curLLdiv$ by the affine geometric
transformation~$\Tgeomvinv$.
Their sum is~1 (\coqelkLdi{LagPd1_sum}).
Since they satisfy the property (\coqelkLdi{LagPd1_barycF})
\begin{equation*}
  \forall \xx \in \Rd,\quad
  \sum_{i \in [0..d]} \curLdiv_i (\xx) \vv_i = \xx,
\end{equation*}
they are also the barycentric coordinates in the system of vertices~$\vv$,
{\ie} $\curLdiv=\bc{\vv}$ (see \coqelkLdi{LagPd1_bcF}), they are affine
mappings (\coqelkLdi{LagPd1_am}), and they interpolate at the vertices,
$\curLdiv_j(\vv_i)=\delta_{ij}$ (see \coqelkLdi{LagPd1_kron_vtx}).

Another property is that~$\curLLdiv$ also forms a basis of the space~$\matPdi$
of $d$-variate affine polynomials (see Section~\ref{sec:P1k}).

\subsection{Geometric Transformations and Face Hyperplanes}
\label{sec:Tgeom-and-Hface}

Faces play an important role in the numerical resolution of~PDE as they carry
the boundary conditions of the problem to solve, and also the continuity
conditions between elements.

We need to consider both a \emph{hyperface} of a $d$-simplex (that is a
$(d-1)$-simplex), and a \emph{face hyperplane}, the hyperplane that contains a
given hyperface.
For instance, when $d=3$, a hyperface of the tetrahedron~$K$ is a triangle.
The affine plane containing this triangle is a face hyperplane,
see Figure~\ref{fig:geo-map-perm-trsp0-d=3-hyperplane}.
In Section~\ref{sec:LagP-unisolv}, both steps of the proof of unisolvence
involve face hyperplanes.

Affine mappings transport affine subspaces, and when they are injective, the
affine dimension is preserved (see Section~\ref{sec:FinDimAS}), thus the image
of a (face) hyperplane is a (face) hyperplane.

Reference and current face hyperplanes are first discussed in
Section~\ref{sec:Hface}, then we study two variants of the geometric
transformation based on the way they deal with face hyperplanes in
Sections~\ref{sec:Inj-Hface} and~\ref{sec:Tgeom-swap}.

\subsubsection{Face Hyperplanes}
\label{sec:Hface}

Face hyperplanes are referred to by the vertex opposite them.
The reference face hyperplane~\Notation{$\hHpl_i$} opposite the reference
vertex~$\hvv_i$ is the hyperplane generated by the other reference
vertices~$(\hvv_j)_{j\neq i}$ (see
Figure~\ref{fig:geo-map-perm-trsp0-d=3-hyperplane} on the left).
It is formalized as
\begin{lstlisting}
Definition Hface_ref?\lkLdiref{Hface_ref}? (i : [0..d]) : '\R^d -> Prop := aff_span_ms (skipF i vtx_ref).
\end{lstlisting}
An important property is that the reference face hyperplanes are the kernels of
the reference affine Lagrange polynomials, $\hHpl_i=\ker\refLdi_i$, in {\Rocq},
\begin{lstlisting}
Lemma Hface_ref_is_Ker?\lkLdiref{Hface_ref_is_Ker}? : \forall i, Hface_ref i = Ker (LagPd1_ref i).
\end{lstlisting}
\begin{proof}
Since the kernel of an affine mapping is an affine subspace and
$\refLdi_i(\hvv_j)=0$ for~$j\neq i$, it is sufficient to show the second
inclusion $\ker\refLdi_i\subset\hHpl_i$.
Let~$\hxx$ such that $\refLdi_i(\hxx)=0$.
Then, from~\eqref{eq:LagPd1-ref-baryc}, we can skip the zero weight in the
sum/barycenter, and~$\hxx$ is a barycenter of the sole other reference
vertices~$(\hvv_j)_{j\neq i}$.
It belongs to their span~$\hHpl_i$.
\end{proof}

A noticeable property is that the hyperfaces of a simplex are in affine
bijection with simplices in one dimension less.
Extended to hyperplanes, all face hyperplanes are in bijection with the whole
space with one dimension less, see Figure~\ref{fig:geo-hyper-map_d32}.
Given~$d>0$, let~\Notation{$\InjhHdi$} be the injection from~$\Rdmi$ to~$\Rd$,
whose range is~$\hHpl_i$, and that is defined for all $\hxxp\in\Rdmi$ by
\begin{align}
  \label{eq:inj:Hdo}
  &&\InjhHdo (\hxxp)
  &\eqdef \left( 1 - \sum_{j \in [0..d - 1)} \hxp_j, \hxxp \right),&&\\
  \label{eq:inj:Hdi}
  &&\forall i \in [1..d],\quad
  \InjhHdi (\hxxp) &\eqdef (\hxp_0, \ldots, \hxp_{i - 2}, 0,
    \hxp_{i - 1}, \ldots, \hxp_{d - 2}).&&
\end{align}
Then, $\InjhHdi$ is a bijective affine mapping from~$\Rdmi$ onto~$\hHpl_i$.
This is formalized as
\begin{lstlisting}
Context {d1 : nat}.
Let d := d1.+1.

Definition inj_Hface_ref?\lkLdiref{inj_Hface_ref}? (i : [0..d]) : '\R^d1 -> '\R^d :=
  match ord_eq_dec i ord0 with
  | left _ => part1F0
  | right Hi => insert0F (lower_S Hi)
  end.
Lemma inj_Hface_ref_am?\lkLdiref{inj_Hface_ref_am}? : \forall i, aff_map_ms (inj_Hface_ref i).
Lemma inj_Hface_ref_bijS?\lkLdiref{inj_Hface_ref_bijS}? : \forall i, bijS fullset (Hface_ref i) (inj_Hface_ref i).
\end{lstlisting}
where the \code{Let} command is for local definitions (see
Section~\ref{sec:Rocq-syntax}).
Note that the positivity of~$d$ is granted structurally to avoid building a
function with a type \coqe{'\R^d.-1 -> '\R^d} depending on the proof
that~$d>0$.
Note also that \coqelkOrd{lower_S}~\coqe{Hi} is indeed of type \coqe{[0..d)}.
The operators~\coqelkRingF{part1F0} and \coqelkMonF{insert0F} state that the
output is in the correct kernel, or reference face hyperplane.
Indeed, the former is the insertion in front of the complement to~1 (see
Section~\ref{sec:Ring}), thus its output is of sum~1, and the latter is the
insertion of~0 in the specified location (see Section~\ref{sec:Monoid}), thus
its output is in~$\ker\refLdi_i$.

\myskip

The face hyperplanes of current simplices are obtained through the affine
geometric transformation.
Given~$d+1$ vertices~$\vv$, the face hyperplane opposite vertex~$\vv_i$ for
$i\in[0..d]$ is defined as \Notation{$\Hplv_i$}$\eqdef\Tgeomv(\hHpl_i)$.
And when the vertices are affinely independent, $\Tgeomv$ is bijective, and we
also have
\begin{equation*}
  \ker \curLdiv_i = \ker \left( \refLdi_i \circ \Tgeomvinv \right)
  = \left( \Tgeomvinv \right)^{-1} (\ker \refLdi_i)
  = \Tgeomv (\hHpl_i) = \Hplv_i.
\end{equation*}
Moreover, the restriction of~$\Tgeomv$ to~$\hHpl_i$ is a bijection
onto~$\Hplv_i$.
This is formalized as
\begin{lstlisting}
Context {d : nat}.

Definition Hface?\lkLdi{Hface}? (vtx : [0..d] -> '\R^d) (i : [0..d]) : '\R^d -> Prop :=
  image (T_geom vtx) (Hface_ref i).

Context {vtx : [0..d] -> '\R^d}.
Hypothesis Hvtx : aff_indep_ms vtx.

Lemma Hface_is_Ker?\lkLdi{Hface_is_Ker}? : \forall i, Hface vtx i = Ker (LagPd1 Hvtx i).
Lemma T_geom_bijS_Hface?\lkLdi{T_geom_bijS_Hface}? : \forall i, bijS (Hface_ref i) (Hface vtx i) (T_geom vtx).
\end{lstlisting}
The hyperplane~$\Hplv_i$ is also the affine span of the
vertices~$(\vv_j)_{j\neq i}$ (\coqelkLdi{Hface_aff_span}).

\subsubsection{Injection onto a Face Hyperplane}
\label{sec:Inj-Hface}

\begin{figure}[ht]
  \centering
  \resizebox{0.95\linewidth}{!}{\input{fig_ImTriaFaceTet_k3}}
  \caption[Injection onto a face hyperplane in~3D]{%
    Injection onto a face hyperplane~$\Injvo$ in the case~$d=k=3$.\\
    The reference triangle~$\hK^{d-1}$ is first mapped onto the 0-th face
    of~$\hK^d$, opposite vertex~$\hvv_0$, and then to the 0-th face of~$K^d$,
    that is contained in the face hyperplane~$\Hplv_0$ supported by
    $(\vv_1,\vv_2,\vv_3)$.
    The hyperplanes are depicted in light blue and the faces in a darker
    blue.\\
    After reading Section~\ref{sec:LagP-nodes} about nodes, note that the
    correspondence between the reference nodes in the triangle ($d-1=2$) and
    the face nodes of the tetrahedron ($d=3$) is illustrated by the colors.
    We have $\Injvo(\haa^{d-1}_{(i,j)})=\aa^{\vv}_{(3-(i+j),i,j)}$, for all
    $(i,j)\in\calA^3_2$.}
  \label{fig:geo-hyper-map_d32}
\end{figure}
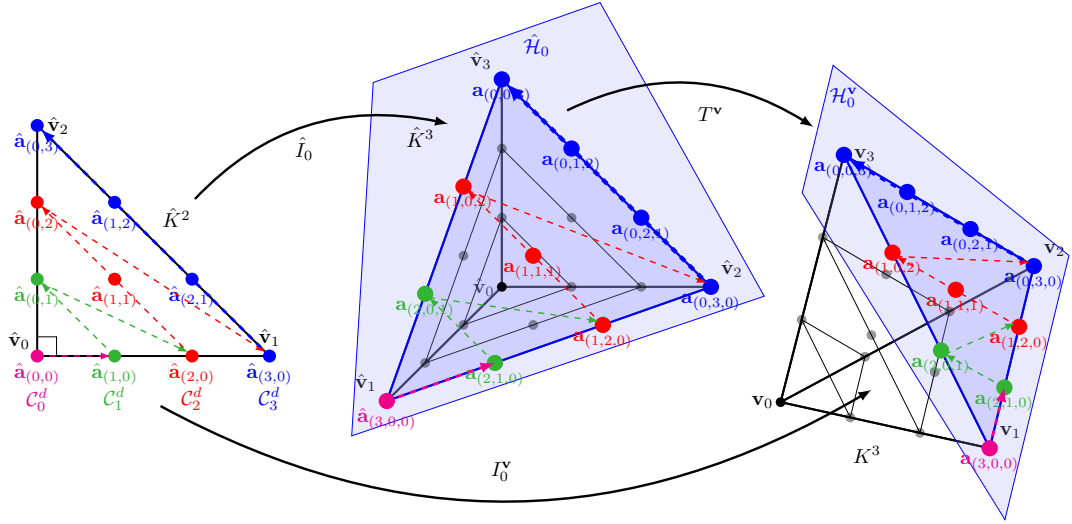

The first step of the inductive proof of unisolvence in
Section~\ref{sec:LagP-unisolv} deals with the increment of the dimension for a
current simplex: a result on the reference rectangle unit simplex in one
dimension less is transported to the first face hyperplane~$\Hplv_0$ opposite
the first vertex~$\vv_0$.
It is thus necessary to transport information on some simplex in
dimension~$d-1$ to a face hyperplane~$\Hplv_i$ in dimension~$d$ (with~$d>0$ and
$i\in[0..d]$).
This is simply obtained by using the composition
\Notation{$\Injvi$}$\eqdef\Tgeomv\circ\InjhHdi$, see an illustration for~$d=3$
in Figure~\ref{fig:geo-hyper-map_d32}.
In {\Rocq},
\begin{lstlisting}
Context {d1 : nat}.
Let d := d1.+1.
Variable vtx : [0..d] -> '\R^d.

Definition inj_Hface?\lkTgeom{inj_Hface}? (i : [0..d]) : '\R^d1 -> '\R^d := T_geom vtx \o inj_Hface_ref i.
\end{lstlisting}
where the positivity of~$d$ is granted structurally because of
\coqelkLdiref{inj_Hface_ref}.

As a composition of affine mappings, $\Injvi$ is an affine mapping
(\coqelkTgeom{inj_Hface_am}).
Moreover, as expected, when the vertices~$\vv$ are affinely independent,
$\Injvi$~is bijective from~$\R^{d-1}$ onto~$\Hplv_i$, as the composition of
bijective functions.
In {\Rocq},
\begin{lstlisting}
Lemma inj_Hface_bijS?\lkLdi{inj_Hface_bijS}? : \forall i, bijS fullset (Hface vtx i) (inj_Hface vtx i).
\end{lstlisting}

\subsubsection{Geometric Transformation Swapping Vertices}
\label{sec:Tgeom-swap}

\begin{figure}[ht]
  \centering
  \resizebox{0.85\linewidth}{!}{\input{fig_TetToTet_k3_perm_trsp0}}
  \caption[Geometric transformation with a transposition in~3D]{%
    Geometric transformation~$\TgeomTranspov$ in the case~$d=k=3$.\\
    The transposition~$\taudo$ maps $(0,1,2,3)$ onto $(3,1,2,0)$.
    The reference simplex~$\hK^d$ is mapped onto the current simplex~$K$.
    The reference vertices are mapped onto the current vertices with a change
    of indices: for all $i\in[0..d]$, we have
    $\TgeomTranspov(\hvv_i)=\vv_{\taudo(i)}$.
    Note that, as~$\taudo(d)=0$, the reference face hyperplane~$\hHpl_d$
    (opposite reference vertex~$\hvv_d$) is mapped onto~$\Hplv_0$ (opposite
    vertex~$\vv_0$).\\
    In this figure, only the visible nodes are depicted.}
  \label{fig:geo-map-perm-trsp0-d=3-hyperplane}
\end{figure}
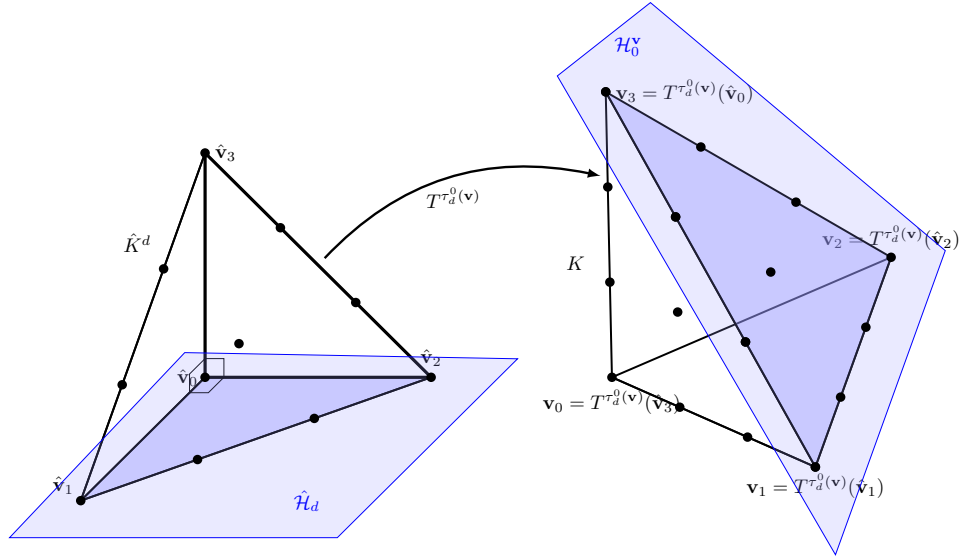

The first step of the proof of unisolvence in Section~\ref{sec:LagP-unisolv} is
also based on a factorization result established in Section~\ref{sec:fact-pol}.
This result involves the first face hyperplane~$\Hplv_0$ opposite the first
vertex~$\vv_0$, but it is more convenient to first prove it for the last
reference face hyperplane~$\hHpl_d$ opposite last reference vertex~$\hvv_d$
(for~$d>0$).

More generally, it can be useful to transport information on some face of some
simplex to any face of any other simplex (see an illustration for~$d=3$ in
Figure~\ref{fig:geo-map-perm-trsp0-d=3-hyperplane}).
To that purpose, given a permutation~$\pi$ of~$[0..d]$ ({\ie} a bijection
from~$[0..d]$ onto itself), we consider the permuted geometric
transformation~$\TgeomPermut{\vv}$, where the notation~\Notation{$\pi(\vv)$}
actually means~$(\vv_{\pi(i)})_{i\in[0..d]}$, and the specialization to the
transposition~$\taudi$ swapping the last index~$d$ with any~$i\in[0..d]$.
This is formalized as
\begin{lstlisting}
Context {d : nat}.

Definition T_geom_permutF?\lkTgeom{T_geom_permutF}? (vtx : [0..d] -> '\R^d) (pi_d : [0..d] -> [0..d]) : '\R^d -> '\R^d :=
  T_geom (permutF pi_d vtx).
Definition T_geom_transpF?\lkTgeom{T_geom_transpF}? vtx (i0 : [0..d]) : '\R^d -> '\R^d :=
  T_geom_permutF vtx (transp_ord ord_max i0).
\end{lstlisting}
where \coqelkOrd{transp_ord} implements any transposition and \coqe{ord_max} is
defined in Section~\ref{sec:FF}.

As expected, when the vertices are affinely independent, the permuted geometric
transformation~$\TgeomPermut{\vv}$ is a bijective affine mapping
(\coqelkTgeom{T_geom_permutF_bij}, and \coqelkTgeom{T_geom_permutF_am}), and it
maps~$\hHpl_i$ onto~$\Hplv_{\pi(i)}$.
We also have, for all $i\in[0..d]$,
$\curLdiv_i=\refLdi_{\pi^{-1}(i)}\circ\TgeomPermutvinv$.
In {\Rocq},
\begin{lstlisting}
Context {vtx : [0..d] -> '\R^d}.
Hypothesis Hvtx : aff_indep_ms vtx.

Context {pi_d : [0..d] -> [0..d]}.
Hypothesis Hpi_d : injective pi_d.
Let pi_d_bij := injF_bij Hpi_d.
Let pi_d_inv := f_inv pi_d_bij.

Lemma T_geom_permutF_Hface?\lkLdi{T_geom_permutF_Hface}? :
  \forall i, image (T_geom_permutF vtx pi_d) (Hface_ref i) = Hface vtx (pi_d i).
Lemma LagPd1_eq_permutF?\lkLdi{LagPd1_eq_permutF}? : \forall i, LagPd1 Hvtx i =
  (permutF pi_d_inv LagPd1_ref i) \o T_geom_permutF_inv Hvtx Hpi_d.
\end{lstlisting}
where \coqe{injF_bij} is a result from the {\MathComp} library extrapolating
bijectivity from injectivity for functions from a finite type to itself.
These results are also available for the specialization to transpositions
swapping the last index, where the inverse of the permutation becomes the
transposition itself as it is involutive (\coqelkLdi{T_geom_transpF_Hface}, and
\coqelkLdi{LagPd1_eq_transpF}).

\subsection{Geometric Transformation of a Finite Element}
\label{sec:TFE}

The geometric transformation defined in Section~\ref{sec:Tgeom} allows to
change the geometry of an element.
We study now how the other two items in the FE~triple, the approximation space
and the degrees of freedom (see Sections~\ref{sec:fe-triple}
and~\ref{sec:FE:Coq}), are affected.

\myskip

In a more general setting, given types~\coqe{T1, T2, A}, consider any function
\coqe{\fhi\sp : T1 -> T2}.
Then, \coqelkFun{pullback}~\coqe{\fhi\sp f\sp := f \o \fhi} is a function that
maps ({\ie} {\em pulls back}) the functions~\coqe{T2 -> A} to~\coqe{T1 -> A}.
When the type~\coqe{A} is endowed with a module space structure, then
\coqe{pullback \fhi} is linear (\coqelkLM{pullback_lm}).
Moreover, given a proof~\coqe{H\fhi\sp : bijective \fhi}, then
\coqe{pullback \fhi} is bijective too (\coqelkFun{pullback_bij}).
Given an additional type~\coqe{B}, the dual transformation maps ({\ie}
{\em pushes forward}) functions~$\sigma$ of type \coqe{(T1 -> A) -> B} to
\coqe{(T2 -> A) -> B}, it is defined by
\coqelkFun{pushforward}~\coqe{\fhi\sp \sigma\sp f\sp := \sigma\sp (pullback \fhi\sp f)}.
Again, when the types~\coqe{A} and~\coqe{B} represent module spaces,
\coqe{pushforward \fhi\sp \sigma} is linear as soon as~\coqe{\sigma} is linear
(\coqelkLM{pushforward_lm}).

\myskip

Given the spatial dimension~\coqe{d}, assuming that types~\coqe{T1, T2}
are~\coqe{'\R^d} and~\coqe{A, B} are the reals~\coqe{\R} (thus~\coqe{A, B} are
module spaces), and given such a bijective transformation
function~\coqe{\fhi\sp : '\R^d -> '\R^d}, we want to express the transformation
of any input~FE into a new~FE of the same type, with the transformed vertices.
\begin{lstlisting}
Context {d : nat}.
Context {\fhi : '\R^d -> '\R^d}.
Hypothesis H\fhi : bijective \fhi.
Variable fe_in : FE d.
\end{lstlisting}
For that, we need to provide all the items of the new instance~\coqe{TFE} (for
``transformed finite element'') of the record~\coqe{FE} defined in
Section~\ref{sec:FE:Coq}, proofs included.
The fields~\coqe{shape} and~\coqe{ndof}, and thus~\coqe{nvtx}, are kept
unchanged.
As wanted, the transformed vertices are the images of the input vertices
($\vvout_i\eqdef\fhi(\vvin_i)$, for all $i\in[0..$\coqe{nvtx}$)$).
The transformed approximation space is defined as the preimage of the input
approximation space by \coqe{pullback \fhi}, or equivalently the input
approximation space is the pullback of the transformed approximation space
($\Pout\eqdef\{\pin\circ\fhi^{-1}\st\pin\in\Pin\}$, {\eg}
see~\cite[Sec.~3.4]{bs:mtf:08} in the affine case, as below).
And the transformed degrees of freedom are the images of the input degrees of
freedom by \coqe{pushforward \fhi}
(that is $\sigmaout_i(f)\eqdef\sigmain_i(f\circ\fhi)$, for all
$i\in[0..$\coqe{ndof}$)$, and for all $f:\ArRdR$).
In {\Rocq},
\begin{lstlisting}
Definition TFE_K_vertices?\lkTFE{TFE_K_vertices}? : [0..nvtx fe_in) -> '\R^d := mapF \fhi\sp (K_vertices fe_in).
Definition TFE_P_approx?\lkTFE{TFE_P_approx}? : FRd d -> Prop := preimage (pullback \fhi) (P_approx fe_in).
Definition TFE_S_dof?\lkTFE{TFE_S_dof}? : [0..ndof fe_in) -> FRd d -> \R\sp :=
  mapF (pushforward \fhi) (S_dof fe_in).
\end{lstlisting}
The proof \coqelkTFE{TFE_S_dof_lm} of linearity of the transformed degrees of
freedom is basically \coqe{pushforward}\linebreak[0]\coqe{_lm} applied to
\coqe{S_dof_lm fe_in}.
The proof \coqelkTFE{TFE_P_approx_has_dim} that the transformed approximation
space is a vector space of the same dimension~\coqe{ndof fe_in} follows from
the fact that \coqe{pullback \fhi} is a bijective linear mapping, and that such
a transformation preserves the dimension of vector subspaces
(\coqelkFDLM{lm_has_dim_rev}).
And finally, the proof \coqelkTFE{TFE_unisolvence_inj} of unisolvence directly
follows from the injectivity of \coqe{pullback \fhi} and the unisolvence of the
input~FE.
Then, the transformed~FE is
\begin{lstlisting}
Definition TFE?\lkTFE{TFE}? : FE d := mk_FE (shape fe_in) TFE_K_vertices (ndof fe_in)
  TFE_P_approx_has_dim TFE_S_dof_lm TFE_unisolvence_inj.
\end{lstlisting}

\myskip

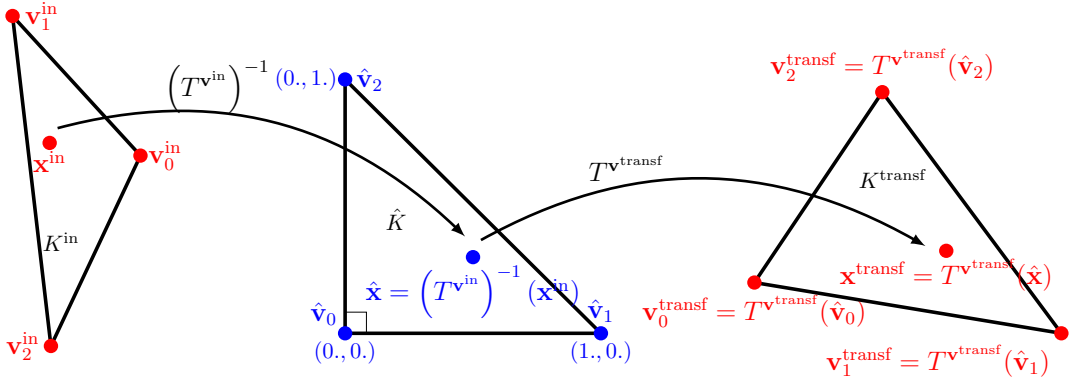
\begin{figure}[ht]
  \centering
  \resizebox{0.95\textwidth}{!}{\input{fig_TFE}}
  \caption[Affine transformation of a finite element in~2D]{%
    Affine transformation of a finite element, from vertices to vertices.\\
    This figure explains the chosen~$\fhi$ in order to construct
    \protect\coqelkTFE{TFE_vtx}.}
  \label{fig:TFE}
\end{figure}

In the simplicial case, the transformation function~\coqe{\fhi} is an affine
mapping.
It can be fully specified by the vertices of the input~FE and those desired for
the transformed~FE.
Indeed, given an input simplicial~FE with affinely independent
vertices~$\vvin$, and output vertices~$\vvout$ that are also affinely
independent, then the geometric transformations~$\Tgeom{\vvin}$
and~$\Tgeom{\vvout}$ defined in Section~\ref{sec:Tgeom} are both bijective, and
the composition $\fhi\eqdef\Tgeom{\vvout}\circ\left(\Tgeom{\vvin}\right)^{-1}$
is a bijective affine transformation that maps~$\vvin$ onto~$\vvout$ as
illustrated in Figure~\ref{fig:TFE}.
The corresponding~FE transformation \coqelkTFE{TFE_vtx} allows to transport any
simplicial~FE of type~\coqe{FE d} to any simplicial geometry.

As a consequence, when the input vertices are the reference vertices~$\hvv$ of
Section~\ref{sec:geom-ref}, for any given affinely independent vertices~$\vv$,
the transformation function simplifies into~$\fhi\eqdef\Tgeomv$, and the
corresponding FE~transformation \coqelkTFE{TFE_vtx_from_ref} allows to
transport any reference simplicial~FE of type~\coqe{FE d} to any simplicial
geometry.

The various transformations with their inputs/hypotheses are given in
Table~\ref{tab:TFE}.

\begin{figure}[ht]
  \centering
  \begin{tabular}{|lcl|l|}
    \hline
    \multicolumn{3}{|c|}{\bf Input / Hypothesis} &
    \multicolumn{1}{|c|}{\bf Output}\\
    \hline
    \code{fe\_in : FE d} && $\fhi:\ArRdRd$ bijective &
    \code{TFE : FE d}\\
    \hline
    \code{fe\_in : FE d} && \code{shape fe\_in = Simplex} &
    \code{TFE\_vtx : FE d}\\
    $\vvin$ aff. indep. && $\vvout$ aff. indep. &
    $\left(\mbox{with }
      \fhi=\Tgeom{\vvout}\circ\left(\Tgeom{\vvin}\right)^{-1}\right)$\\
    \hline
    \code{fe\_in : FE d} && \code{shape fe\_in = Simplex} &
    \code{TFE\_vtx\_from\_ref : FE d}\\
    $\vvin=\hvv$ && $\vv$ aff. indep. & (with $\fhi=\Tgeomv$)\\
    \hline
  \end{tabular}
  \caption[Table of inputs and hypotheses for the finite element
  transformation]{%
    Table of inputs and hypotheses for the finite element
    transformation~\protect\coqelkTFE{TFE}, and its successive
    specializations~\protect\coqelkTFE{TFE_vtx}
    and~\protect\coqelkTFE{TFE_vtx_from_ref}.}
  \label{tab:TFE}
\end{figure}

\section{Multi-indices and Polynomial Space $\matPdk$}
\label{sec:poly-space-Pdk}

Following the definition of the {\Rocq} record for a~FE of
Section~\ref{sec:FE:Coq}, we have detailed in the previous section results
about the geometry and the vertices.
We now focus on the approximation space.
Indeed, we need to define a finite dimensional function space, in which the
approximate solution is sought in each element.
The most common one is the space of polynomials of bounded total degree, that
we define in Section~\ref{sec:Pdk}.
The subtlety here is that we have  {\multivariate} polynomials, that is to say
polynomials from~$\Rd$ to~$\R$.
We rely on {\multiindices} defined in Section~\ref{sec:MultiIndex} to define
a basis of {\multivariate} polynomials of bounded degree.
The definition and properties of these {\multiindices} is key to the following,
as they are also used for numbering the nodes in Section~\ref{sec:LagP-nodes}.

\subsection{{\Multiindices}}
\label{sec:MultiIndex}

Given a dimension~$d\in\N$, a {\multiindex}~$\aalpha$ is a family of size~$d$
of natural numbers, that is an element of~$\N^d$.
Its sum is denoted by $\len{\aalpha}\eqdef\sum_{i=0}^{d-1}\alpha_i\in\N$.
Given a maximal sum value~$k\in\N$, $\calAdk$ is defined in
Section~\ref{sec:conform:FE} as the subset of {\multiindices} of size~$d$ and
of sum at most~$k$, that is~\eqref{eq:multi-ind-Adk},
\begin{equation*}
  \calAdk \eqdef \left\{ \aalpha \in \N^d \st \len{\aalpha} \leq k \right\}.
\end{equation*}
This is a versatile object.
Its intrinsic combinatorial nature can serve several purposes.
Interpreted as exponents, {\multiindices} of~$\calAdk$ can represent
multivariate monomials of bounded total degree (see~\eqref{eq:X:alpha}), and
are a clever tool for the proof of freedom of the monomials in
Section~\ref{sec:Pdk}.

\begin{figure}[ht]
  \centering
  \resizebox{0.9\linewidth}{!}{\input{fig_Tet-Tria_k3}}
  \caption[Lagrange nodes of the reference simplex in~2D and~3D]{%
    Lagrange nodes~$(\haa_{\aalpha})_{\aalpha\in\calAdk}$ of the reference
    simplex~$\hK_d$ when~$d\in\{2,3\}$ and~$k=3$.\\
    Each node is depicted as a colored ball, and corresponds to a unique
    element of~$\calAdiii$.
    The colors correspond to degrees~$l\leq3$ of polynomials, or equivalently
    to the sizes of {\multiindices} ({\ie} in~$\calCdl$ for~$l\leq3$).
    In magenta, the node~$\haa_\zzero$ corresponds to constant polynomials
    (with degree~0) in~$\matPdo$, and to the {\multiindex}~$\zzero$ in the
    singleton~$\calCdo$.
    In green, the nodes correspond to {\nonconstant} affine polynomials (with
    degree~1), and to the {\multiindices} $\ddelta_1,\ldots,\ddelta_d$
    in~$\calCdi$.
    In red, the nodes correspond to {\nonaffine} quadratic polynomials (with
    degree~2), and to {\multiindices} in~$\calCdii$.
    In blue, the nodes correspond to {\nonquadratic} cubic polynomials (with
    degree~3), and to {\multiindices} in~$\calCdiii$.
    We observe in this picture that
    $\calAdiii=\calCdo\uplus\calCdi\uplus\calCdii\uplus\calCdiii$.}
  \label{fig:lag-k3-d2-d3}
\end{figure}

It also has a geometric usage as the indices of the nodes in the
simplicial Lagrange~FE.
More precisely, see Figures~\ref{fig:Lag:geom} and~\ref{fig:lag-k3-d2-d3}.
The Lagrange nodes are the dots of any color; they are defined such that they
are evenly distributed in the simplex.
Their formal definition is given in Section~\ref{sec:LagP-ref-cur-nodes}
(see~\eqref{eq:lagP-nodes}), but let us give here an intuition.
Each reference node~$\haa_\aalpha$ has coordinates
$(\frac{\alpha_i}{k})_{i\in[0..d)}$.
It is thus natural to represent a Lagrange node in a simplex by its
{\multiindex} $\aalpha$ in~$\calAdk$.
The correspondences between a {\multiindex}, a node and a monomial are used
throughout the document.

Section~\ref{sec:Adk} formally defines~$\calAdk$ and~$\calCdk$ (all the
families of size~$d$ of sum equal to~$k$, see~\eqref{eq:Cdk-Adk}) as
enumerating functions.
Section~\ref{sec:Adk:order} focuses on the chosen order for~$\calAdk$.
Section~\ref{sec:bij-multi-ind} defines an inverse for the function~\coqe{Adk}.
Section~\ref{sec:face-multi-ind} defines various {\multiindices} bijections
related to faces.

\subsubsection{Formal Definition of {\Multiindices}}
\label{sec:Adk}

We formalize~$\calAdk$ in an extensional way (see Section~\ref{sec:FF})
as a family of {\multiindices} of type \coqe{[0..pbinom d k] -> '\nat^d},
recall the definition of \coqelkBinom{pbinom} in Section~\ref{sec:Binom}.
It is indeed known that the set~$\calAdk$ is of size~$\binom{d+k}{d}$, so we
choose to define $\calAdk$ as a family of size~$\binom{d+k}{d}$ of
{\multiindices} of size~$d$.
The family~$\calAdk$ is defined by layers: the {\multiindices} of sum~0, then
of sum~1, and so on until the {\multiindices} of sum~$k$.
More precisely,
\begin{equation}
  \label{eq:Cdk-Adk}
  \Notation{$\calCdk$} \eqdef \{ \aalpha \in \N^d \st \len{\aalpha} = k \},
  \AND
  \calAdk = \biguplus_{l = 0}^k \calCdl.
\end{equation}
In words, $\calCdk$ is the family of {\multiindices} of size~$d$ and of sum
equal to~$k$.
Then, $\calAdk$ is the (disjoint) union of $(\calCdl)_{l\in[0..k]}$, built in
{\Rocq} by concatenation of the~$\calCdl$'s.

\myskip

Now let us formalize\thedefofJ~$\calCdk$.
We choose not to represent the case~$d=0$ as~$\calC^0_k$ is empty
when~$k\neq0$.
We therefore only consider~$\calC^{d_1+1}_k$ for $d_1\in\N$.
Its type is \coqe{[0..pbinom d1 k] -> '\nat^d1.+1}.

It is constructed as a fixpoint on~$d_1$.
For~$d_1=0$, there is a single vector of size~1 and of sum~$k$ that is~$\{k\}$.
Let us assume we have computed all the elements of~$\calC^{d_1}_i$.
Then, the family~$\calC^{d_1+1}_k$ is made of {\multiindices} of size~$d_1+1$.
Let $\aalpha\in\calC^{d_1+1}_k$; then we split it as
$\aalpha=(\alpha_0,\aalphap)$ and we have that $0\le\alpha_0\le k$ and
$\alpha_0+\len{\aalphap}=k$.
So we deduce that the set~$\calC^{d_1+1}_k$ is the (disjoint) union, for
$0\le i\le k$, of $\{(k-i,\aalphap)\st\aalphap\in\calC^{d_1}_i\}$.
By using the \coqelkConcat{concatnF} function of Section~\ref{sec:Monoid}, we
define~\coqe{CSdk d1 k} that is~$\calC^{d_1+1}_k$,
\begin{lstlisting}
Fixpoint CSdk?\lkMI{CSdk}? (d1 k : nat) : [0..pbinom d1 k] -> 'nat^d1.+1 :=
  match d1 with
  | O => fun _ _ => k
  | S d2 => castF _ (concatnF (fun i => mapF (insertF0 (k - i)) (CSdk d2 i)))
  end.
\end{lstlisting}
The \coqelkFF{insertF0} function only concatenates a value (here~$k-i$) as a
first element in a family of {\multiindices} of size~$s$, making a family (of
the same size) of {\multiindices} of size~$s+1$.
This construction is also illustrated in Figure~\ref{fig:lag-k3-d3-slices}.

\begin{figure}[ht]
  \centering
  \resizebox{0.7\linewidth}{!}{\input{fig_Tet_k3_slice}}
  \caption[Vertical slices of the reference simplex in~3D]{%
    Partition
    $\calC^d_k=\biguplus_{i=0}^k\{(k-i,\bbeta)\st\bbeta\in\calC^{d-1}_i\}$,
    with~$d=k=3$.\\
    Geometrically, $\{(k-i,\bbeta)\st\bbeta\in\calC^{d-1}_i\}$ is the
    intersection of the hyperplane of equation~$\alpha_0=k-i$ (depicted in
    light pink) and of~$\calC^{d}_k$ (of equation
    $\alpha_0+\alpha_1+\alpha_2=k$).\\
    The reference Lagrange node~$\haa_{(\alpha_1,\alpha_2,\alpha_3)}$ belonging
    to the face~$\hHpl_0$ is in blue and corresponds to the element
    $(\alpha_1,\alpha_2,\alpha_3)\in\calC^{d}_k$.
    For instance, the set $\{(1,2,0),(1,1,1),(1,0,2)\}$ is depicted by the blue
    nodes linked by a dashed arrow.}
  \label{fig:lag-k3-d3-slices}
\end{figure}

Note that the \emph{set}~$\calCdk$ defined in~\eqref{eq:Cdk-Adk} may have been
defined otherwise.
For instance, we could have concatenated the $(i,\bbeta)$ for
$\bbeta\in\calC^{d-1}_{k-i}$, or the $(\bbeta,k-i)$ for
$\bbeta\in\calC^{d-1}_i$, and so on.
All those are mathematically correct and the \emph{sets} are equivalent.
As we define~$\calCdk$ in an extensional way (see Section~\ref{sec:FF}), we
have the advantage of choosing the ordering of the values in $\calCdk$, see
Section~\ref{sec:Adk:order}.

\myskip

The {\Rocq} definition of~$\calAdk$ is then straightforward, it is the
concatenation of the~$\calCdl$ as in~\eqref{eq:Cdk-Adk},
\begin{lstlisting}
Definition Adk?\lkMI{Adk}? (d k : \nat) : '\nat^{(pbinom d k).+1,d} :=
  match d with
  | 0 => 0
  | S d1 => castF _ (concatnF (CSdk d1))
  end.
\end{lstlisting}
As for~\coqe{CSdk} above, the underscore here is a rather complex proof
(compared to most proofs inside casts) as it is \coqe{pbinomS_rising_sum_l}
(see~\eqref{eq:binom-rising-sum} in Section~\ref{sec:Binom}) to ensure that the
sum of the size of the~$\calCdl$'s makes the expected size of~$\calAdk$.

Interesting results include the values for~$k=1$ (\coqelkMI{Ad1_eq}) and~$d=1$
(\coqelkMI{A1k_eq}), and the fact that the last layer of~$\calAdk$ corresponds
to~$\calCdk$, that is to say the maximal sum is at the end of the family,%
\footnote{Note that when positivity of natural numbers is not needed
  structurally, we choose to use {\noncancellation} instead of positiveness.}
\begin{lstlisting}
Lemma Adk_last_layer?\lkMI{Adk_last_layer}? : \forall d k idk, d <> 0 -> k <> 0 ->
  \sum\sp (Adk d k idk) = k <-> (pbinom d k.-1).+1 <== idk.
\end{lstlisting}

\subsubsection{Ordering {\Multiindices}}
\label{sec:Adk:order}

There are several ways to construct~$\calAdk$ in an extensional way.
The choice of the ordering of these {\multiindices} leads to the expression of
the enumerating function.

With their interpretation as exponents of {\multivariate} monomials,
recall~\eqref{eq:X:alpha}, ordering {\multiindices} can be seen as ordering
monomials.
This is a known problem with many solutions, that is to say many possible
orderings, {\eg} see~\cite[Chap.~2, Sec.~2]{clo:iva:15}.
A common ordering is the lexicographic order and another is the graded reverse
lexicographic order ({\aka} grevlex order), often used when Gröbner bases are
concerned.
But neither suited us for reasons explained later in
Section~\ref{sec:face-multi-ind}.
When the implementation of the~FEM will be tackled, we may need other
orderings, but program verification is out of the scope of this paper.

\paragraph{Ordering Definition: Grsymlex.}

Let us now define the chosen order, graded symmetric lexicographic, or simply
``grsymlex'' order,
\begin{equation}
  \label{eq:grsymlex}
  \aalpha \ltgrsymlex \bbeta \EQUIVDEF
  \left\{
    \begin{array}{l}
      \len{\aalpha} < \len{\bbeta}, \mbox { or}\\
      \len{\aalpha} = \len{\bbeta} \CONJ \bbeta \ltlex \aalpha.
    \end{array}
  \right.
\end{equation}
This amounts to first compare the sum of {\multiindices}, and in case of
equality, use the symmetric of the lexicographic order.
In other words, when $\len{\aalpha}=\len{\bbeta}$, we have
$\aalpha\ltgrsymlex\bbeta$ iff $\beta_i<\alpha_i$ for the first index~$i$
where~$\alpha_i$ and~$\beta_i$ differ.
An equivalent recursive point of view is
\begin{equation*}
  \aalpha  \ltgrsymlex \bbeta \EQUIV
  \left\{
    \begin{array}{l}
      \len{\aalpha} < \len{\bbeta}, \mbox { or}\\
      \len{\aalpha} = \len{\bbeta}
      \CONJ d \geq 2 \CONJ \aalphap \ltgrsymlex \bbetap \\
      \mbox{where we denote } \aalpha = (\alpha_0, \aalphap) \mbox { and }
      \bbeta = (\beta_0, \bbetap).
    \end{array}
  \right.
\end{equation*}
This is defined for a generic order~$<$, and we use it with~\coqe{lt}, the
strict order on~$\N$.

The {\Rocq} definition is exactly~\eqref{eq:grsymlex}, but we have also the
previous equivalence on~$\N$,
\begin{lstlisting}
Lemma grsymlex_lt_S?\lkMO{grsymlex_lt_S}? : \forall {n : nat} (x y : 'nat^n.+1),
  grsymlex_lt x y <-> \sum\sp x < \sum\sp y \/ (\sum x = \sum\sp y /\ grsymlex_lt (skipF0 x) (skipF0 y)).
\end{lstlisting}
We prove that this is a strict and total order (\coqelkMO{grsymlex_lt_sto}),
and a monomial order
(\coqe{grsymlex_}\linebreak[0]\coqe{br_plus_compat_l}\lkMO{grsymlex_br_plus_compat_l}).
This ordering is shown in Figure~\ref{fig:tet-tria_k3_grsymlex} on~$\calAdiii$
for $d\in\{2,3\}$, where examples of values ordered by grsymlex are provided.

\begin{figure}[ht]
  \centering
  \resizebox{0.75\linewidth}{!}{\input{fig_Tet-Tria_k3_grsymlex}}
  \caption[Grsymlex ordering of the {\multiindices} in~2D and~3D]{%
    Grsymlex ordering of~$\calAdk$ when~$d\in\{2,3\}$ and~$k=3$.\\
    Dashed arrows represent the increase in the order.\\
    For~$\calA^2_3$ ($d=2$), we have
    $(0,0)\ltgrsymlex(1,0)\ltgrsymlex(0,1)\ltgrsymlex(2,0)\ltgrsymlex(1,1)\ltgrsymlex(0,2)\ltgrsymlex(3,0)\ltgrsymlex(2,1)\ltgrsymlex(1,2)\ltgrsymlex(0,3)$.\\
    When~$d\geq3$, we have for instance for~$\calC^3_3$
    $(3,0,0)\ltgrsymlex(2,1,0)\ltgrsymlex(2,0,1)\ltgrsymlex(1,2,0)\ltgrsymlex(1,1,1)\ltgrsymlex(1,0,2)\ltgrsymlex(0,3,0)\ltgrsymlex(0,2,1)\ltgrsymlex(0,1,2)\ltgrsymlex(0,0,3)$.}
  \label{fig:tet-tria_k3_grsymlex}
\end{figure}

\paragraph{Ordering $\calAdk$.}

There is left to prove that the~$\calAdk$'s are sorted in grsymlex order.
\begin{lstlisting}
Lemma Adk_sortedF?\lkMI{Adk_sortedF}? : \forall d k, sortedF grsymlex_lt (Adk d k).
\end{lstlisting}
where \coqelkOrd{sortedF}~\coqe{leT A} is the predicate stating that the
family~\coqe{A : 'T^n} is ordered according to the binary
relation~\coqe{leT : T -> T -> Prop}.

As~$\calAdk$ is built by concatenation of the~$\calCdl$'s, we rely on the
\coqelkConcat{concatnF_sortedF} lemma of Section~\ref{sec:Monoid}: each layer
is ordered, the set of layers is ordered (as~$\calAdk$ is built by layers of
increasing sum), and all of them are {\nonempty}.
The most complex proof is that each~$\calCdl$ (layer) is ordered.
This is done by induction, relying on our choice to concatenate the
$(k-i,\bbeta)$ for $\bbeta\in\calC^{d+1}_i$ for~$i$ from~$0$ to~$k$.

\subsubsection{Bijection and {\Multiindices}}
\label{sec:bij-multi-ind}

Now let us focus on the correctness of~$\calAdk$.
Another needed property is that~$\calAdk$ is a bijection between the integers
from~0 to~$\binom{d+k}{d}-1$ and the {\multiindices} of size~$d$ and of sum
smaller or equal to~$k$.
\begin{lstlisting}
Lemma Adk_inj?\lkMI{Adk_inj}? : \forall d k, injective (Adk d k).
Lemma Adk_surj?\lkMI{Adk_surj}? : \forall d k (b : 'nat^d), \sum\sp b <== k -> { idk | b = Adk d k idk }.
\end{lstlisting}
The injectivity is deduced from the previous lemma: as~$\calAdk$ is sorted with
a strict order, it is injective.
The surjectivity (all {\multiindices} of sum smaller or equal to~$k$ belong
to~$\calAdk$) relies on the surjectivity of~$\calCdk$, that is done by
induction.
There is no {\Rocq} specific difficulty except a tedious handling of casts.

From Lemma~\coqe{Adk_surj} that constructs a value, we then define an inverse
of~\coqe{Adk} that takes a {\multiindex} of size~$d$ and returns an integer
between~0 and~$\binom{d+k}{d}-1$,
\begin{lstlisting}
Definition Adk_inv?\lkMI{Adk_inv}? (d k : \nat) : '\nat^d -> [0..pbinom d k] :=
  fun b => match le_dec (\sum b) k with
    | left H => proj1_sig (Adk_surj d k b H)
    | right _ => ord0
    end.
\end{lstlisting}
We choose to make it a total function, that takes any \multiindex.
When the sum of the {\multiindex} is too large, a default value (zero) is
returned.
We proved the correctness of \coqe{Adk_inv}, with a hypothesis on the
{\multiindex} when needed,
\begin{lstlisting}
Lemma Adk_inv_correct_r?\lkMI{Adk_inv_correct_r}? : \forall d k b, \sum\sp b <== k -> Adk d k (Adk_inv d k b) = b.
Lemma Adk_inv_correct_l?\lkMI{Adk_inv_correct_l}? : \forall d k idk,  Adk_inv d k (Adk d k idk) = idk.
\end{lstlisting}

\subsubsection{Face {\Multiindices}}
\label{sec:face-multi-ind}

As {\multiindices} are related to Lagrange nodes (see
Figure~\ref{fig:lag-k3-d2-d3} and Section~\ref{sec:LagP-nodes}), it is useful
to characterize the {\multiindices} associated with the nodes on the faces and
to construct bijections similar to the geometric mappings of
Section~\ref{sec:LagPd1}.
When~$d>0$, for $i\in[0..d]$, there exist bijections between the family of
{\multiindices}~$\calAdmik$ and either the layer~$\calCdk$ (case~$i=0$), or
the {\multiindices} of~$\calAdk$ with a zero component.

More precisely, we define
\begin{equation}
  \label{eq:Adki}
  \Notation{$\calAdko$} \eqdef \calCdk,
  \AND
  \forall i\in[1..d],\quad
  \Notation{$\calAdki$} \eqdef \{ \aalpha \in \calAdk \st \alpha_{i - 1} = 0 \}.
\end{equation}
The {\Rocq} definition is the property \coqelkMI{in_ASdki} on the
index~\coqe{idk} such that \coqe{Adk d k idk} belongs to~$\calAdki$.
Then, the aforementioned mappings are defined for all $\aalphap\in\calAdmik$ as
\begin{align}
  \label{eq:inj:adki0}
  &&\Notation{$\InjAdko$} (\aalphap)
  &\eqdef (k - |\aalphap|, \aalphap) \in \calAdko,&&\\
  \label{eq:inj:adki}
  &&\forall i \in [1..d],\quad
  \Notation{$\InjAdki$} (\aalphap) &\eqdef
    (\alphap_0, \dots, \alphap_{i - 2}, 0, \alphap_{i - 1}, \dots,
    \alphap_{d - 2}) \in \calAdki.&&
\end{align}

We define \coqelkMI{inj_ASdki} of type
\coqe{[0..pbinom d1 k] -> [0..pbinom d k]} that takes the index of~$\aalphap$
and provides the index of the result of either~$\InjAdko$ or~$\InjAdki$
(\eqref{eq:inj:adki0} and~\eqref{eq:inj:adki} are merged as a single
definition).
Note that this reminds~\eqref{eq:inj:Hdo} and~\eqref{eq:inj:Hdi} related to
faces.
This is exploited later in Section~\ref{sec:LagP-nodes-faces} that details in
which faces are the Lagrange nodes.

\myskip

Now, we have all the definitions to give and ensure the ordering requirements,
that are the reasons we choose the \coqe{grsymlex_lt} order.
We want the order to be:
\begin{enumerate}[(i)]
\item a monomial order, {\ie} a total order that is compatible with the monoid
  structure of the monomials, that is for all
  $\aalpha,\bbeta,\ggamma\in\calAdk$, $\XX^\aalpha<\XX^\bbeta$ implies
  $\XX^\aalpha\XX^\ggamma<\XX^\bbeta\XX^\ggamma$, which means that
  $\aalpha<\bbeta$ implies $\aalpha+\ggamma<\bbeta+\ggamma$.
  It is proved in Section~\ref{sec:Adk:order}.
  It is also generally required that for all $\aalpha\neq\zzero$,
  $1=\XX^\zzero<\XX^\aalpha$, that is $\zzero<\aalpha$ (see~(iv) below).

\item increasing with respect to the degree (from~$k$ to~$k+1$), {\ie}
  {\multiindices} of sum at most~$k$ should be numbered before those of
  sum~$k+1$, that is for all $\aalpha\in\calAdk$ and for all
  $\bbeta\in\calC^d_{k+1}$, we should have~$\aalpha<\bbeta$.
  It allows us to easily sort monomials with respect to their total degree and
  is proved by Lemma \coqelkMI{Adk_last_layer} in Section~\ref{sec:Adk}

\item increasing with respect to the dimension (from~$d-1$ to~$d$), {\ie}
  ``natural'' bijections between $(d-1)$-{\multiindices} of sum at most~$k$ and
  $d$-{\multiindices} of sum~$k$ or with a zero entry ($\InjAdki$ for
  $i\in[0..d]$ defined in~\eqref{eq:inj:adki0} and~\eqref{eq:inj:adki}), should
  be increasing.
  It helps relate the face nodes and the volume nodes during the computations
  and is proved by Lemma \coqelkMI{inj_ASdki_monot}.

\item the {\multiindices} corresponding to the reference vertices~$\hvv$ (see
  Section~\ref{sec:geom-ref}) are increasing, {\ie}
  $\zzero<k\ddelta_0<k\ddelta_1<\ldots<k\ddelta_{d-1}$.
  It allows to have \emph{positive} simplices ({\ie} positively oriented) when
  computing the integrals.
  This is formally proved by two lemmas.
  First, \coqelkMO{grsymlex_lt_zl} states that $\aalpha\neq\zzero$
  implies~$\zzero<\aalpha$.
  Second, we prove that~$k\ddelta$ is sorted for the grsymlex order (for a
  positive~$k$).
\begin{lstlisting}
Lemma grsymlex_sortedF_itemF?\lkMO{grsymlex_sortedF_itemF}? :
  \forall {n} k, k <> 0 -> sortedF grsymlex_lt (fun i => itemF n i k).
\end{lstlisting}
\end{enumerate}

\subsection{Polynomial Space $\matPdk$}
\label{sec:Pdk}

In this section, we formalize the polynomial space~$\matPdk$ defined
in~\eqref{eq:pol-space-Pkd} and its properties.
It contains all the polynomials on~$\Rd$ with total degree at most~$k$.
In Section~\ref{sec:Pdk-def}, we define this subset and its first properties,
including the fact that it is a finite-dimensional vector space and its
dimension.
Section~\ref{sec:P1k} is about specific values of~$d$ and~$k$.
Section~\ref{sec:fact-pol} is about the factorization of polynomials.

Note, however, that we do not need to formalize the (total) degree of
polynomials {\perse}.
In particular, we do not have to deal with the degree of the zero polynomial,
which usual value is~$-\infty$.

\subsubsection{$\matPdk$ Definition and Basis}
\label{sec:Pdk-def}

We have defined in Section~\ref{sec:MultiIndex} the {\multiindices}~$\calAdk$,
that are all the families of size~$d$ with sum at most~$k$.
We therefore consider each of these {\multiindex} as the exponent of a monomial
in order to construct each monomial.
The polynomials are then all the linear combinations of these monomials.

Following~\eqref{eq:X:alpha}, monomials are first formalized as
\begin{lstlisting}
Definition Monom_dk?\lkPdk{Monom_dk}? d k : [0..pbinom d k] -> FRd d :=
  fun idk x => powF_P (Adk d k idk) x.
\end{lstlisting}
where \coqelkMonom{powF_P} is the product of univariate monomials.
We remind that \coqe{FRd d} is the vector space of functions mapping from the
real space~$\Rd$ to~$\R$.
Therefore, \coqe{Monom_dk d k} is a family of such functions.
The size of this family is the cardinal of~$\calAdk$, that is~$\binom{d+k}{k}$,
{\ie} \coqe{(pbinom d k).+1}.
Then, following~\eqref{eq:pol-space-Pkd}, $\matPdk$ is formalized as their
linear span,
\begin{lstlisting}
Definition Pdk?\lkPdk{Pdk}? d k : FRd d -> Prop := lin_span (Monom_dk d k).
\end{lstlisting}

As a linear span, it is trivially a vector space.
We proved many lemmas, including
\begin{lstlisting}
Lemma Pdk_pullback_am?\lkPdk{Pdk_pullback_am}? : \forall {d n} k (p : FRd d) (f : '\R^n -> '\R^d),
  aff_map_ms f -> Pdk d k p -> Pdk n k (pullback f p).
\end{lstlisting}
The composition of a polynomial of degree at most~$k$ with an affine mapping is
also a polynomial of degree at most~$k$.
Note that the affine mapping~$f$ is from~$\R^n$ to~$\Rd$, therefore~$p$ is a
polynomial on~$\Rd$ while $p\circ f$ is on~$\R^n$.
The proof relies on two lemmas.
First, the product of polynomials has degree at most the sum of the degrees of
the polynomials.
And second, an affine mapping to the power~$k$ belongs to~$\matPdk$.

\myskip

The most complex proof of this section is that the monomials
\coqe{Monom_dk d k} form a linearly independent family.
This is required to prove that it is a basis of~$\matPdk$, and then its
cardinal is the dimension of~$\matPdk$.

For this proof, we rely on analytic derivatives.
More precisely, we rely on the derivative of real functions of a real variable
provided by the {\Coquelicot} library~\cite{BLM15}.
We first define the derivative with respect to a given variable index
$i\in[0..d-1]$, then an iterated derivative and finally the derivative with
respect to a {\multiindex}~$\aalpha$.
The next challenge is then to switch the derivatives with respect to several
variables.
This holds when the function is regular enough, but a generic theory of
multi-variable differentiation was out of scope of this article, so we decided
to only derive polynomials (of any degree).
On this subset, we proved we could switch iterative derivatives at will and
compute any derivative with respect to a {\multiindex}.
Actually, we do not need the real value of the derivative, only whether it is
zero or not.

We proved that, for all $\aalpha,\bbeta\in\N^d$, there exists
$C_{\aalpha,\bbeta}\in\R$ such that
\begin{equation}
  \label{eq:deriv-X-alpha}
  \Notation{$\p^\bbeta \XX^\aalpha$}
  \eqdef \frac{\p^{\len{\bbeta}} \XX^\aalpha}{%
    \p x_0^{\beta_0}\dots \p x_{d-1}^{\beta_{d-1}}}
  = C_{\aalpha, \bbeta} \XX^{\aalpha - \bbeta}
  \AND
  (C_{\aalpha, \bbeta} = 0 \Equiv \exists i, \alpha_i < \beta_i),
\end{equation}
where~$\aalpha-\bbeta$ denotes the family $(\alpha_i-\beta_i)_{i\in[0..d)}$,
with the subtraction being the {\Rocq} total function~\coqe{sub} on~$\N$, that
returns~0 when~$\alpha_i<\beta_i$ (but in this case~$\XX^{\aalpha-\bbeta}$ is
useless because $C_{\aalpha,\bbeta}=0$).

Now let us go back to prove that $(\XX^{\aalpha})_{\aalpha\in\calAdk}$ is linearly
independent.
Let us assume that we have a family of scalars~$\LL$ such that
$\sum_{\aalpha \in \calAdk}L_\aalpha\XX^\aalpha=0$ (a function equality).
Given any $\bbeta\in\calAdk$, we have to prove that~$L_\bbeta=0$.
By linearity of differentiation with respect to~$\bbeta$, we obtain
$\sum_{\aalpha\in\calAdk}L_\aalpha\p^{\bbeta}\XX^\aalpha=0$.
From~\eqref{eq:deriv-X-alpha}, we deduce that
\begin{equation*}
  \sum_{\substack{\aalpha \in \calAdk \\ \forall i, \beta_i \le \alpha_i}}
  L_\aalpha C_{\aalpha, \bbeta} \XX^{\aalpha - \bbeta}
  = L_\bbeta C_{\bbeta, \bbeta} +
  \sum_{\substack{\aalpha \in \calAdk \\ \forall i, \beta_i \le \alpha_i \\
      \exists j, \beta_j < \alpha_j}}
  L_\aalpha C_{\aalpha, \bbeta} \XX^{\aalpha - \bbeta} = 0.
\end{equation*}
Now let us evaluate these functions at zero, we get
\begin{equation*}
  L_\bbeta C_{\bbeta, \bbeta} +
  \sum_{\substack{\aalpha \in \calAdk \\ \forall i, \beta_i \le \alpha_i \\
    \exists j, \beta_j < \alpha_j}}
  L_\aalpha C_{\aalpha, \bbeta} \zzero^{\aalpha - \bbeta}
  = L_\bbeta C_{\bbeta, \bbeta} = 0.
\end{equation*}
As $C_{\bbeta,\bbeta}\neq0$, we deduce that~$L_\bbeta=0$ and that the
family~$(\XX^\aalpha)_{\aalpha\in\calAdk}$ is linearly independent.
Since~$\matPdk$ is the span of the monomials, it is then straightforward that
they form a basis, and the dimension is the cardinal of~$\calAdk$,
\begin{lstlisting}
Lemma Monom_dk_basis?\lkPdk{Monom_dk_basis}? : basis (Pdk d k) (Monom_dk d k).
Lemma Pdk_has_dim?\lkPdk{Pdk_has_dim}? : has_dim (Pdk d k) (pbinom d k).+1.
\end{lstlisting}

\myskip

For this proof, we first tried to use a formal derivative, but surprisingly it
failed.
The sketch is similar to the previous proof by taking the $\beta$-formal
derivative at point~1.
Similarly, from a $p\in\matPdk$, as the monomials are generators, we write~$p$
as $p=\sum_{\aalpha\in\calAdk}L_\aalpha\XX^\aalpha$, using Hilbert's epsilon to
get $\LL$, and we define its formal derivative as expected.
With the same proof as previously, we prove that the formal derivative, with
respect to a given $\beta$ at point 1 is~$L_\beta$.
But the operation from $p$ to $L_\beta$ is \emph{not} a function.
As we have not yet proved that the monomials are free, there may be another
family $\LL^\prime$ such that $p=\sum_{\aalpha\in\calAdk}L_\aalpha\XX^\aalpha
=\sum_{\aalpha\in\calAdk}L^\prime_\aalpha\XX^\aalpha$.
So this formal derivative may be~$L_\beta$ or~$L^\prime_\beta$ and the Hilbert
epsilon axiom we used does not provide their equality.

\subsubsection{$\matPdk$ for Small Values of $d$ and $k$}
\label{sec:P1k}

Now that we have defined a generic~$\matPdk$, it is useful to focus on the
small values of~$d$ and~$k$ for specific properties and links with the
polynomials defined earlier.

\paragraph{Small $d$.}

First, for~$d=0$, \coqe{'\R^0}~being the unit type~$\{0\}$ (see
Section~\ref{sec:Monoid}), all functions of type~\coqe{FRd 0} are constant
functions, and thus they are also polynomials.
\begin{lstlisting}
Lemma P0k_eq?\lkPdk{P0k_eq}?: \forall k (p : FRd 0), Pdk 0 k p <-> p = fun _ => p 0.
Lemma P0k_full?\lkPdk{P0k_full}?: \forall k, Pdk 0 k = fullset.
\end{lstlisting}

Then, for~$d=1$, we have the standard polynomials of a single variable.
In addition to the canonical basis \coqe{Monom_dk}, we define another
basis~\cite[p.~8]{eg:tpf:04}.
Let~$\aa$ be a family of~$k+1$ (pairwise distinct) reals called \emph{nodes}.
Then, the Lagrange polynomial basis, called \coqelkLik{LagP1k} in {\Rocq}, is
informally defined as
\begin{equation*}
  \forall i \in [0..k],\;
  \forall x \in \R,\quad
  \Notation{$\lagLik_i$} (x) \eqdef
  \prod_{j \in [0..k] \setminus \{ i \}} \frac{x - a_j}{a_i - a_j}.
\end{equation*}
The {\Rocq} definition is a total function that does not require the
injectivity of the nodes~$\aa$.
However, the interesting properties hold when they are pairwise distinct:
each~$\lagLik_i$ evaluates to~1 at its corresponding node~$a_i$, and to~0 at
all others (\coqelkLik{LagP1k_kron});
these polynomials are linearly independent and span the polynomial
space~$\matPik$, forming a basis (\coqelkLik{LagP1k_basis}).
Moreover, the components of any polynomial on this basis are its values taken
at the nodes,
\begin{lstlisting}
Hypothesis Hnode : injective node.
Lemma LagP1k_decomp?\lkLik{LagP1k_decomp}? : \forall {p}, Pdk 1 k p -> p = lin_comb (mapF p node) (LagP1k node).
\end{lstlisting}

\paragraph{Small $k$.}

First, for~$k=0$, zero degree polynomials are constant functions, but
contrarily to the case~$d=0$, they do not span the entire set of functions.
\begin{lstlisting}
Lemma Pd0_eq?\lkPdk{Pd0_eq}? : \forall {d} (p : FRd d), Pdk d 0 p <-> p = fun _ => p 0.
\end{lstlisting}

The most interesting case is~$k=1$.
It corresponds to the affine mappings,
\begin{lstlisting}
Lemma Pd1_am_1_equiv?\lkPdk{Pd1_am_1_equiv}? : \forall {d} (f : FRd d), Pdk d 1 f <-> aff_map_ms f.
\end{lstlisting}
We also provide two other bases for \coqe{Pdk d 1}.
More precisely, the reference Lagrange polynomials~$\refLLdi$ of
Section~\ref{sec:LagPd1-ref} (in \Rocq, \coqelkLdiref{LagPd1_ref}) form a basis
of \coqe{Pdk d 1}.
It is also the case for the current Lagrange polynomials~$\curLLdiv$ of
Section~\ref{sec:LagPd1} (in \Rocq, \coqelkLdi{LagPd1}), provided the
vertices~$\vv$ are affinely independent.
Moreover, just as for~$d=1$, the components of any polynomial on this latter
basis are its values taken at the vertices, that are also the nodes in this
case (see Section~\ref{sec:LagP-ref-cur-nodes}),
\begin{lstlisting}
Hypothesis Hvtx : aff_indep_ms vtx.
Lemma LagPd1_decomp_node?\lkNode{LagPd1_decomp_node}? :
  \forall {p}, Pdk d 1 p -> p = lin_comb (mapF p (node vtx)) (castF _ (LagPd1 Hvtx)).
\end{lstlisting}

\subsubsection{Factorization of Polynomials}
\label{sec:fact-pol}

Polynomials vanishing on a face hyperplane are factorizable by the
corresponding Lagrange polynomial.
These results are based on the Euclidean division of any polynomial by a
monomial.

\myskip

For $d,k\in\N$, given a polynomial~$p$ in~$\matPdpikpi$, its Euclidean division
by the monomial~$X_d$ writes $p=p_0+X_d*p_1$ where~$p_0$ belongs
to~$\matPdkpi$, {\ie} does not depend on the last variable~$x_d$, and~$p_1$
belongs to~$\matPdpik$, {\ie} has a total degree one less than that of~$p$.
More precisely, \coqelkPdk{PSdSk_split} states that,
\[
  \forall p \in \matPdpikpi,\;
  \exists p_0 \in \matPdkpi,\;
  \exists p_1 \in \matPdpik,\;
  \forall \xx \in \R^{d + 1},\quad
  p(\xx) = p_0 (\txx) + x_d * p_1 (\xx),
\]
where $\xx=(\txx,x_d)$ with $\txx\in\Rd$.
The proof path we followed was not found in textbooks, and is original up to
our knowledge.
The idea is first that the existence of these two polynomials is transported by
linear combination.
As~$\matPdpikpi$ is the span of monomials, there is left to prove that the
property holds for monomials.
For $\aalpha\in\calAdpikpi$, we have two cases.
If~$\alpha_d=0$, then we pose $p_0=\XX^{\aalpha}$ and~$p_1=0$.
If~$\alpha_d>0$, then we pose~$p_0=0$ and $p_1=\XX^{\aalpha-\ddelta_{d}}$.

This lemma is interesting as it could have been the definition of~$\matPdk$.
Instead of defining~$\matPdk$ as the span of the monomials (see
Section~\ref{sec:Pdk-def}) defined from the {\multiindices}, we could have
defined~$\matPdk$ by induction as $\matPdpikpi=\matPdkpi+X_d*\matPdpik$, with
initializations from Section~\ref{sec:P1k}.

\myskip

When~$d>0$, the following lemma states that any polynomial of degree at
most~$k+1$ ({\ie} in~$\matPdkpi$) cancels on the last reference face
hyperplane~$\hHpl_d$ (see Section~\ref{sec:Hface}) iff it factors
into~$\refLdi_d*q$ for some polynomial~$q$ of degree at most~$k$ ({\ie}
in~$\matPdk$).
More formally,
\[
  \forall p \in \matPdkpi,\quad
  (\forall \hxx \in \hHpl_d,\; p (\hxx) = 0) \IMPLIES
  \exists q \in \matPdk,\; \ p = \refLdi_d * q.
\]
In {\Rocq},
\begin{lstlisting}
Lemma factor_zero_on_last_Hface_ref?\lkLdiref{factor_zero_on_last_Hface_ref}? : \forall {p}, Pdk d k.+1 p ->
  (\forall\nsp x, Hface_ref ord_max x -> p x = 0) -> \exists q, Pdk d k q /\ p = LagPd1_ref ord_max * q.
\end{lstlisting}
\begin{proof}
The proof relies on the Euclidean division \coqe{PdSk_split} above.
Given~$p\in\matPdkpi$, there exists~$\tpo$\inRR$\matPdmik$ and~$q\in\matPdk$
such that $p=\tpo+X_{d-1}*q=\tpo+\refLdi_d*q$.
Thus, for any~$\txx\in\Rdmi$, using \coqelkLdiref{Hface_ref_is_Ker} from
Section~\ref{sec:Hface}, we have
\[
  \xx \eqdef (\txx, 0) \in \hHpl_d = \ker \refLdi_d \subset \Rd
  \AND
  0 = p (\xx) = \tpo (\txx) + 0 * q (\xx).
\]
Hence, $\tpo=0$ and $p=\refLdi_d*q$.
\end{proof}

The slight technical point is the handling of the injection and projection
between~$\Rdmi$ and~$\Rd$, respectively using \coqe{insertF _ 0 ord_max} and
\coqe{widenF_S} (see Section~\ref{sec:FF}).

\myskip

The previous result is actually valid for any reference face hyperplane, but
such generality is more profitable on a current element.
Given vertices~\coqe{vtx} and a proof \coqe{Hvtx : aff_indep_ms vtx} of their
affine independence, a similar result holds for cancellation on any face
hyperplane of the current element,
\begin{lstlisting}
Lemma factor_zero_on_Hface?\lkLdi{factor_zero_on_Hface}? : \forall {p} i, Pdk d k.+1 p ->
  (\forall\nsp x, Hface vtx i x -> p x = 0) -> \exists q, Pdk d k q /\ p = LagPd1 Hvtx i * q.
\end{lstlisting}
\begin{proof}
The proof (both on paper and in {\Rocq}) uses the affine geometric
mapping~$\TgeomTranspiv$ that maps the reference vertices~$\hvv$ onto the
transposition of vertices~$\vv$ where the vertex~$\vv_i$ is swapped with the
last one~$\vv_d$ (see Section~\ref{sec:Tgeom-swap}).
Given~$p\in\matPdkpi$ vanishing on~$\Hplv_i$,
let~$\hp\eqdef p\circ\TgeomTranspiv$.
Then, $\hp\in\matPdkpi$ because~$\TgeomTranspiv$ is affine, and it vanishes
on~$\hHpl_d$ because~$\TgeomTranspiv(\hHpl_d)=\Hplv_i$ (see
Section~\ref{sec:Tgeom-swap}).
Hence, the previous result states the existence of~$\hq\in\matPdk$ such that
$\hp=\refLdi_d*\hq$.
Finally, the bijectivity of~$\TgeomTranspiv$ provides
\begin{equation*}
  p = \hp \circ (\TgeomTranspiv)^{-1}
  = \refLdi_d \circ (\TgeomTranspiv)^{-1} \ *\ \hq \circ (\TgeomTranspiv)^{-1}
  = \curLdiv_i * \hq \circ (\TgeomTranspiv)^{-1},
\end{equation*}
and~$q\eqdef\hq\circ(\TgeomTranspiv)^{-1}\in\matPdk$ answers the question.
\end{proof}

\section{$\FElagPdk$ Simplicial Lagrange Finite Element}
\label{sec:FELagP}

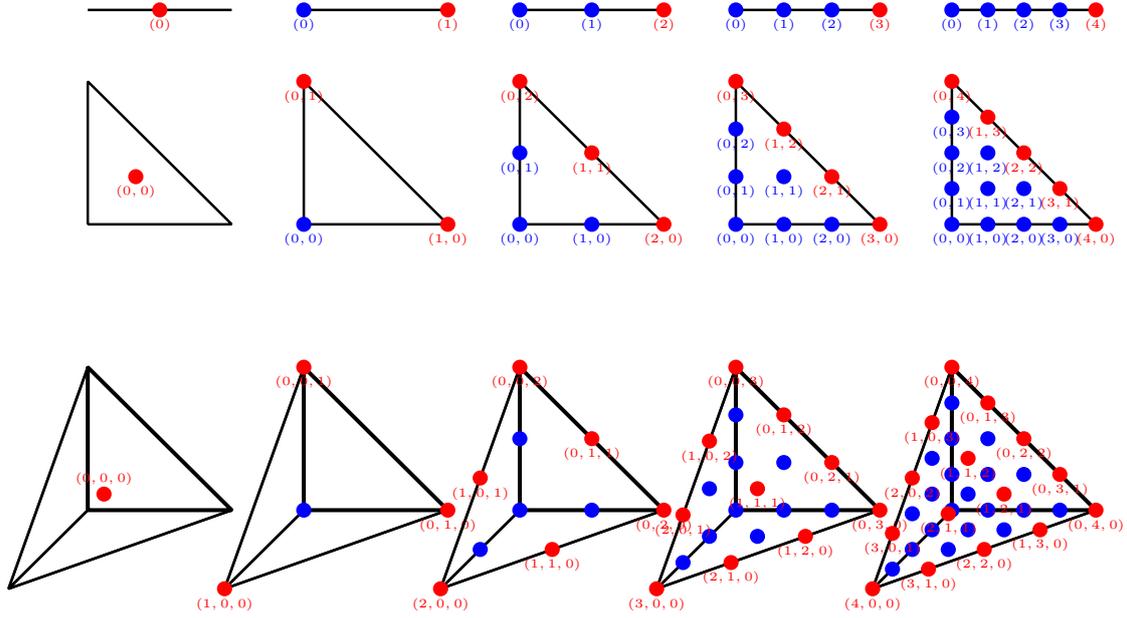
\begin{figure}[ht]
  \centering
  \resizebox{1\linewidth}{!}{\input{fig_SegTriaTet_1to4}}
  \caption[Lagrange nodes in~1D, 2D and~3D]{%
    Lagrange nodes of the reference simplex for $d\in\{1,2,3\}$ (from
    top to bottom), and~$k$ in~$\{0,1,2,3,4\}$ (from left to right).
    The nodes corresponding to the highest degree are depicted in red, and the
    others in blue.
    For readability, the {\multiindices} are indicated in all cases
    when~$d=1,2$, and only for the highest degree when~$d=3$.}
  \label{fig:Lag:nodes}
\end{figure}

The Lagrange~FE are nodal~FE.
This means that the degrees of freedom~$\SSigma$ are evaluations at specific
points in the geometric element, called \emph{nodes}.
These nodes are the interpolation points where the approximate solution of the
PDE is computed.
The Lagrange nodes are characterized by the fact that they are evenly
distributed over the element, thus, given a spatial dimension~$d>0$, they are
fully defined by choosing the vertices~$\vv$ and the approximation degree~$k$,
as illustrated in Figure~\ref{fig:Lag:nodes}.
The simplicial Lagrange~FE correspond to simplicial geometries,
and to the approximation space~$\matPdk$ of polynomials in~$d$ space variables
and of (total) degree at most~$k$ (see Section~\ref{sec:Pdk}).

For {\nonzero}~$k$, Section~\ref{sec:LagP-nodes} is dedicated to the
Lagrange nodes and their properties, Section~\ref{sec:LagP-dof} to the Lagrange
degrees of freedom, and Section~\ref{sec:LagP-unisolv} to the proof of
unisolvence.
The degenerate case~$k=0$ is treated separately in
Section~\ref{sec:LagP-degen}.
And finally, the~$\FElagPdk$ simplicial Lagrange~FE is built in
Section~\ref{sec:FELagP-constr} for any~$d$ and~$k$, and
Section~\ref{sec:face-unisolvence} is devoted to the face unisolvence
property.

\subsection{Simplicial Lagrange Nodes}
\label{sec:LagP-nodes}

One of the strengths of~FEM is to bring all computations in mesh cells of
various size and aspect ratio back to the sole case of the reference unit cell.
The same operating mode can be used in the process of building~FE
such as the simplicial Lagrange~FE: definitions and properties are first
established in the case of the reference simplex (see
Section~\ref{sec:geom-ref}), and then transported to the current case by means
of the bijective affine geometric transformation~$\Tgeomv$ (see
Section~\ref{sec:Tgeom}).

Following Section~\ref{sec:geometry}, reference and current vertices are
respectively denoted~\Notation{$\hvv$} and~\Notation{$\vv$}.
The reference vertices are always affinely independent, while this is assumed
for~$\vv$ when necessary.
Section~\ref{sec:LagP-ref-cur-nodes} is dedicated to the definition of the
Lagrange nodes.
The reference nodes only depend on~$d$ and~$k$.
They are denoted~\Notation{$\haa$}, and sometimes~\Notation{$\haa^d$},
or~\Notation{$\haa^k$}, to emphasize the dimension, or the degree of
approximation.
Given (current) vertices~$\vv$, the corresponding nodes are
denoted~\Notation{$\aa^\vv$}.
The interaction of nodes with face hyperplanes is discussed in
Section~\ref{sec:LagP-nodes-faces}.
Section~\ref{sec:LagP-nodes-incr-dim} focuses on the increment in the spatial
dimension, and Section~\ref{sec:LagP-sub-nodes} on the increment in the degree
of approximation, as shown in Figure~\ref{fig:Lag:nodes}.
In the latter section, reference and current {\subvertices} are respectively
denoted~\Notation{$\uhvv$} and~\Notation{$\uvv$}, and reference and current
{\subnode}s~\Notation{$\uhaa$} and~\Notation{$\uaa^\vv$}.
Recall from Section~\ref{sec:MultiIndex} that {\multiindices} used to number
nodes are denoted~\Notation{$\aalpha$}.
Usually, $\aalpha$ belongs to~$\calAdk$, and~\Notation{$\aalphap$} with a prime
is used to denote a ``smaller'' {\multiindex}, either in~$\calAdmik$ or
in~$\calAdkmi$.

In this section, the spatial dimension~$d$ and the approximation degree~$k$ are
assumed positive, and sometimes, we need~$k>1$.
However, in the {\Rocq} formalization, all definitions are total functions,
also set for~$d=0$ or~$k=0$.
For instance, the inverse of {\noninvertible} elements in rings (such as~0
in~$\R$) is set to~0, but this special value is not exploited, while some
properties may need additional hypotheses.

\subsubsection{Lagrange Reference and Current Nodes}
\label{sec:LagP-ref-cur-nodes}

The Lagrange reference nodes (see Figure~\ref{fig:Lag:nodes}) and the current
nodes associated with a family~$\vv$ of~$d+1$ vertices in~$\Rd$ are defined as
\begin{equation}
  \label{eq:lagP-nodes}
  \forall \aalpha \in \calAdk,\quad
  \haa_\aalpha \eqdef \Scalinvk \aalpha,
  \AND
  \aa^\vv_\aalpha\eqdef\Tgeomv(\haa_\aalpha),
\end{equation}
where~$\Scal{\frac{1}{k}}$ is the scaling by~$\frac{1}{k}$, see
Section~\ref{sec:Tgeom}.
In the sequel, the latter is simplified under the compact
form~$\aa^\vv=\Tgeomv(\haa)$.
In {\Rocq},
\begin{lstlisting}
Definition node_ref?\lkNoderef{node_ref}? : [0..pbinom d k] -> '\R^d :=
  fun idk => scal (/ INR k) (mapF INR (Adk d k idk)).
Definition node?\lkNode{node}? (vtx : [0..d] -> '\R^d) : [0..pbinom d k] -> '\R^d :=
  mapF (T_geom vtx) node_ref.
\end{lstlisting}
where \coqe{INR : \nat -> \R} is the injection of~$\N$ into~$\R$.
The finite family operator \coqelkFF{mapF} (see Section~\ref{sec:FF}), and
\coqe{scal} (from the {\Coquelicot} library, see also
Section~\ref{sec:ModuleSpace}), allow higher-level proofs, directly on the
whole family of points, just as in pen-and-paper proofs.
Since the geometric transformation for the reference vertices~$\hvv$ is the
identity map (see Section~\ref{sec:Tgeom}), we have~$\aa^\hvv=\haa$.

From the definition of the reference Lagrange polynomials (see
Section~\ref{sec:LagPd1-ref}) and of the geometric transformation~$\Tgeomv$
(see Section~\ref{sec:Tgeom}), we have for all $\aalpha\in\calAdk$,
\begin{gather}
  \label{eq:LagPd1-node-ref}
  \refLdi_0 (\haa_\aalpha) = 1 - \frac{\len{\aalpha}}{k},
  \qquad
  \forall i \in [1..d],\;
  \refLdi_i (\haa_\aalpha) = \frac{\alpha_{i - 1}}{k},\\
  \nonumber
  \mbox{and}\quad
  \aa^\vv_\aalpha =
  \left( 1 - \frac{\len{\aalpha}}{k} \right) \vv_0 +
  \sum_{i = 0}^{d - 1} \frac{\alpha_i}{k} \vv_{i + 1}.
\end{gather}
Thus, vertices are obviously specific nodes: $\vv_0=\aa^\vv_\zzero$ and
$\vv_{i+1}=\aa^\vv_{\ddelta_i}$ (see Figure~\ref{fig:Lag:nodes}).
Moreover, when~$k=1$, the nodes are exactly the vertices
(\coqelkNode{node_vtx_d1}).

The reference nodes~$\haa$ are pairwise distinct
(\coqelkNoderef{node_ref_inj}).
Thus, there are~$\card(\calAdk)=\binom{d+k}{d}$ distinct reference nodes.
When the vertices~$\vv$ are affinely independent, the weights
$1-\frac{\len{\aalpha}}{k}$ and $(\frac{\alpha_i}{k})_{i\in[0..d)}$ are the
barycentric coordinates of the nodes with respect to the vertices (their sum
is~1, see Section~\ref{sec:LagPd1-ref}).
Moreover, in this case, the geometric transformation is bijective, the Lagrange
nodes are also pairwise distinct, and their number is also given by
\begin{lstlisting}
Definition nnode_LagPdk?\lkFELagP{nnode_LagPdk}? (d k : \nat) : \nat\sp:= (pbinom d k).+1.
\end{lstlisting}
It stands as the number of Lagrange degrees of freedom in
Section~\ref{sec:LagP-dof}.

\myskip

The main characteristic of these nodes is that they are evenly distributed.
The proof path for unisolvence is much simpler in this case as subnodes are
also nodes (see Sections~\ref{sec:LagP-sub-nodes} and~\ref{sec:LagP-unisolv}).
Although unisolvence clearly holds for less regular distributions of nodes
(more in \AppSection~\ref{sec:FE:zoo}), it is nevertheless difficult to express
this hypothesis: for instance, we require the nodes not to be all clumped in
one face or one hyperplane.

\subsubsection{Lagrange Nodes and Faces}
\label{sec:LagP-nodes-faces}

Some Lagrange nodes live on the (hyper)faces ({\ie} the boundaries) of the
simplex, and thus on the face hyperplanes that carry the faces (see
Section~\ref{sec:Hface} and Figure~\ref{fig:geo-hyper-map_d32}).
These nodes can simply be characterized by their {\multiindex}: either of
sum~$k$, or with a zero component.

The face hyperplanes being the kernels of the Lagrange polynomials (see
Section~\ref{sec:Hface}), this is straightforward
from~\eqref{eq:LagPd1-node-ref}, and the definition of~$\calAdki$ (see
Section~\ref{sec:face-multi-ind}) for the reference nodes~$\haa$.
And then, when the vertices~$\vv$ are affinely independent, the
bijective~$\Tgeomv$ transports reference vertices, face hyperplanes and nodes
to their current counterparts.
For all $i\in[0..d]$ and $\aalpha\in\calAdk$, we have
\begin{equation}
  \label{eq:node-face}
  \aa^\vv_\aalpha \in \Hplv_i
  \EQUIV \Tgeomv (\haa_\aalpha) \in \Tgeomv (\hHpl_i)
  \EQUIV \haa_\aalpha \in \hHpl_i \EQUIV \aalpha \in \calAdki.
\end{equation}
This is formalized in {\Rocq} as
\begin{lstlisting}
Context {d1 k : nat}.
Let d := d1.+1.
Hypothesis Hk : k <> 0.

Context {vtx : [0..d] -> '\R^d}.
Hypothesis Hvtx : aff_indep_ms vtx.

Lemma node_Hface?\lkNode{node_Hface}? : \forall i (idk : [0..pbinom d k]),
  Hface vtx i (node vtx idk) <-> in_ASdki i idk.
\end{lstlisting}
Note that~$d>0$ is granted here structurally by \coqe{d := d1+1}, which is
needed for the predicate \coqelkMI{in_ASdki}.

\subsubsection{Incrementing the Spatial Dimension}
\label{sec:LagP-nodes-incr-dim}

We present now results used in Section~\ref{sec:LagP-unisolv} for the proof of
unisolvence in the induction step incrementing the spatial dimension (and to
get face unisolvence in Section~\ref{sec:face-unisolvence}).

In Section~\ref{sec:Hface}, the injection $\InjhHdi:\ArRdmiRd$ for
$i\in[0..d]$ maps~$\Rdmi$ onto~$\hHpl_i$, either by inserting in front one
minus the sum ($i=0$), or by inserting~0 at position~$i-1$ ($i\neq0$).
It also maps the reference nodes in~$\Rdmi$ onto the reference nodes in~$\Rd$
belonging to~$\hHpl_i$ (see Figure~\ref{fig:geo-hyper-map_d32} for the
case~$i=0$),
\begin{equation*}
  \forall \aalphap \in \calAdmik,\quad
  \InjhHdi (\haa^{d-1}_\aalphap) = \haa^d_{\InjAdki (\aalphap)},
\end{equation*}
where the~$^{d-1}$ and~$^d$ superscripts indicate the dimension in which the
reference nodes live.
In Section~\ref{sec:face-multi-ind}, the injection~$\InjAdki:\ArNdmiNd$
maps~$\calAdmik$ onto~$\calAdki$, either by inserting in front~$k$ minus the
sum ($i=0$), or by inserting~0 at position~$i-1$ ($i\neq0$).
The proof is merely a matter of commutativity with the scaling
by~$\frac{1}{k}$.
From~\eqref{eq:lagP-nodes}, \eqref{eq:inj:Hdo}, and~\eqref{eq:inj:adki0},
when~$i=0$, we have
\begin{equation*}
  \InjhHdo (\haa^{d - 1}_\aalphap)
  = \left( 1 - \sum_{i=0}^{d-1} \frac{\alphap_i}{k}, \Scalinvk \aalphap \right)
  = \Scalinvk (k - \len{\aalphap}, \aalphap)
  = \haa^d_{\InjAdko (\aalphap)}.
\end{equation*}
And this is even simpler when~$i\neq0$ as the insertion of~0 is linear.

Then, by applying the geometric transformation on both sides, we obtain
\begin{equation}
  \label{eq:Inj-Hface-node}
  \forall \aalphap \in \calAdmik,\quad
  \Injvi (\haa^{d-1}_\aalphap) = \aa^\vv_{\InjAdki (\aalphap)},
\end{equation}
and in {\Rocq},
\begin{lstlisting}
Context {d1 k : nat}.
Let d := d1.+1.
Hypothesis Hk : k <> 0.
Variable vtx : [0..d] -> '\R^d.

Lemma inj_Hface_node?\lkNode{inj_Hface_node}? : \forall {i} (idk : [0..pbinom d1 k]),
  inj_Hface vtx i (node_ref idk) = node vtx (inj_ASdki i idk).
\end{lstlisting}
Note that the vertices~\coqe{vtx} do not need here to be affinely independent.

\subsubsection{Incrementing the Degree of Approximation: Lagrange {\Subnode}s}
\label{sec:LagP-sub-nodes}

\begin{figure}[ht]
  \centering
  \resizebox{0.5\linewidth}{!}{\input{fig_Tet_k3_k2}}
  \caption[{\Subvertices} and Lagrange {\subnode}s in~3D]{%
    Passage from the reference {\subnode}s of~$\FElagP{3}{3}$ (degree~$\leq3$)
    to the reference nodes of~$\FElagP{3}{2}$ (degree~$\leq2$) in the
    case~$d=k=3$.
    Plot of {\subvertices}~$(\uhvv_i)_{i\in[0..d]}$ and
    {\subnode}s~$(\uhaa_{\aalphap})_{\aalphap\in\calA^3_2}$.\\
    The nodes~$(\haa_{\aalpha})_{\aalpha\in\calA^3_3}$ are set in the tetrahedron
    defined by the vertices $(\hvv_0,\hvv_1,\hvv_2,\hvv_3)$.\\
    The {\subvertices} are
    $(\hvv_0,\haa_{2\ddelta_0},\haa_{2\ddelta_1},\haa_{2\ddelta_2})$.
    The {\subnode}s~$(\uhaa_{\aalphap})_{\aalphap\in\calA^3_2}$ are defined with
    respect to the tetrahedron whose vertices are~$(\uhvv_i)_{i\in[0..d]}$.
    These are the original nodes~$(\haa_{\aalpha})_{\aalpha\in\calA^3_3}$,
    except the (small blue) nodes corresponding to~$k=3$ (with indices
    in~$\calC^3_3$).\\
    Thus, geometrically, passing from~$\matPdk$ to~$\matPdkmi$ amounts to
    remove the face hyperplane~$\hHpl_0$ from the tetrahedron defined
    by~$(\hvv_0,\hvv_1,\hvv_2,\hvv_3)$.}
  \label{fig:sub-vert-sub-nodes_d3-k3}
\end{figure}

{\Subnode}s are the key ingredient in Section~\ref{sec:LagP-unisolv} for the
proof of unisolvence in the induction step incrementing the degree of
approximation.
And this works well because Lagrange nodes are evenly distributed.

What we call reference and current ``{\subvertices}''~$\uhvv$ and~$\uvv$, and
reference and current ``{\subnode}s''~$\uhaa$ and~$\uaa$ are defined by
(see Figure~\ref{fig:sub-vert-sub-nodes_d3-k3})
\begin{align}
  \label{eq:sub-vtx}
  \uhvv &\eqdef \Scalratio \hvv, &\quad
  \uvv &\eqdef \Tgeomv \uhvv,\\
  \label{eq:sub-nodes}
  \forall \aalphap \in \calAdkmi,\quad
  \uhaa_\aalphap &\eqdef \Scalratio \haa^{k - 1}_\aalphap, &\quad
  \uaa^\vv_\aalphap &\eqdef \Tgeomv \uhaa_\aalphap,
\end{align}
where the~$^{k-1}$ superscript indicates that these reference nodes correspond
to the degree of approximation~$k-1$.
And in {\Rocq}, with vertices~\coqe{vtx} of type~\coqe{[0..d] -> '\R^d},
\begin{lstlisting}
Definition sub_vtx_ref?\lkNoderef{sub_vtx_ref}? k : [0..d] -> '\R^d := mapF (scal (INR k.-1 / INR k)) vtx_ref.
Definition sub_vtx?\lkNode{sub_vtx}? k vtx : [0..d] -> '\R^d := mapF (T_geom vtx) (sub_vtx_ref k).
Definition sub_node_ref?\lkNoderef{sub_node_ref}? : [0..pbinom d k.-1] -> '\R^d :=
  mapF (scal (INR k.-1 / INR k)) node_ref.
Definition sub_node?\lkNode{sub_node}? k vtx : [0..pbinom d k.-1] -> '\R^d :=
  mapF (T_geom vtx) sub_node_ref.
\end{lstlisting}

Consider now~$k>1$.
The reference {\subvertices}~$\uhvv$ are not reference vertices, but they are
affinely independent.
In the same way, when the vertices~$\vv$ are affinely independent
(and~$\Tgeomv$ is bijective), the {\subvertices}~$\uvv$ are also affinely
independent.
On the contrary, the reference {\subnode}s~$\uhaa$ are some of the reference
nodes, and the {\subnode}s~$\uaa^\vv$ are some of the nodes~$\aa^\vv$.
Indeed, from~\eqref{eq:lagP-nodes} and~\eqref{eq:sub-nodes}, and since~$k>1$,
we have for all~$\aalphap\in\calAdkmi$,
\begin{align}
  \nonumber
  \uhaa_\aalphap &= \Scalratio \Scalinvkmi \aalphap
  = \Scalinvk \aalphap = \haa^{k}_\aalphap,\quad\mbox{and}\\
  \label{eq:subnodes-some-nodes-k}
  \uaa^\vv_\aalphap &= \Tgeomv \uhaa_\aalphap
  = \Tgeomv \haa^k_\aalphap = \aa^{\vv, k}_\aalphap.
\end{align}
This is formalized in {\Rocq} as
\begin{lstlisting}
Lemma sub_node_eq?\lkNode{sub_node_eq}? : sub_node k vtx = widenF _ (node vtx).
\end{lstlisting}
where \coqelkFF{widenF}~\coqe{_} only keeps the first {\multiindices}
in~$\calAdkmi$ (see Section~\ref{sec:FF}).
Moreover, the {\subnode}s~$\uaa^\vv$ are also the nodes~$\aa^{\uvv,k-1}$, of
degree~$k-1$, associated with the {\subvertices},
\begin{lstlisting}
Lemma sub_node_node?\lkNode{sub_node_node}? : sub_node k vtx = node (sub_vtx k vtx).
\end{lstlisting}
Indeed, from~\eqref{eq:lagP-nodes}, \eqref{eq:sub-vtx}, \eqref{eq:sub-nodes},
\eqref{eq:Tgeom_v_circ_Tgeom_w} and~\eqref{eq:Tgeom-scal}, (the latter with
$\frac{k-1}{k}\neq0$), we have
\begin{align}
  \nonumber
  \forall \aalphap \in \calAdkmi,\quad
  \aa^{\uvv, k - 1}_\aalphap &= \Tgeom{\Tgeomv \uhvv} \haa^{k - 1}_\aalphap
  = \Tgeomv \circ \Tgeomuhv \haa^{k - 1}_\aalphap
  = \Tgeomv \circ \Tgeom{\Scalratio \hvv} \haa^{k - 1}_\aalphap\\
  \label{eq:subnodes-nodes-km1}
  &= \Tgeomv \Scalratio \haa^{k - 1}_\aalphap = \uaa^\vv_\aalphap.
\end{align}

And finally, when the vertices~$\vv$ are affinely independent, the
{\subnode}s~$\uaa^{\vv}$ are not in the face hyperplane~$\Hplv_0$ opposite the
vertex~$\vv_0$,
\begin{lstlisting}
Context {vtx : [0..d] -> '\R^d}.
Hypothesis Hvtx : aff_indep_ms vtx.

Lemma sub_node_out_Hface_0?\lkNode{sub_node_out_Hface_0}? : \forall idk, ~ Hface Hvtx ord0 (sub_node k vtx idk).
\end{lstlisting}
Indeed, from~\eqref{eq:node-face}, the reference nodes $\haa\in\hHpl_0$
correspond to {\multiindices} in~$\calCdk$, which is disjoint
from~$\calAdkmi$, see~\eqref{eq:Cdk-Adk}.
Thus, for all~$\aalphap\in\calAdkmi$, $\uhaa_\aalphap\nin\hHpl_0$, and
from~\eqref{eq:node-face}, we have $\uaa^\vv_\aalphap\nin\Hplv_0$.

\subsection{Lagrange Degrees of Freedom}
\label{sec:LagP-dof}

The nodal Lagrange degrees of freedom are the point evaluation at the Lagrange
nodes defined above.
They are expressed in {\Rocq} as
\begin{lstlisting}
Context {d : nat}.
Variable k : nat.
Variable vtx : [0..d] -> '\R^d.

Definition FE_LagPSdSk_S_dof?\lkFELagP{FE_LagPSdSk_S_dof}? : [0..nnode_LagPdk d k) -> FRd d -> \R :=
  fun idk p => p (node vtx idk).
\end{lstlisting}
The linearity of the Lagrange degrees of freedom
(\coqelkFELagP{FE_LagPSdSk_S_dof_lm}) simply comes from the generic result
\coqelkLM{pt_eval_lm} for any point evaluation function.

\myskip

From Section~\ref{sec:FE:Coq}, the unisolvence property to establish for the
construction of an instance of type~\coqelkFE{FE} is that the restriction
to~$\matPdk$ of the linear mapping~$\Phi_{\SSigma^{\FElagPdk}}$ has a zero
kernel.
Mathematically, this property is expressed as the following
predicate~$\PropPP$, for all $d,k\in\Nstar$ (zero values are handled
otherwise; note also that the degrees of freedom~$\SSigma^{\FElagPdk}$ actually
depend on the vertices~$\vv$),
\begin{equation}
  \label{eq:def-unisolv-inj-dSk}
  \Notation{$\PropPP$} (d, k) \eqdef
  \forall \vv \in (\Rd)^{d + 1} \mbox{ affinely  independent, }
    \ker \Phi_{\SSigma^{\FElagPdk}} \cap \matPdk = \{ 0 \}.
\end{equation}
We recall that for a function~$f:E\to F$ and~$P\subset E$, the property
$\ker f\cap P=\{0\}$ is formalized in {\Rocq} as
\coqelkMonS{KerS0}~\coqe{P f}.
Moreover, it is convenient to start unisolvence proofs by applying
\coqelkMonS{KerS0_gather_equiv} to expose the internals of
\coqe{KerS0 _ (gather _)} (see Sections~\ref{sec:Ker} and~\ref{sec:FE:Coq}).
This means that to prove~$\PropPP(d,k)$, one has simply to prove that for any
$p\in\matPdk$ such that $\forall\aalpha\in\calAdk$, $p(\aa_{\aalpha})=0$, we
have~$p=0$.

\subsection{Unisolvence: Double Induction for {\Nonzero} $d$ and $k$}
\label{sec:LagP-unisolv}

When~$d$ and~$k>0$, the unisolvence property stated
in~\eqref{eq:def-unisolv-inj-dSk} is formalized in {\Rocq} as
\begin{lstlisting}
Let UP d k := \forall vtx, aff_indep_ms vtx ->
  KerS0 (Pdk d k) (gather (FE_LagPSdSk_S_dof k vtx)).
Theorem FE_LagPSdSk_unisolvence_inj?\lkFELagP{FE_LagPSdSk_unisolvence_inj}? : \forall {d k}, d <> 0 -> k <> 0 -> UP d k.
\end{lstlisting}

We mostly follow the proofs from~\cite[Sec.~7.4, pp.~79--80]{eg:FE1:21}.
The proof is conducted as a double induction on~$d$ and~$k$, both starting at~1
(see Section~\ref{sec:FE-degenerate} for $d=0$, and
Section~\ref{sec:LagP-degen} for $k=0$).
Thus, it is sufficient to prove three lemmas, first the two base cases
\begin{lstlisting}
Lemma FE_LagP1Sk_unisolvence_inj?\lkFELagP{FE_LagP1Sk_unisolvence_inj}? : \forall {k}, k <> 0 -> UP 1 k.
Lemma FE_LagPSd1_unisolvence_inj?\lkFELagP{FE_LagPSd1_unisolvence_inj}? : \forall {d}, d <> 0 -> UP d 1.
\end{lstlisting}
and then the (double) induction step
\begin{lstlisting}
Lemma FE_LagPSdSk_unisolvence_inj_ind?\lkFELagP{FE_LagPSdSk_unisolvence_inj_ind}? :
  \forall {d k}, d <> 0 -> k <> 0 -> UP d.+1 k -> UP d k.+1 -> UP d.+1 k.+1.
\end{lstlisting}

\myskip

When~$d=1$ and~$k\neq0$, or when~$k=1$ and~$d\neq0$, both proofs follow
directly from the fact that the components of any polynomial in~$\matPik$, or
in~$\matPdi$, on the corresponding Lagrange basis are simply given by its
values at the Lagrange nodes, see lemmas \coqelkLik{LagP1k_decomp} and
\coqelkNode{LagPd1_decomp_node}, both in Section~\ref{sec:P1k}.
Thus, canceling the values at the nodes is sufficient to cancel the polynomial
everywhere.
Note that when~$d=1$, the unisolvence proof is actually established for any~$k$
in the more general case of any pairwise distinct nodes
(\coqelkFELagP{FE_LagP1k_unisolvence_inj}), which is the case for the evenly
distributed Lagrange nodes.

\myskip

When~$d$ and~$k>0$, let~$\vv$ be~$d+2$ affinely independent vertices
in~$\Rdpi$, and let~$p\in\matPdpikpi$ such that $p(\aa^\vv_\aalpha)=0$ for
all~$\aalpha\in\calAdpikpi$ ($\aa^\vv$ are the Lagrange nodes associated with
the vertices~$\vv$, see Section~\ref{sec:LagP-ref-cur-nodes}).

\paragraph{Step~1: Factorization.}

\begin{proof}
Let~$p_0\eqdef p\circ\Injvo\in\FRd$,
where~$\Injvo:\Arrow{\Rd}{\Hplv_0}\subset\Rdpi$ is the bijective affine
transformation mapping the reference simplex~$\hKd$ onto the face included
in~$\Hplv_0$, opposite vertex~$\vv_0$ (see its definition and main properties
in Section~\ref{sec:Inj-Hface}).

Then, $p_0\in\matPdkpi$ (from \coqelkTgeom{inj_Hface_comp}), and
\coqelkNode{inj_Hface_node} states that~$\Injvo$ transports the reference nodes
of~$\hKd$ onto the nodes of~$\Hplv_0$, see~\eqref{eq:Inj-Hface-node}.
Hence, $p_0$ vanishes on those reference nodes, and applying the injectivity
induction hypothesis~$\PropPP(d,k+1)$ provides its cancellation on the
whole~$\Rd$.

Then, the bijectivity of~$\Injvo$ provides
$\restr{p}{\Hplv_0}=p_0\circ(\Injvo)^{-1}=0$, and there exists~$q$
in~$\matPdpik$ such that $p=\curLdiv_0*q$ (from
\coqelkLdi{factor_zero_on_Hface}, see Section~\ref{sec:fact-pol}).
\end{proof}

\myskip

Note that the vanishing of~$p$ on the face hyperplane~$\Hplv_0$ is actually
established at no extra cost for any face hyperplane~$\Hplv_i$
for~$i\in[0..d]$ (\coqelkFELagP{unisolvence_inj_Hface}), since it is needed
in Section~\ref{sec:face-unisolvence} for face unisolvence.

\paragraph{Step~2: Cancellation.}

\begin{proof}
By hypothesis, $p=\curLdiv_0*q$ vanishes on all Lagrange
nodes~$\aa^\vv_\aalpha$, {\ie} for all~$\aalpha$ in~$\calAdpikpi$.
Hence, $q$~vanishes on all nodes not in~$\ker\curLdiv_0=\Hplv_0$, thus
corresponding to the {\multiindices}
in~$\calAdpik(=\calAdpikpi\setminus\calC^{d+1}_{k+1})$,
see~\eqref{eq:node-face}, \eqref{eq:Adki}, and~\eqref{eq:Cdk-Adk}.
This means that~$q$ vanishes on the nodes~$\aa^{\vv,k+1}_\aalphap$ for all
$\aalphap\in\calAdpik$.

The {\subvertices}~$\uvv$, and the associated
{\subnode}s~$(\uaa^{\vv}_{\aalphap})_{\aalphap\in\calAdpik}$ defined in
Section~\ref{sec:LagP-sub-nodes} are precisely designed to allow the
application of the injectivity induction hypothesis~$\PropPP(d+1,k)$ to~$q$.
Indeed, these {\subnode}s~$\uaa^{\vv}_{\aalphap}$ are in the same time the
above-mentioned nodes~$\aa^{\vv,k+1}_\aalphap$ (where $q$\inRR$\matPdpik$
vanishes), see~\eqref{eq:subnodes-some-nodes-k}, and the
nodes~$\aa^{\uvv,k}_\aalphap$ of degree~$k$ for the {\subvertices}~$\uvv$ (for
which~$\PropPP(d+1,k)$ can be applied), see~\eqref{eq:subnodes-nodes-km1}.
And thus, we obtain the cancellation of~$q$, and of~$p$.
\end{proof}

As the induction step is done, we have proved that~$\PropPP$ (that is~\coqe{UP}
in {\Rocq}) holds for {\nonzero}~$d$ and~$k$.
This hard proof is required to build the Lagrange~FE in
Section~\ref{sec:FELagP-constr}.

\subsection{Degenerate Case $k=0$}
\label{sec:LagP-degen}

The expression and proof path of unisolvence being different in the case~$k=0$,
it is treated separately.
When~$k=0$, the approximation polynomials are constant, thus any point in the
geometric element could serve as the single node.
The isobarycenter of the vertices is chosen for symmetry reasons, and because
it is not on the boundary, thus making less interactions between neighbor
elements when considering a mesh (more in \AppSection~\ref{sec:mesh}).
Thus, vertices are no longer nodes, and the affine independence hypothesis on
the vertices is no longer needed for the unisolvence property.

The nodes and degrees of freedom are formalized as
\begin{lstlisting}
Definition node_d0?\lkNode{node_d0}? (vtx : [0..d] -> '\R^d) : '\R^d := isobarycenter_ms vtx.
Definition FE_LagPd0_S_dof?\lkFELagP{FE_LagPd0_S_dof}? (vtx : [0..d] -> '\R^d) :
  [0..nnode_LagPdk d 0) -> FRd d -> \R\sp := fun _ p => p (node_d0 vtx).
\end{lstlisting}
And the predicate~$\PropPP$ to grant unisolvence simplifies into
\begin{equation*}
  \forall d > 0,\quad
  \Notation{$\PropPP$} (d, 0) \eqdef
  \ker \Phi_{\SSigma^{\FElagPdo}} \cap \matPdo = \{ 0 \}.
\end{equation*}
In {\Rocq},
\begin{lstlisting}
Let UP0 d := \forall vtx, KerS0 (Pdk d 0) (gather (FE_LagPd0_S_dof vtx)).
Lemma FE_LagPd0_unisolvence_inj?\lkFELagP{FE_LagPd0_unisolvence_inj}? : \forall {d}, UP0 d.
\end{lstlisting}
In this case, the approximation space~$\matPdo$ is the one-dimensional space of
constant functions.
The proof follows directly from the fact that a constant function vanishes iff
it is zero at some point.

\subsection{Construction of the $\FElagPdk$ Simplicial Lagrange Finite Element}
\label{sec:FELagP-constr}

Once the unisolvence properties are established, we can build the records
corresponding to Lagrange FE~$\FElagPdk$.
First, for~$d,k>0$,
\begin{lstlisting}
Context {d k : nat}.
Hypothesis Hd : d <> 0.
Hypothesis Hk : k <> 0.
Context {vtx : [0..d] -> '\R^d}.
Hypothesis Hvtx : aff_indep_ms vtx.

Definition FE_LagPSdSk?\lkFELagP{FE_LagPSdSk}? : FE d :=
  mk_FE Simplex vtx (nnode_LagPdk d k) (Pdk_has_dim d k)
    (FE_LagPSdSk_S_dof_lm k vtx) (FE_LagPSdSk_unisolvence_inj Hd Hk _ Hvtx).
\end{lstlisting}
This is the main construction of~$\FElagPdk$, based on the Lagrange degrees of
freedom defined in Section~\ref{sec:LagP-dof} and the unisolvence proved in
Section~\ref{sec:LagP-unisolv}.

Then, for any~$d$ and for~$k=0$ from Section~\ref{sec:LagP-degen}, where the
vertices no longer need to be affinely independent,
\begin{lstlisting}
Context {d : nat}.
Variable vtx : [0..d] -> '\R^d.

Definition FE_LagPd0?\lkFELagP{FE_LagPd0}? : FE d :=
  mk_FE Simplex vtx (nnode_LagPdk d 0) (Pdk_has_dim d 0)
    (FE_LagPd0_S_dof_lm vtx) (FE_LagPd0_unisolvence_inj vtx).
\end{lstlisting}
However, we have to keep in mind that {\nondegenerate} simplices will need
affinely independent vertices.

And for~$d=0$ from Section~\ref{sec:FE-degenerate}, where the only existing
vertex does not need to be specified and~$k$ is no longer relevant,
\begin{lstlisting}
Definition FE_LagP0k?\lkFELagP{FE_LagP0k}? : FE 0 := FE_d_0 Simplex one_not_zero_R.
\end{lstlisting}

Finally, for the case~$d=1$ for any~$k$, the nodes need only be pairwise
distinct, and need not be evenly distributed,
\begin{lstlisting}
Context {k : nat}.
Context {node_1k : [0..nnode_LagPdk 1 k) -> '\R^1}.
Hypothesis Hnode : injective node_1k.
Variable vtx : '\R^{2,1}.

Definition FE_LagP1k?\lkFELagP{FE_LagP1k}? : FE 1 :=
  mk_FE Simplex vtx (nnode_LagPdk 1 k) (Pdk_has_dim 1 k)
    (FE_LagP1k_S_dof_lm node_1k) (FE_LagP1k_unisolvence_inj Hnode).
\end{lstlisting}
This construction holds for any couple of vertices (in~$\R$).
However, we note that {\nondegenerate} geometric elements will need affinely
independent vertices ({\ie} distinct).
Moreover, in practice, it will be wise to assume that the nodes are located
between vertices and include them.
This will be useful for the construction of Lagrange~FE on cuboids, see
Section~\ref{sec:Persp}.

Note also that the arguments to build the Lagrange~FE are not the same in all
cases: either none, vertices, or nodes, with some property or not.
Consequently, the unification of all definitions into a single function proves
to be an attractive misconception.

\myskip

Another seemingly nice idea would have been to construct only the reference
Lagrange FE~$\FElagPdkref$ by a similar induction and build~$\FElagPdk$ using a
geometric transformation of Section~\ref{sec:TFE}.
This proof path would have failed as the second step of the unisolvence proof
applies the induction hypothesis to {\subvertices}, which are not reference
vertices.
Said otherwise, we need the induction hypothesis on current vertices to be
applicable.

However, we can check for any~$d,k>0$ and independent vertices~$\vv$ that the
trans\-formation induced by~$\Tgeomv$ of the Lagrange~FE specialized to the
reference vertices gives the corresponding Lagrange~FE back, see
Figure~\ref{fig:TFE}, {\ie} in {\Rocq}  (where the \Notation{\coqe{@@}}
annotation allows access to all arguments, including implicit ones, here~$d$)
\begin{lstlisting}
Context {d k : nat}.
Hypothesis Hd : d <> 0.
Hypothesis Hk : k <> 0.

Definition FE_LagPSdSk_ref?\lkFELagP{FE_LagPSdSk_ref}? := FE_LagPSdSk Hd Hk (@@vtx_ref_aff_indep d).

Lemma FE_LagPSdSk_eq_from_ref?\lkFELagP{FE_LagPSdSk_eq_from_ref}? : \forall {vtx : [0..d] -> '\R^d} (Hvtx : aff_indep_ms vtx),
  FE_LagPSdSk Hd Hk Hvtx = TFE_vtx_from_ref FE_LagPSdSk_ref Hvtx.
\end{lstlisting}
where the transformed~FE \coqelkTFE{TFE_vtx_from_ref} is presented in
Table~\ref{tab:TFE}.
The proof directly follows from the extensionality result \coqelkFE{FE_ext} for
\coqelkFE{FE} records, which stipulates that it is sufficient to check that the
{\nonproof} fields match, namely vertices, approximation spaces and degrees of
freedom.
This is quite straightforward since the geometric transformation~$\Tgeomv$ is
affine and bijective, thus it preserves barycenters (such as vertices and
nodes), and the polynomial spaces~$\matPdk$.
The slight difficulty is the handling of the casts due to different expressions
for the same dimension when they are not structurally equal.
The proof is actually the same when~$k=0$
(\coqelkFELagP{FE_LagPd0_eq_from_ref}), and from the uniqueness result
\coqelkFE{FE_d_0_uniq} of Section~\ref{sec:FE-degenerate} (with the scaling
factor~1), it is trivial when~$d=0$.

\subsection{Face Unisolvence}
\label{sec:face-unisolvence}

Assume that~$d>1$ and~$k>0$.
The {\em face unisolvence property} states that
\begin{equation*}
  \forall i \in [0..d],\;
  \forall p \in \matPdk,\quad
  \left( \forall \aalpha \in \calAdk,\;
  \aa_\aalpha \in \Hplv_i \Implies p (\aa_\aalpha) = 0 \right)
  \EQUIV \restr{p}{\calH_i} = 0.
\end{equation*}
This property states that a polynomial in~$\matPdk$ vanishes on a face
hyperplane if and only if it is zero at the Lagrange nodes that lie in this
face hyperplane.
This is useful to enforce inter-element continuity using \emph{discrete}
conditions.
Indeed, let~$f$ be a function that is in~$\matPdk$ in two adjacent mesh
cells~$K_1$ and~$K_2$, sharing the ({\nonempty}) face $\calF=K_1\cap K_2$.
Assume that face unisolvence holds on the hyperplane containing~$\calF$.
Thus, by linearity of the evaluation of functions at some point, in order
for~$f$ to be continuous, it suffices to enforce that the two
polynomials~$\restr{f}{K_1}$ and~$\restr{f}{K_2}$ take the same values at the
Lagrange nodes in the face~$\calF$.

Face unisolvence holds for~$\FElagPdk$, thanks to the fact that the
Lagrange nodes are evenly distributed on each face.
As we do not need here to structurally grant the bounds on~$d$ and~$k$, this is
formalized as
\begin{lstlisting}
Context {d k : nat}.
Hypothesis Hd : 1 < d.
Hypothesis Hk : k <> 0.
Context {vtx : [0..d] -> '\R^d}.
Hypothesis Hvtx : aff_indep_ms vtx.

Lemma Hface_unisolvence?\lkFELagP{Hface_unisolvence}? : \forall {i} p, Pdk d k p ->
  (\forall\nsp (idk : [0..pbinom d k]), Hface vtx i (node vtx idk) -> p (node vtx idk) = 0) <->
  (\forall\nsp x, Hface Hvtx i x -> p x = 0).
\end{lstlisting}
Backwards implication is immediate.
The proof of the forward one directly derives from
\coqe{unisol}\linebreak[0]\coqe{vence_inj_Hface}\lkFELagP{unisolvence_inj_Hface}
established in Section~\ref{sec:LagP-unisolv}, where the induction
hypothesis is fulfilled here by the full unisolvence result.

\section{Conclusion and Perspectives}
\label{sec:ConclPersp}

After an insight of the contributions, we describe some difficulties and design
choices, before giving perspectives.

\subsection{Contributions}
\label{sec:Contrib}


After some mathematical background on finite elements~(FE), we have described
our {\Rocq} formalization on this topic.
To the best of our knowledge, this is the first time~FE have been fully
formalized and instantiated by defining the Lagrange~FE.
As usual, we had to choose the proof path to be formalized among several, here
depending mostly on the formally available mathematicaltools.
Crucially, this contribution is a cornerstone of the future formalization of
the finite element method, that involves going from a local computation on a
single element to a global computation on the whole mesh (more in
\AppSection~\ref{sec:lin:system}).

To give an idea, the contributions presented in this article amount to about a
hundred files, more than 7,000 statements (such as lemmas, definitions, and
canonical structures), for about 45~kloc of {\Rocq}.

All this is free and available at
\begin{center}
  \url{https://lipn.univ-paris13.fr/rocq-num-analysis/tree/2.2.0/}
\end{center}
as the {\Opam} packages {\RNASubset},%
\footnote{\url{https://rocq-prover.org/p/rocq-num-analysis-subset/2.2.0}}
{\RNAAlgebra},%
\footnote{\url{https://rocq-prover.org/p/rocq-num-analysis-algebra/2.2.0}}
and {\RNAFEM}.%
\footnote{\url{https://rocq-prover.org/p/rocq-num-analysis-fem/2.2.0}}
The respective dependencies of these packages are shown in
Figure~\ref{fig:perspective}.
This development has led to a major reorganization of the previous
{\CoqNumAnalysis} package.

This formalization includes both generic and dedicated components.
Among the generic components that could be reused for other projects, we find
the finite families ({\aka} vectors of a given size), affine spaces,
{\substructure}s and restrictions of functions (that belong to the {\RNASubset}
and {\RNAAlgebra} packages), but also {\multiindices}.
The dedicated components are the definition of what a~FE is with a precise
\code{Record} including both values and requirements, and the instantiation of
this record in the (popular) case of simplicial Lagrange~FE.
We have been especially careful for statements and proofs to remain fairly
accessible, even for {\nonexpert}s in {\Rocq}.
This is particularly the case for the main results on finite-dimensional linear
and affine algebra (see Sections~\ref{sec:FinDimMS} and~\ref{sec:FinDimAS}),
and ultimately even for the long proof of unisolvence of the Lagrange~FE (see
Section~\ref{sec:LagP-unisolv}).

\subsection{Difficulties}
\label{sec:Diff}

The main difficulties we encountered are of two types: either due to
mathematics or to formalization.
As far as mathematics are concerned, the proofs are mostly based
on~\cite{eg:FE1:21}, but with adaptation.
For instance, we decided to work with almost no tools of analysis, and to
employ purely affine arguments, only based on barycenters, for almost all steps
of the construction of the simplicial Lagrange~FE, see
Sections~\ref{sec:math:ref:FEM} and~\ref{sec:AffineAlgebra}, and~\cite{CM24}.
We nevertheless relied on derivation in order to establish the linear
independence of monomials in the space of {\multivariate} polynomials, see
Section~\ref{sec:Pdk-def}.

The most difficult mathematical proof is the unisolvence of the Lagrange~FE
(see Section~\ref{sec:LagP-unisolv}).
It relies on a complex double induction with a polynomial factorization on a
face.
The formalization nicely follows the mathematical proof, but it still requires
many steps.
Another difficult mathematical proof is the freedom of
monomials~$(\XX^\aalpha)_{\aalpha \in \calAdk}$ (see
Section~\ref{sec:Pdk-def}).
There exist several proof paths and we chose one with the greatest number of
already formalized tools.
So, based on the derivative of \Coquelicot, we defined derivatives of
{\multivariate} polynomials at any order to derive~$\XX^\aalpha$ and proved
that it is a linearly independent family.

\myskip

Another kind of difficulty is that of formalization choices.
Given a mathematical object, how can we define it in {\Rocq}, so that it is
usable and even user-friendly?
An example is the {\multiindices} and more precisely the~$\calAdk$'s (see
Section~\ref{sec:Adk}), that can be sets, or lists, but we chose to make them
finite families of the expected size and we proved an injectivity property.
There are several ways to build them (with different orderings), and we chose
what we called the grsymlex ordering, forcing the definition to fit this
ordering.
Contrary to mathematics, we made heavy use of the indexing of the
{\multiindices}.

Another example is how to formally handle {\substructure}s.
We have several subsets (such as polynomials, face hyperplanes) that preserve
the algebraic structure endowed by their support type.
For instance, functions to~$\R$ are a \coqe{ModuleSpace}, and so are
polynomials.
This subspace can be represented as a predicate \coqe{E -> Prop} and the
properties by an assumption.
But from these, we can also define, with a dependent type, a set that can be
instantiated as \coqe{ModuleSpace} itself.
This is formalized and was initially used (notably for unisolvence), but we
consider more readable to handle subsets as additional parameters of, for
instance, bijectivity, see Section~\ref{sec:FunSub}.

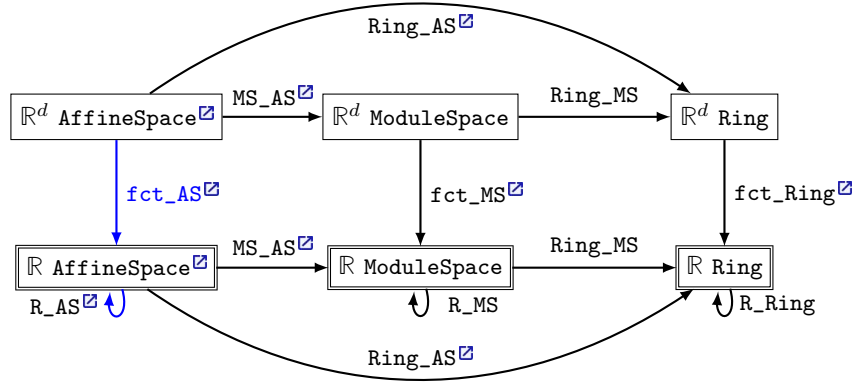
\begin{figure}[ht]
  \centering
  \begin{tikzpicture}
    \tikzset{%
      terminal/.style = {draw, rectangle, double},
      initial/.style = {draw, rectangle},
      canonical/.style = {->, >=latex, thick},
      annot/.style = {midway},
      annotAL/.style = {annot, above left},
      annotA/.style = {annot, above},
      annotR/.style = {annot, right},
      annotAR/.style = {annot, above right},
      annotBR/.style = {annot, below right},
    }

    \node[terminal] (RR)   at (8,0) {$\R$ \coqe{Ring}};
    \draw[canonical] (RR) edge [loop below] (RR);
    \node at (8.7,-.5) {\coqe{R_Ring}};

    \node[terminal] (RMS)  at (4,0) {$\R$ \coqe{ModuleSpace}};
    \draw[canonical] (RMS) edge [loop below] (RMS);
    \node at (4.7,-.5) {\coqe{R_MS}};

    \node[initial]  (RdR)  at (8,2) {$\Rd$ \coqe{Ring}};
    \node[initial]  (RdMS) at (4,2) {$\Rd$ \coqe{ModuleSpace}};
    \draw[canonical] (RMS) -- (RR)
      node[annotA] {\coqe{Ring_MS}};
    \draw[canonical] (RdR) -- (RR)
      node[annotR] {\coqelkH{fct_Ring}};
    \draw[canonical] (RdMS) -- (RMS)
      node[annotR] {\coqe{fct_MS}%
        \lk{Algebra.Hierarchy_compl}{fct_ModuleSpace}};
    \draw[canonical] (RdMS) -- (RdR)
      node[annotA] {\coqe{Ring_MS}};

    \node[terminal] (RAS)  at (0,0) {$\R$ \coqelkAS{AffineSpace}};
    \draw[canonical,blue] (RAS) edge [loop below] (RAS);
    \node at (-.7,-.5) {\coqe{R_AS}%
      \lk{Algebra.AffineSpace.AffineSpace_def}{R_AffineSpace}};

    \node[initial]  (RdAS) at (0,2) {$\Rd$ \coqelkAS{AffineSpace}};
    \draw[canonical] (RAS) edge [out=-35, in=-145] (RR);
    \node at (4,-1.2) {\coqe{Ring_AS}%
        \lk{Algebra.AffineSpace.AffineSpace_def}{Ring_AffineSpace}};
    \draw[canonical] (RAS) -- (RMS)
      node[annotA] {\coqe{MS_AS}%
        \lk{Algebra.AffineSpace.AffineSpace_def}{ModuleSpace_AffineSpace}};
    \draw[canonical,blue] (RdAS) -- (RAS)
      node[annotR] {\coqe{fct_AS}%
        \lk{Algebra.AffineSpace.AffineSpace_def}{fct_AffineSpace}};
    \draw[canonical] (RdAS) -- (RdMS)
      node[annotA] {\coqe{MS_AS}%
        \lk{Algebra.AffineSpace.AffineSpace_def}{ModuleSpace_AffineSpace}};
    \draw[canonical] (RdAS) edge [out=35, in=145] (RdR);
    \node at (4,3.2) {\coqe{Ring_AS}%
        \lk{Algebra.AffineSpace.AffineSpace_def}{Ring_AffineSpace}};
  \end{tikzpicture}
  \caption[Various ways to equip $\Rd$ with the affine space structure]{%
    Various ways to equip~$\Rd$ with the affine space structure.\\
    Double-edged boxes denote actual instances of canonical structures in the
    {\Rocq} system for~$\R$, that is to say possible terminal points for the
    canonical structure search.
    Rectangular boxes denote the intent to equip a space (either~$\R$ or~$\Rd$)
    with an algebraic structure.
    Arrows correspond to generic canonical structures in our library, or in the
    {\Coquelicot} library (when there is no link annotation).
    Names are abbreviated, \coqe{Ring_AS} actually means
    \coqe{Ring_AffineSpace}, an instance of \coqe{AffineSpace}, and
    \coqe{fct_MS} means \coqe{fct_ModuleSpace}, an instance of
    \coqe{ModuleSpace}.
    Then, the possibilities are all the paths from the wanted space and
    algebraic structure to a  double-edged box.\\
    For instance, there is only one way to equip~$\Rd$ with the ring structure:
    by applying the \coqe{fct_Ring} canonical structure.
    $\Rd$~is then of type \coqe{fct_Ring [0..d) R_Ring}, which is an instance
    of \coqe{Ring}.
    This is automatic.\\
    For the module space~$\Rd$, there is a true diamond problem, with two
    distinct mathematical interpretations: either~$\Rd$ is a module space over
    itself (\coqe{Ring_ModuleSpace (fct_Ring [0..d) R_Ring)}), or it is the
    regular module over~$\R$
    (\coqe{fct_ModuleSpace [0..d) (Ring_ModuleSpace R_Ring)}).
    But, there is also a third path,
    \coqe{fct_ModuleSpace [0..d) R_ModuleSpace}, which is the direct one, and
    is automatic.\\
    Finally, for the affine space~$\Rd$, there are eight distinct paths, the
    chosen one being the blue one.
    Again, the automatic path is the shortest one
    \coqe{fct_AffineSpace [0..d) R_AffineSpace} (blue arrows).
    But this path prevents us from seeing~$\Rd$ as a \coqe{ModuleSpace},
    preventing us from using some lemmas.\\
    On top of all this complexity, affine spaces can also be mathematically
    equipped with a module space structure by choosing some origin and making
    it the zero vector.
    Fortunately, our proofs were not based on such construction, as it makes
    {\Rocq} loop.}
  \label{fig:diamond-pbs}
\end{figure}

A last example is the definition of affine spaces (see
Section~\ref{sec:AlgStruct}).
This is a known geometric object (such as a hyperplane, shifted from the
origin), but it ended up creating formalization hardships.
The reason is that of canonical structures and the fact that there may be
several paths to create such a structure.
Consider the example of Figure~\ref{fig:diamond-pbs} that describes the eight
paths to build the fact that~$\Rd$ is an instance of the \coqe{AffineSpace}
structure.
The mechanism of canonical structures forces us to choose one of these
paths (the blue one), but the path we need may be different depending on the
proof at stake ({\eg} a \coqe{ModuleSpace} instance).
It is of course much worse in the case of~$\left(\Rd\right)^{d+1}$ which is
very common in our application (for the vertices of the geometrical element,
{\ie}~$d+1$ affinely independent points in~$\Rd$).
For a given definition on an affine space, such as \coqe{barycenter} or the
predicate \coqe{aff_indep}, we add another definition with the
suffix~\coqe{_ms} equal to the preceding one, but typing only on a module
space.
This means duplicating some code, but helps choose the canonical structure path
we need in several proofs.
This problem is now known as {\em non-forgetful inheritance}~\cite{ACKMRS20},
and should be solvable with a {\Coquelicot} hierarchy built upon
{\HB}~\cite{CST20}.

\subsection{Design Choices}
\label{sec:Design}

Now let us discuss our design choices.
First of all, one of the guiding principles of our formalization work is to
target both experts and {\nonexpert}s in {\Rocq}, including applied
mathematicians and students.
The challenge is then to find the right balance between, on the one hand, the
legibility of the statements (and to a lesser extent of the proofs) and, on the
other hand, the level of abstraction of mathematical concepts as well as the
use of numerous compact notations.
As a consequence, we have chosen to rely on classical logic (see
Section~\ref{sec:Classical}) as this seems the best choice for the mathematical
proofs we want to formalize.
For instance, constructive analysis can lead statements of the form
$\neg\neg(\exists x,\neg\neg P(x))$ instead of $\exists x,P(x)$, and that
would not convince mathematicians of the maturity of our tools.

From a more technical perspective, subsets are usually formalized through a
predicate or a sigma-type.
An uncommon alternative is extensional representation through the range of an
enumerating function from some type of indices.
This allows a direct access to elements, and for instance, the transport of a
possible order on the indices to the elements ({\eg} see
Section~\ref{sec:MultiIndex}).
Moreover, enumerating functions from a finite type of indices can represent
vectors, thus sharing the same definition and results.
We claim that both approaches are complementary and useful.

On the specific matter of finite elements, note that we do not require the
geometric element~$K$ to have a {\nonempty} interior (see
Section~\ref{sec:fe-triple}).
Most of the constructed~FE (in Sections~\ref{sec:FE:Coq} and~\ref{sec:FELagP})
are suitable as the simplex~$K$ is built from affinely independent vertices. We
will prove that affine independence indeed ensures the {\nondegeneracy} in the
simplicial case.
Another perspective is then to require that a {\nonempty} ball is included
in~$K$, but another appealing solution would be to build~$K$ by a regular
enough function from reference vertices (be they simplex or cuboid reference
vertices).
In the record, the (current) vertices would be replaced by the function and the
rest of the record would be kept (unisolvence included) and most of this
development would still apply, a notable exception being the geometric
transformation of a~FE of Section~\ref{sec:TFE}.
This perspective would also allow us to deal with general hexahedra in~3D,
whose faces are not necessarily planar and that are not convex in general.
Such {\nonconvex} geometric elements can be used for instance in metal
forming~\cite{fm:07:hexafem}.

\subsection{Perspectives}
\label{sec:Persp}

A straightforward perspective is to consider cuboid~FE and their properties.
This will require to factorize part of our development between simplex and
cuboid~FE.
Cuboid Lagrange~FE are often defined as the tensorization of~1D Lagrange~FE
({\ie} taking~$\FElagPik$ for each variable).
Their unisolvence proof would be simpler and by induction on the dimension
only.
What is more delicate is the fact that the geometric transformation is no
longer affine in general, see~\cite[Sec.~13.5]{eg:FE1:21}.

Another important and challenging perspective is to handle face or edge
elements, {\ie}~FE of dimension~$l$ that live in~$\Rd$ with~$l<d$
(typically~$l=d-1$).
They can be used to apply the boundary conditions of the~PDE, but also for the
treatment of fractures~\cite{mjr:05:frac}, or for the boundary finite element
methods, see for instance~\cite{jr:16:bem} and references therein.
These boundary elements do not fit in our formalization so far.

It would also be interesting to extend the record~\coqe{FE} to~FE whose
approximation space \coqe{P_approx} takes its value in~$\R^q$, with~$q\geq1$
(often~$q\in\{1,d\}$).
This is a necessary step for the construction of Raviart--Thomas flux~FE,
see~\cite[Chap.~14]{eg:FE1:21} or \AppSection~\ref{sec:FE:zoo}, that requires a
significant development of integral, differential calculus and geometric
components (such as face integrals, normals, and {\nonaffine} geometric
transforms).

\myskip

\begin{figure}[ht]
  \centering
  \begin{tikzpicture}[every text node part/.style={align=left}]
\node[draw,rounded corners=3pt,fill=\colorC] (Reals) at (4.5,7) 
     {Real numbers library~\cite{May01}};

\node[draw,rounded corners=3pt,fill=\colorC] (Coquelicot) at (4.5,6) 
     {\Coquelicot~\cite{BLM15}};
\draw[->,>=latex,thick] (Reals) to (Coquelicot);

\node[draw,rounded corners=3pt,fill=\colorC] (MC) at (-0.25,7) 
     {\MathComp~\cite{MathComp_Ref}};

\node[draw,rounded corners=3pt,fill=\colorC] (flocq) at (9,7)
     {\Flocq~\cite{BolMel11}};
\node[draw,rounded corners=3pt,fill=\colorB, very thick] (RNA-sub) at (4.5,4.5)
     {Subsets, Sec.~\ref{sec:FunSub}, \ref{sec:FF}\\ (\RNASubset)};
\draw[->,>=latex,thick] (Coquelicot) to (RNA-sub);
\draw[->,>=latex,thick] (MC) to (RNA-sub);

\node[draw,rounded corners=3pt,fill=\colorB,very thick] (RNA-alg) at (1.5,3)
     {Algebra,  Sec.~\ref{sec:AlgStruct}--\ref{sec:Binom}
       and~\ref{sec:AffineAlgebra}\\
     (\RNAAlgebra)};
\draw[->,>=latex,thick,bend left=20] (MC) to (RNA-alg);
\draw[->,>=latex,thick] (RNA-sub) to (RNA-alg);

\node[draw,rounded corners=3pt,fill=\colorC] (RNA-leb) at (7.5,3)
     {Lebesgue integral~\cite{BCF22,BCM23}\\ (\RNALebesgue)};
\draw[->,>=latex,thick] (RNA-sub) to (RNA-leb);
\draw[->,>=latex,thick] (flocq) to (RNA-leb);

\node[draw,rounded corners=3pt,fill=\colorC] (RNA-lm) at (5.7,1.3)
     {{\LM} theorem~\cite{BCF17}\\ (\RNALM)};
\draw[->,>=latex,thick] (RNA-alg) to (RNA-lm);

\node[draw,rounded corners=3pt,fill=\colorB,very thick] (RNA-fem) at (0.5,1.3)
     {FE def. + simplicial Lagrange FE\\
     Sec.~\ref{sec:FE:Coq}, \ref{sec:geometry}, \ref{sec:poly-space-Pdk},
     and~\ref{sec:FELagP}\\ (\RNAFEM)};
\draw[->,>=latex,thick] (RNA-alg) to (RNA-fem);

\draw (-2.4,5.2) rectangle (10,.4);
\node at (-.8,4.9) {\RocqNumAnalysis};
\node[draw,rounded corners=3pt,fill=\colorD] (quad) at (-0.5,-.2)
     {Cuboid Lagrange FE};
\draw[->,>=latex,thick,dotted] (RNA-fem) to (quad);

\node[draw,rounded corners=3pt,fill=\colorD] (Sobolev) at (4,-.2)
     {Sobolev spaces ({\eg} $\HioOm$)};
\draw[->,>=latex,thick,dotted,bend left=60] (RNA-leb) to (Sobolev.north east);
\draw[->,>=latex,thick,dotted] (RNA-lm) to (Sobolev);

\node[draw,rounded corners=3pt,fill=\colorD] (qform) at (8.5,-.2)
     {Quadrature formulas};
\draw[->,>=latex,thick,dotted] ($(RNA-leb.south)+(1,0)$) to (qform);

\node[draw,rounded corners=3pt,fill=\colorD] (fem) at (4,-1.2)
     {Finite element method};
\draw[->,>=latex,thick,dotted] (quad) to (fem);
\draw[->,>=latex,thick,dotted] (Sobolev) to (fem);
\draw[->,>=latex,thick,dotted] (qform) to (fem);

\node[draw,rounded corners=3pt,fill=\colorD] (prog) at (4,-2.2)
          {Program verification};
\draw[->,>=latex,thick,dotted] (fem) to (prog);
\draw[thick,dotted,bend left=15] (flocq.south east) to (9.7,-2.2);
\draw[->,>=latex,thick,dotted] (9.7,-2.2) to (prog.east);
  \end{tikzpicture}
  \caption[Dependency graph of our packages and of perspectives]{%
    Dependency graph, including the literature (in green), our contributions
    presented in this article (in yellow), and perspectives (in blue).
    Plain arrows are current dependencies between libraries, while dotted
    arrows are expected dependencies.}
  \label{fig:perspective}
\end{figure}
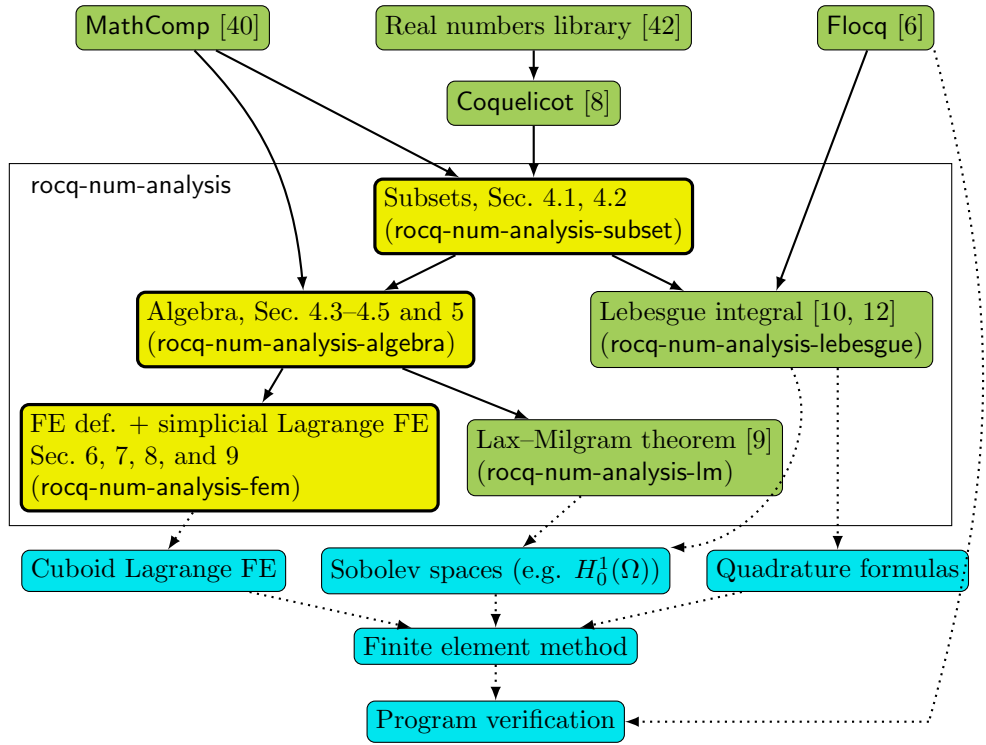

More than other~FE, we have several long-term perspectives described in
Figure~\ref{fig:perspective}.
We want to formally prove the finite element method~(FEM), meaning the full
algorithm to solve a~PDE.
As explained in more details in \AppSection~\ref{sec:FEM:Poisson}, the~PDE is
solved on each~FE and glued together.
This so-called assembly procedure is a necessary step to be formalized.
Another one is the bounding of the approximation error, which requires to
formalize Sobolev spaces (and prove they are Hilbert spaces), see
\AppSection~\ref{sec:fun:space}.

The ultimate goal is to guarantee some (part of a) library implementing
the~FEM, such as {\XLiFE}.\footnote{\url{https://xlifepp.pages.math.cnrs.fr/}}
It will be a different kind of proofs: more than proving the~FEM as a
mathematical algorithm, it requires to prove it complies with the {\Cpp}
program.
A specific point out of the literature is to ensure that the floating-point
errors do not mess too much with the results, relying on automatic provers or
on the {\Flocq} library~\cite{BolMel11}.

\appendix

\section{Brief Mathematical Presentation of the Finite Element Method}
\label{sec:Maths:FEM}

For the reader who is unfamiliar with the finite element \emph{method}~(FEM),
it might be interesting to present its principle in a simple, standard
example.
We illustrate how the Finite Element~(FE), that is formalized in this paper,
comes into play in this method.

We limit this presentation to the Poisson problem~\eqref{eq:poisson:strong}
below, the standard stationary diffusion equation, that models for instance the
equilibrium of heat transfer in a homogeneous material.

Note that many extensions of the~FEM (such as Discontinuous Galerkin methods)
derive from this framework, however, they are outside the scope of this study.
We do not study either the Finite Volume Methods.

After setting some notations on functional spaces in
\AppSection~\ref{sec:fun:space}, the steps of the~FEM are described in
\AppSection~\ref{sec:FEM:Poisson}: the continuous Poisson formulation is
presented, then discretized with a given mesh of the domain; this allows to
write the linear system whose solution represents the approximation we are
looking for.
Beyond the Lagrange~FE, a short overview of various~FE is given in
\AppSection~\ref{sec:FE:zoo}.

\subsection{Functional Spaces}
\label{sec:fun:space}

Let us introduce some notations and definitions that are commonly used for the
study of~PDE on Hilbert functional spaces.

We recall that the vector quantities in mathematical expressions are written
with {\bf bold type}.
For instance, we use $\xx\eqdef(x_0,\dots,x_{d-1})$ for points in~$\Rd$, and
the gradient is denoted by \Notation{$\nnabla u$}%
$\eqdef(\frac{\p u}{\p x_0},\dots,\frac{\p u}{\p x_{d-1}})^T$ for a scalar
function $u\in\FRd$, {\ie} from~$\Rd$ to~$\R$.

In this appendix, we assume that the space dimension~$d$ lies in $\{1,2,3\}$.
Let~$\Omega\subset\Rd$ be an open bounded connected domain, and
let~\Notation{$\Gamma$}$\eqdef\p\Omega$ be its smooth boundary.
In the sequel, for the sake of simplicity, we assume that~$\Omega$ is
polyhedral.

The most commonly used functional spaces are
\begin{gather*}
  \Notation{$\HoOm$} \eqdef \Notation{$\LiiOm$}
  \eqdef \{ f : \ArOmR \st \normo{f} < +\infty \},\\
  \Notation{$\HiOm$}
  \eqdef \{ f \in \LiiOm \st \nnabla f \in (\LiiOm)^d \}, \AND
  \Notation{$\HioOm$} \eqdef \{ f \in \HiOm \st \restr{f}{\Gamma} = 0 \},
\end{gather*}
where~$\LiiOm$ is the Lebesgue space of measurable, square integrable, real
functions over~$\Omega$, it is equipped with the $\Lii$-norm defined by
\Notation{$\normo{f}$}$\eqdef\sqrt{\int_{\Omega}|f|^2}$.
Note that, as it is usual in this context, norms are named after their
corresponding~$H^m$ space.
$\HiOm$~is the Sobolev space made up of~$\LiiOm$ functions that have a gradient
in~$(\LiiOm)^d$, and~$\HioOm$ is the subspace of functions vanishing on the
boundary~$\Gamma$.
They are equipped with the seminorms defined by \Notation{$\normio{f}$}%
$\eqdef\sqrt{\sum_{i\in[0..d)}\normo{\frac{\p f}{\p x_i}}^2}$, and
\Notation{$\normi{f}$}$\eqdef\sqrt{\normo{f}^2+\normio{f}^2}$.
The spaces $(\LiiOm,\normo{\cdot})$, $(\HiOm,\normi{\cdot})$, and
$(\HioOm,\normio{\cdot})$ are Hilbert spaces ({\ie} complete normed vector
spaces, with an inner product, {\eg} see~\cite[Cor.~IX.19, p.~174]{bre:af:83}).

\subsection{The Finite Element Method on a Simple Example}
\label{sec:FEM:Poisson}

Let us present the~FEM on a simple example, the Poisson problem.

\subsubsection{Continuous Poisson Problem}
\label{sec:cont:poisson}
Let~$f\in\LiiOm$.
The Poisson problem is defined in its strong form as
\begin{equation}
  \label{eq:poisson:strong}
  \mbox{find } u, \mbox{ such that }
  \left\{
    \begin{array}{rcll}
      - \Delta u & = & f & \mbox{in } \Omega\\
      u & = & 0 & \mbox{on } \Gamma,
    \end{array}
  \right.
\end{equation}
where the Laplace operator~$\Delta$ is defined by
\Notation{$\Delta u$}$\eqdef\sum_{i=0}^{d-1}\frac{\p^2 u}{\p x_i^2}$.
If we multiply by a regular test function~$v$, integrate over~$\Omega$,
integrate by parts, and take the boundary condition into account, we obtain the
weak form of the Poisson equation,
\begin{equation}
  \label{eq:poisson:weak}
  \mbox{find } u \in \HioOm, \mbox{ such that }
  \forall v \in \HioOm,\;
  \int_\Omega \nnabla u \cdot \nnabla v = \int_\Omega f \, v.
\end{equation}
Setting \Notation{$\blf{u}{v}$}$\eqdef\int_\Omega\nnabla u\cdot\nnabla v$,
\Notation{$\lf{v}$}$\eqdef\int_\Omega f\, v$, and~\Notation{$V$}$\eqdef\HioOm$,
problem~\eqref{eq:poisson:weak} takes the general form
\begin{equation}
  \label{eq:general-problem}
  \mbox{find } u \in V, \mbox{ such that }
  \forall v \in V,\; \blf{u}{v} = \lf{v}.
\end{equation}

{\TheLMt} states that problem~\eqref{eq:general-problem} admits a unique
solution, assuming some hypotheses: continuity and coercivity of~$\Blf$,
continuity of~$\Lf$, and that~$(V,\normV{\cdot})$ is a Hilbert space.
This theorem has already been formalized in {\Rocq}~\cite{BCF17}.

Note that more general results exist, taking different Banach spaces for the
test function and the solution (Petrov--Galerkin framework and {\BNB} theorem,
{\eg} see~\cite[Chap.~2]{eg:tpf:04}).

The {\LM} hypotheses are fulfilled in the case of~\eqref{eq:poisson:weak}
({\eg} see~\cite[Chap.~1, pp.~35--36]{hna:cia:91}), proving that the weak
Poisson formulation admits a unique solution.
It is called the \emph{continuous} solution ({\ie} living in a continuous
space), in contrast to the \emph{discrete} solution (living in a finite
dimensional space) that is constructed in the sequel.

\subsubsection{Discrete Poisson Problem}
\label{sec:disc:poisson}

The conforming Galerkin method consists in replacing~$V$ by a
finite-dimensional space~$\Vh\subset V$.
The index ``$h$'' is commonly added in the notations to indicate their discrete
nature.
Assuming that the bilinear and linear forms remain unchanged, the discrete
counterpart of~\eqref{eq:general-problem} reads
\begin{equation}
  \label{eq:general-problem-discrete}
  \mbox{find } \uh \in \Vh, \mbox{ such that }
  \forall \vh \in \Vh,\; \blf{\uh}{\vh} = \lf{\vh}.
\end{equation}
{\TheLMt} also applies to~\eqref{eq:general-problem-discrete}, and {\Cea}'s
Lemma provides an estimate on the error,
\[
  \normV{u - \uh} \leq C \inf_{\vh \in \Vh} \normV{u - \vh},
\]
which stresses the importance of interpolating~$u$, {\ie} finding a ``good''
approximation space~$\Vh$ to be as close as possible to~$u$.
The~FEM is a way of constructing such~$\Vh$, for wide ranges of problems.

We carry on with the discrete version of the simple
model~\eqref{eq:poisson:weak}, that reads
\begin{equation}
  \label{eq:poisson:weak:discr}
  \mbox{find } \uh \in \Vh, \mbox{ such that }
  \forall \vh \in \Vh,\;
  \int_\Omega \nnabla \uh \cdot \nnabla \vh = \int_\Omega f \, \vh.
\end{equation}

\subsubsection{Mesh and Approximation Space}
\label{sec:mesh}

As written in Section~\ref{sec:ref-cur-fe-geomap}, we assume that a
mesh~$\calTh$ of $\Omega$ is given, satisfying
\begin{gather*}
  \forall K \in \calTh,\; \Interior{K} \neq \emptyset,\quad
  (\forall K_1, K_2 \in \calTh,\;
  K_1 \neq K_2 \Implies \Interior{K_1} \cap \Interior{K_2} = \emptyset),\\
  \nonumber
  \mbox{and}\quad \Omegab = \bigcup_{K \in \calTh} K.
\end{gather*}
The index ``$h$'', that also denotes the maximum diameter of the cells
of~$\calTh$, is going down to zero in convergence studies.
For the standard~FEM, one requires in addition that the mesh is
\emph{conforming} (or matching), {\ie} the intersection of two neighboring mesh
cells is exactly an hyperface of both cells when it is indeed of
dimension~$d-1$.
Note that, as the mesh cells have disjoint interior, the intersection of two
neighboring cells only happens on their boundary.
For instance, when~$d=3$, the intersection of two distinct mesh cells~$K_1$
and~$K_2$ is either empty, a vertex, an edge, or a face for both~$K_1$ and
for~$K_2$.

Then, a global FE~space~$\Vh$ can be defined as the space of continuous
functions whose restriction to any mesh cell is in~$\matPdk$,
\[
  \Notation{$\Vh$} \eqdef
  \{ \vh \in \calC^0 (\Rd, \R) \st
  \forall K \in \calT,\; \restr{\vh}{K} \in \matPdk \}.
\]
The global continuity condition ensures that~$\Vh\subset V$.
Such a space is called \emph{Lagrange}~$\FElagPdk$ FE~space.
It can be shown that it suffices to define~$\Vh$ by the global Larange node
values, the continuity over~$\Omega$ is recovered by applying discrete
inter-element conditions (see face unisolvence in
Section~\ref{sec:face-unisolvence}).
For computational efficiency, it is convenient to build a basis for~$\Vh$ where
each basis function has a small support.
Typical FE~basis functions are {\nonzero} only on elements sharing a face, an
edge, or a vertex.

\subsubsection{Building the Linear System}
\label{sec:lin:system}

Let~$(\phi_i)_{i\in[0..\Ndof)}$ be a basis of~$\Vh$, where~$\Ndof$ is the total
number of ``degrees of freedom'' ({\ie} the dimension) of~$\Vh$.
Take successively~$\vh=\phi_i$ in~\eqref{eq:poisson:weak:discr} and write
\begin{multline*}
  \mbox{find } \uh = \sum_{j \in [0..\Ndof)} U^h_j \phi_j \in \Vh,
  \mbox{ such that}\\
  \forall i \in [0..\Ndof),\quad
    \sum_{j \in [0..\Ndof)} \left(
      \int_\Omega \nnabla \phi_j \cdot \nnabla \phi_i \right) U^h_j
    = \int_\Omega f \, \phi_i,
\end{multline*}
which can be seen as the linear system
\begin{align}
  \label{eq:lapl:weak:discr:linsys:Axb}
  &\mbox{find } \Notation{$U^h$}
    \eqdef [U^h_0, \ldots, U^h_{\Ndof-1}]^\top \in \R^\Ndof,
  \mbox{ such that }AU^h = b,\\
  \nonumber
  &\mbox{where } \left\{
    \begin{array}{l}
      \dps \forall i, j \in [0..\Ndof),\quad
      \Notation{$A_{ij}$}
      \eqdef \int_\Omega \nnabla \phi_j \cdot \nnabla \phi_i,\\
      \dps \forall i \in [0..\Ndof),\quad
      \Notation{$b_i$} \eqdef \int_\Omega f \, \phi_i.
    \end{array}
  \right.
\end{align}
It then suffices to build~$A$ and~$b$, and apply a linear solver.
Let~$\JacTgeomv$ be the invertible Jacobian matrix of the geometric
transformation~$\Tgeomv$, see Section~\ref{sec:ref-cur-fe-geomap}.
To compute~$A$, one usually approximates the integrals on the reference
cell~$\hK$, using a change of variable and a quadrature formula (at~$N_q$
points~$(\hxxi_{i_q})_{i_q\in[0..N_q)}$ with
weights~$(\homega_{i_q})_{i_q\in[0..N_q)}$),
\begin{align}
  \nonumber
  \int_\Omega \nnabla \phi_j \cdot\nnabla \phi_i
  &= \sum_{K \in \calTh} \int_K \nnabla \phi_j \cdot \nnabla \phi_i
  = \sum_{K \in \calTh} \int_{\hK}
    (\JacTgeomv)^{-T} \hat{\nnabla} \hat{\phi}_j \cdot
    (\JacTgeomv)^{-T} \hat{\nnabla} \hat{\phi}_i \; | \det \JacTgeomv |\\
  \label{eq:int:transf:quad}
  &\approx \sum_{K \in \calTh} \sum_{i_q \in [0..N_q)} \left(
      (\JacTgeomv)^{-T} \hat{\nnabla} \hat{\phi}_j \cdot
      (\JacTgeomv)^{-T} \hat{\nnabla} \hat{\phi}_i \; | \det \JacTgeomv |
    \right) (\hxxi_{i_q}) \; \homega_{i_q}.
\end{align}
Note that for an invertible matrix~$M$, $M^{-T}$ represents the transposed of
its inverse.
Formula~\eqref{eq:int:transf:quad} is used in practice to build the matrix~$A$
piece by piece.
For each element, taking advantage of the small support of the~$\phi_i$'s, one
computes small matrices containing the {\nonzero}
$\int_K\nnabla\phi_j\cdot\nnabla\phi_i$ terms (it is done in the reference
element), then one fills the corresponding components of the matrix.
This procedure is called the {\em assembly} of the matrix, it outputs a sparse
matrix~$A$.
Once~$A$ is built, it remains to solve the (sparse) linear
system~\eqref{eq:lapl:weak:discr:linsys:Axb} using one's favorite library, to
compute~$U^h$, and then to reconstruct~$\uh$.

\myskip

We presented the finite element \emph{method} on a simple but typical example,
from the continuous variational formulation to the construction of the linear
system that needs to be solved.

\subsection{A Zoology of Finite Elements}
\label{sec:FE:zoo}

There exist several families of~FE.
Note that, except the simplicial Lagrange~FE for evenly distributed nodes
described in this article, none of these are formalized in {\Rocq} yet, but
most fit into the generic framework of Section~\ref{sec:FE:Coq}.
The object of this section is merely to show that there exists a large variety
of~FE.

Nodal~FE are the most popular.
Their degrees of freedom ({\ie} the linear forms~$\SSigma$) are the evaluation
operators at given points called nodes, and the discrete solution is
reconstructed from these values.
Among them, the case of evenly distributed ({\ie} Lagrange) nodes, leaning on
the vertices of the geometric element~$K$, is very common.
For a degree~$k\in\N$, they are denoted here as~$\FElagPdk$ for $d$-simplices,
and~$\FElagQ{d}{k}$ for $d$-dimensional hexahedra
(see~\cite[Sec.~7.4 and~6.4]{eg:FE1:21}).
There exists also the Hermite~FE, where the linear forms are the evaluation at
the nodes of both the function and its derivatives ({\eg}
see~\cite{cr:glh:72}), and nodal~FE for which the nodes are not evenly
distributed ({\eg} see~\cite[Sec.~6.3.5, and Remark~7.14]{eg:FE1:21}).

For some (\nonnodal) families of~FE, the degrees of freedom are average values
or moments in the geometric element~$K$.
For instance, the Crouzeix--Raviart~FE on simplices is based on~$\matPdi$
polynomials, and its degrees of freedom are the mean values on the faces
of~$K$, inducing discontinuous functions over~$\Omega$
(see~\cite[Sec.~7.5]{eg:FE1:21}).

For some families where the approximated solution~$\restr{\uu_h}{K}$ is
$d$-vectorial, the normal fluxes through the faces of~$K$ are computed, and in
others it is the tangential integral along the edges.
Examples of face-flux driven~FE are the Raviart--Thomas ($\RT$),
Brezzy--Douglas--Marini ($\BDM$), and Brezzi--Douglas--Fortin--Marini ($\BDFM$)
families, both for $d$-simplices, and regular $d$-hexahedra
(see~\cite[Chap.~14]{eg:FE1:21}).
They aim at approximating~$\Hdiv$ functions, that is vector functions~$\uu$
in~$(L^2)^d$, whose divergence $\nnabla\cdot\uu\eqdef%
\sum_{i\in[0..d)}\frac{\p u_i}{\p x_i}$ is in~$L^2$.
For instance, for a tetrahedron~$K$, whose faces are denoted
by~$(F_i)_{i\in[0..3]}$ and have a normal unit vector~$\nn_i$, the
space~$\RT_0(K)$ is represented by the quantities
$\frac{1}{|F_i|}\int_{F_i}\uu_h\cdot\nn_i$, where~$|F_i|$ is the area of~$F_i$.
The Nedelec family is an example of~FE approximating~$\Hcurl$ functions, that
is vector functions~$\uu$ in~$(L^2)^d$, whose curl~$\nnabla\ttimes\uu$ is
in~$(L^2)^d$, where~$\ttimes$ is the cross product.
Its degrees of freedom are the tangential component on the edges when $d=3$.

%% file: fig_Lag_2to3.tex
  \begin{tikzpicture}[scale=2,math3d] 
        \def\kkmax{2}

        \def\colo{magenta}
        \def\coli{green}
        \def\colii{blue}
        \def\coliii{blue}
        \def\hsq{8}
        \pgfmathparse{1-1/\hsq}\let\hhsq\pgfmathresult
        \pgfmathparse{1-2/\hsq}\let\hhhsq\pgfmathresult

        \foreach \K in {1,...,\kkmax}  {
	  \coordinate (AA) at ($ (0,0,0) + 3*(0,\K,0) $);
	  \coordinate (CC) at  ($ (AA) + (0,1,0) $);
	  \coordinate (DD) at  ($ (AA) + (0,0,1) $);
	  \draw[line width=1.0pt,rounded corners=0.5pt] (AA) -- (CC) -- (DD) -- cycle;
          \node[color=\coliii,below] (Nxyz) at (AA) {$(0.,0.)$};
          \node[color=\coliii,below] (Nxyz) at (CC) {$(1.,0.)$};
          \node[color=\coliii,left] (Nxyz) at (DD) {$(0.,1.)$};

	  \coordinate (hCC) at  ($ \hhsq*(AA) + 1/\hsq*(CC) $);
	  \coordinate (hDD) at  ($ \hhsq*(AA) + 1/\hsq*(DD) $);
	  \coordinate (hCD) at  ($ \hhhsq*(AA) + 1/\hsq*(CC) + 1/\hsq*(DD) $);
	  \draw[line width=0.4pt, color=\colo] (hCC) -- (hCD) -- (hDD);

          \foreach \k in {0,1,...,\K}  {
            \newcount\y
            \foreach \x in {\k,...,0}  {
              \pgfmathsetcount{\y}{\k-\x} 
              \coordinate (Axyz) at  ($ (AA) + 1/\K*(0,\x,\y) $) ;
              \ifthenelse{\k<\K}
                         {\fill[color=\coliii] (Axyz) circle (1.5pt);
                           \node[color=\coliii,below] (Nxyz) at (Axyz) {};}
                         {\fill[color=\colii] (Axyz) circle (1.5pt);
                           \node[color=\colii,below] (Nxyz) at (Axyz) {};}
            }
          }
        }
%
        \foreach \K in {1,...,\kkmax}  {
	  \coordinate (A) at ($ (0,0,-1.5) + 3*(0,\K,0) $);
	  \coordinate (B) at  ($ (A) + (1,0,0) $);
	  \coordinate (C) at  ($ (A) + (0,1,0) $);
	  \coordinate (D) at  ($ (A) + (0,0,1) $);
	  \draw[line width=1.0pt,rounded corners=0.5pt] (A) -- (B) -- (D) -- cycle;
	  \draw[line width=1.0pt,rounded corners=0.5pt] (A) -- (C) -- (B) -- cycle;
	  \draw[line width=1.0pt,rounded corners=0.5pt] (B) -- (C) -- (D) -- cycle;
	  \draw[line width=1.6pt,rounded corners=0.5pt] (A) -- (C) -- (D) -- cycle;

          \node[color=\coliii,below] (Nxyz) at (A) {$(0.,0.,0.)$};
          \node[color=\coliii,below] (Nxyz) at (B) {$(1.,0.,0.)$};
          \node[color=\coliii,below] (Nxyz) at (C) {$(0.,1.,0.)$};
          \node[color=\coliii,left] (Nxyz) at (D) {$(0.,0.,1.)$};

          \def\hsqB{5.5}
          \pgfmathparse{1-1/\hsqB}\let\hhsqB\pgfmathresult
          \pgfmathparse{1-1/\hsq-1/\hsqB}\let\hhhsqB\pgfmathresult
	  \coordinate (hBB) at  ($ \hhsqB*(A) + 1/\hsqB*(B) $);
	  \coordinate (hCC) at  ($ \hhsq*(A) + 1/\hsq*(C) $);
	  \coordinate (hDD) at  ($ \hhsq*(A) + 1/\hsq*(D) $);
	  \coordinate (hBC) at  ($ \hhhsqB*(A) + 1/\hsqB*(B) + 1/\hsq*(C) $);
	  \coordinate (hBD) at  ($ \hhhsqB*(A) + 1/\hsqB*(B) + 1/\hsq*(D) $);
	  \coordinate (hCD) at  ($ \hhhsq*(A) + 1/\hsq*(C) + 1/\hsq*(D) $);
	  \draw[line width=0.4pt, color=\colo] (hBB) -- (hBC) -- (hCC);
	  \draw[line width=0.4pt, color=\colo] (hBB) -- (hBD) -- (hDD);
	  \draw[line width=0.4pt, color=\colo] (hCC) -- (hCD) -- (hDD);
          
          \newcount\z
          \foreach \k in {0,1,...,\K}  {
            \foreach \x in {\k,...,0}  {
              \pgfmathparse{\k-\x}\let\YY\pgfmathresult
              \foreach \y in {0,...,\YY} {
                \pgfmathsetcount{\z}{\k-\x-\y} 
                \coordinate (Axyz) at  ($ (A) + 1/\K*(\x,\y,\z) $) ;
                \ifthenelse{\k<\K}
                           {\fill[color=\coliii] (Axyz) circle (1.5pt); }
                           {\fill[color=\colii] (Axyz) circle (1.5pt) ;
                             \node[color=\colii,below] (Nxyz) at  (Axyz) {};}
              }
            }
          }
        }
  \end{tikzpicture}

%% file: fig_Tgeom.tex
\begin{tikzpicture}[scale=4,math3d] 

  \def\kk{3}

  \def\colk{black}
  \def\colo{magenta}
  \def\coli{darkgreen}
  \def\colii{red}
  \def\coliii{blue}

  \def\bcAAX{0.2}
  \def\bcCCX{0.5}
  \def\bcDDX{0.3}
  \def\bcAX{\bcAAX}
  \def\bcCX{\bcCCX}
  \def\bcDX{\bcDDX}

  \def\bcAAK{0.2}
  \def\bcCCK{0.2}
  \def\bcDDK{0.6}
  \def\bcAK{\bcAAK}
  \def\bcCK{\bcCCK}
  \def\bcDK{\bcDDK}

  \coordinate (AA) at (0,0,0);
  \coordinate (CC) at  ($ (AA) + (0,1,0) $);
  \coordinate (DD) at  ($ (AA) + (0,0,1) $);
  \draw (AA) node[color=\coliii,above left] {$\hvv_0$} ;
  \draw (CC) node[color=\coliii,above=1.5pt] {$\hvv_1$} ;
  \draw (DD) node[color=\coliii,right=1.5pt] {$\hvv_2$} ;
  \draw[line width=1.6pt,rounded corners=0.5pt] (AA) -- (CC) -- (DD) -- cycle;

  \node[color=\colk] (KK2) at ($ \bcAAK*(AA) + \bcCCK*(CC)+ \bcDDK*(DD) $) {$\hK$};

  \def\hsq{12}
  \pgfmathparse{1-1/\hsq}\let\hhsq\pgfmathresult
  \pgfmathparse{1-2/\hsq}\let\hhhsq\pgfmathresult
  \coordinate (hCC) at  ($ \hhsq*(AA) + 1/\hsq*(CC) $);
  \coordinate (hDD) at  ($ \hhsq*(AA) + 1/\hsq*(DD) $);
  \coordinate (hCD) at  ($ \hhhsq*(AA) + 1/\hsq*(CC) + 1/\hsq*(DD) $);
  \draw[line width=0.4pt] (hCC) -- (hCD) -- (hDD);

  \coordinate (XX) at ($ \bcAAX*(AA) + \bcCCX*(CC)+ \bcDDX*(DD) $) ;
  \fill[color=\coliii] (XX) circle (0.8pt);
  \node[color=\coliii,below] (NXX) at (XX) {$\hxx$};

  \fill[color=\coliii] (AA) circle (0.8pt);
  \node[color=\coliii,below] (Nxyz) at (AA) {$(0.,0.)$};
  \fill[color=\coliii] (CC) circle (0.8pt);
  \node[color=\coliii,below] (Nxyz) at (CC) {$(1.,0.)$};
  \fill[color=\coliii] (DD) circle (0.8pt);
  \node[color=\coliii,left] (Nxyz) at (DD) {$(0.,1.)$};

  \coordinate (A) at (0,1.6,0.2);
  \coordinate (C) at  ($ (A) + (0,1.2,-0.2) $);
  \coordinate (D) at  ($ (A) + (0,0.5,0.75) $);
  \draw (A) node[color=\colii,left] {$\vv_0=\Tgeomv(\hvv_0)$} ;  
  \draw (C) node[color=\colii,above right=2pt] {$\vv_1=\Tgeomv(\hvv_1)$} ; 
  \draw (D) node[color=\colii,right=1.5pt] {$\vv_2=\Tgeomv(\hvv_2)$} ; 
  \draw[line width=1.6pt,rounded corners=0.5pt] (A) -- (C) -- (D) -- cycle;

  \node [color=\colk] (K2) at ($ \bcAK*(A) + \bcCK*(C)+ \bcDK*(D) $) {$K$};

  \coordinate (X) at ($ \bcAX*(A) + \bcCX*(C)+ \bcDX*(D) $) ;
  \fill[color=\colii] (X) circle (0.8pt);
  \node[color=\colii,below] (NXX) at (X) {$\xx=\Tgeomv(\hxx)$};

  \fill[color=\colii] (A) circle (0.8pt);
  \fill[color=\colii] (C) circle (0.8pt);
  \fill[color=\colii] (D) circle (0.8pt);

  \tikzstyle{fl}=[->,>=latex,very thick];
  \node (KK2D) at  ($ (XX) + (0,0.0,0.05) $) {} ;
  \node (K2D) at  ($ (X) + (0,-0.05,0.) $) {};
  \draw[fl] (KK2D) to[bend left] (K2D);
  \node (geomap) at  ($ (CC) + (0,0.1,0.7) $) {$\Tgeomv$};
\end{tikzpicture}

%% file: fig_ImTriaFaceTet_k3.tex
\begin{tikzpicture}[scale=4,math3d] 

  \def\kk{3}

  \def\colk{black}
  \def\colo{magenta}
  \def\coli{darkgreen}
  \def\colii{red}
  \def\coliii{blue}

  \def\opacity{0.3}
  \def\opacityi{0.4}
  \def\opacityo{0.2}

  \coordinate (AA) at (0,0,0);
  \coordinate (CC) at  ($ (AA) + (0,1,0) $);
  \coordinate (DD) at  ($ (AA) + (0,0,1) $);
  \draw (AA) node[above left] {$\hvv_0$} ;
  \draw (CC) node[above=1.5pt] {$\hvv_1$} ;
  \draw (DD) node[right=1.5pt] {$\hvv_2$} ;
  \draw[line width=1.0pt,rounded corners=0.5pt] (AA) -- (CC) -- (DD) -- cycle;

  \def\hsq{12}
  \pgfmathparse{1-1/\hsq}\let\hhsq\pgfmathresult
  \pgfmathparse{1-2/\hsq}\let\hhhsq\pgfmathresult
  \coordinate (hCC) at  ($ \hhsq*(AA) + 1/\hsq*(CC) $);
  \coordinate (hDD) at  ($ \hhsq*(AA) + 1/\hsq*(DD) $);
  \coordinate (hCD) at  ($ \hhhsq*(AA) + 1/\hsq*(CC) + 1/\hsq*(DD) $);
  \draw[line width=0.4pt] (hCC) -- (hCD) -- (hDD);

  \node[color=\colk] (K2) at ($ (AA) + (0,0.6,0.6) $) {$\hK^2$};

  \fill[color=\colo] (AA) circle (0.8pt);
  \node[color=\colo,below] (Nxyz) at (AA) {$\haa_{(0,0)}$};
  \newcount\y
  \foreach \x in {1,0}  {
    \pgfmathsetcount{\y}{1-\x} 
    \coordinate (Axyz) at  ($ (AA) + 1/\kk*(0,\x,\y) $) ;
    \node[color=\coli,below] (Nxyz) at (Axyz) {$\haa_{(\x,\the\y)}$};
    \fill[color=\coli] (Axyz) circle (0.8pt);
  }
  \newcount\y
  \foreach \x in {2,1,...,0}  {
    \pgfmathsetcount{\y}{2-\x} 
    \coordinate (Axyz) at  ($ (AA) + 1/\kk*(0,\x,\y) $) ;
    \node[color=\colii,below] (Nxyz) at (Axyz) {$\haa_{(\x,\the\y)}$};
    \fill[color=\colii] (Axyz) circle (0.8pt);
  }
  \newcount\y
  \foreach \x in {3,2,...,0}  {
    \pgfmathsetcount{\y}{3-\x} 
    \coordinate (Axyz) at  ($ (AA) + 1/\kk*(0,\x,\y) $) ;
    \node[color=\coliii,below] (Nxyz) at (Axyz) {$\haa_{(\x,\the\y)}$};
    \fill[color=\coliii] (Axyz) circle (0.8pt);
  }

  \coordinate (eps0) at ($1/\kk*(0,-0.05,0)$); 
  \coordinate (eps) at ($1/\kk*(0,0.05,-0.05)$); 
  \node[color=\colo,line width=1.4,below=12pt] (Nxyz) at (AA) {$\calCdo$};
  \coordinate (A10) at ($(AA) + 1/\kk*(0,1,0)$);
  \coordinate (A01) at ($(AA) + 1/\kk*(0,0,1)$);
  \draw[color=\colo,dashed,line width=0.7pt,->,>=latex] (AA) -- ($(A10) + (eps0)$);
  \draw[color=\coli,dashed,line width=0.7pt,->,>=latex] (A10) -- ($(A01) + (eps)$);
  \node[color=\coli,below=12pt] (Nxyz) at (A10) {$\calCdi$};
  \coordinate (A20) at ($(AA) + 1/\kk*(0,2,0)$);
  \coordinate (A02) at ($(AA) + 1/\kk*(0,0,2)$);
  \draw[color=\coli,dashed,line width=0.7pt,->,>=latex] ($(A01) + (eps)$) -- ($(A20) - (eps)$);
  \draw[color=\colii,dashed,line width=0.7pt,->,>=latex] (A20) -- ($(A02) + (eps)$);
  \node[color=\colii,below=12pt] (Nxyz) at (A20) {$\calCdii$};
  \coordinate (A30) at ($(AA) + 1/\kk*(0,3,0)$);
  \coordinate (CD) at ($1/2*(CC) + 1/2*(DD)$);
  \draw[color=\colii,dashed,line width=0.7pt,->,>=latex] ($(A02) + (eps)$) -- ($(A30) - (eps)$);
  \draw[color=\coliii,dashed,line width=1.4pt,->,>=latex] (A30) -- ($(DD) + (eps)$);
  \node[color=\coliii,below=12pt] (Nxyz) at (A30) {$\calCdiii$};

  \coordinate (RA) at (0,2,0.3);
  \coordinate (RB) at  ($ (RA) + (0.9,0.,0.) $);
  \coordinate (RC) at  ($ (RA) + (0.,0.9,0.) $);
  \coordinate (RD) at  ($ (RA) + (0.,0.,0.9) $);
  \draw (RA) node[left] {$\hvv_0$} ;  
  \draw (RB) node[above left=2pt] {$\hvv_1$} ; 
  \draw (RC) node[above right=2pt] {$\hvv_2$} ; 
  \draw (RD) node[above left=2pt] {$\hvv_3$} ; 
  \draw[line width=1.0pt,rounded corners=0.5pt] (RA) -- (RB) -- (RD) -- cycle;
  \draw[line width=1.0pt,rounded corners=0.5pt] (RA) -- (RC) -- (RB) -- cycle;
  \draw[line width=1.0pt,rounded corners=0.5pt] (RA) -- (RC) -- (RD) -- cycle;

  \coordinate (RE) at ($ 0.3 *(RB) - 0.45 *(RC) + 1.15 *(RD) $) ;
  \coordinate (RF) at ($ -0.4 *(RB) + 0.25 *(RC) + 1.15 *(RD) $) ;
  \coordinate (RG) at ($ -0.1 *(RB) + 1.2 *(RC) - 0.1 *(RD) $) ;
  \coordinate (RH) at ($ 1.16 *(RB) - 0.08 *(RC) - 0.08 *(RD) $) ;

  \draw[color=\coliii,line width=0.2pt,fill=\coliii!40,fill
  opacity=\opacityo]  (RE) -- (RF) -- (RG) -- (RH) -- cycle;

  \node [color=\coliii] (RKK3) at ($ (RD) + (0,0.15,0.15) $) {$\hHpl_0$};
  \node[color=\colk] (RK3) at ($ (RD) + (0,-0.35,-0.25) $) {$\hK^3$};

  \pgfmathparse{1-1/\kk}\let\kkp\pgfmathresult
  \coordinate (RB1) at ($ \kkp *(RA) + 1/\kk *(RB) $) ;
  \coordinate (RC1) at ($ \kkp *(RA) + 1/\kk *(RC) $) ;
  \coordinate (RD1) at ($ \kkp *(RA) + 1/\kk *(RD) $) ;
  \pgfmathparse{1-2/\kk}\let\kkp\pgfmathresult
  \coordinate (RB2) at ($ \kkp *(RA) + 2/\kk *(RB) $) ;
  \coordinate (RC2) at ($ \kkp *(RA) + 2/\kk *(RC) $) ;
  \coordinate (RD2) at ($ \kkp *(RA) + 2/\kk *(RD) $) ;

  \draw[color=\colk]   (RB1) -- (RC1) -- (RD1) -- cycle;
  \draw[color=\colk]  (RB2) -- (RC2) -- (RD2) -- cycle;
  \draw[color=\coliii,line width=1pt,fill=\coliii!40,fill opacity=\opacity]  (RB) -- (RC) -- (RD) -- cycle;

  \fill[color=\colk] (RA) circle (0.6pt);
  \foreach \x in {1,0}  {
    \pgfmathparse{1-\x}\let\YY\pgfmathresult
    \foreach \y in {0,...,\YY} {
      \pgfmathparse{1-\x-\y}\let\z\pgfmathresult
      \pgfmathparse{\x/\kk}\let\xkk\pgfmathresult
      \pgfmathparse{\y/\kk}\let\ykk\pgfmathresult
      \pgfmathparse{\z/\kk}\let\zkk\pgfmathresult
      \pgfmathparse{1-\xkk-\ykk-\zkk}\let\okk\pgfmathresult
      \fill[color=\colk,fill opacity=\opacityi]
      ($ \okk *(RA) + \xkk *(RB) + \ykk *(RC) + \zkk *(RD) $) circle (0.6pt);
    }
  }
  \foreach \x in {2,1,...,0}  {
    \pgfmathparse{2-\x}\let\YY\pgfmathresult
    \foreach \y in {0,...,\YY} {
      \pgfmathparse{2-\x-\y}\let\z\pgfmathresult
      \pgfmathparse{\x/\kk}\let\xkk\pgfmathresult
      \pgfmathparse{\y/\kk}\let\ykk\pgfmathresult
      \pgfmathparse{\z/\kk}\let\zkk\pgfmathresult
      \pgfmathparse{1-\xkk-\ykk-\zkk}\let\okk\pgfmathresult
      \fill[color=\colk,fill opacity=\opacityi]
      ($ \okk *(RA) + \xkk *(RB) + \ykk *(RC) + \zkk *(RD) $) circle (0.6pt);
    }
  }
  \newcount\x
  \foreach \y in {0}  {
    \foreach \z in {0}  {
      \pgfmathsetcount{\x}{\kk-\y-\z} 
      \pgfmathparse{\x/\kk}\let\xkk\pgfmathresult
      \pgfmathparse{\y/\kk}\let\ykk\pgfmathresult
      \pgfmathparse{\z/\kk}\let\zkk\pgfmathresult
      \pgfmathparse{1-\xkk-\ykk-\zkk}\let\okk\pgfmathresult
      \coordinate (RAxyz) at  ($ \okk *(RA) + \xkk *(RB) + \ykk *(RC) + \zkk *(RD) $) ;
      \fill[color=\colo] (RAxyz) circle (1pt);
      \node[color=\colo,below] (Nxyz) at (RAxyz) {$\haa_{(\the\x,\y,\z)}$};
    }
  }
  \newcount\x
  \foreach \z in {0,1}  {
    \pgfmathsetcount{\y}{1-\z} 
    \pgfmathsetcount{\x}{\kk-\y-\z} 
    \pgfmathparse{\x/\kk}\let\xkk\pgfmathresult
    \pgfmathparse{\y/\kk}\let\ykk\pgfmathresult
    \pgfmathparse{\z/\kk}\let\zkk\pgfmathresult
    \pgfmathparse{1-\xkk-\ykk-\zkk}\let\okk\pgfmathresult
    \coordinate (RAxyz) at  ($ \okk *(RA) + \xkk *(RB) + \ykk *(RC) + \zkk *(RD) $) ;
    \node[color=\coli,below] (Nxyz) at (RAxyz) {$\aa_{(\the\x,\the\y,\z)}$};
    \fill[color=\coli] (RAxyz) circle (1pt);
  }
  \newcount\x
  \foreach \z in {0,1,2}  {
    \pgfmathsetcount{\y}{2-\z} 
    \pgfmathsetcount{\x}{\kk-\y-\z} 
    \pgfmathparse{\x/\kk}\let\xkk\pgfmathresult
    \pgfmathparse{\y/\kk}\let\ykk\pgfmathresult
    \pgfmathparse{\z/\kk}\let\zkk\pgfmathresult
    \pgfmathparse{1-\xkk-\ykk-\zkk}\let\okk\pgfmathresult
    \coordinate (RAxyz) at  ($ \okk *(RA) + \xkk *(RB) + \ykk *(RC) + \zkk *(RD) $) ;
    \node[color=\colii,below] (Nxyz) at (RAxyz) {$\aa_{(\the\x,\the\y,\z)}$};
    \fill[color=\colii] (RAxyz) circle (1pt);
  }
  \newcount\x
  \foreach \z in {0,1,...,3}  {
    \pgfmathsetcount{\y}{3-\z} 
    \pgfmathsetcount{\x}{\kk-\y-\z} 
    \pgfmathparse{\x/\kk}\let\xkk\pgfmathresult
    \pgfmathparse{\y/\kk}\let\ykk\pgfmathresult
    \pgfmathparse{\z/\kk}\let\zkk\pgfmathresult
    \pgfmathparse{1-\xkk-\ykk-\zkk}\let\okk\pgfmathresult
    \coordinate (RAxyz) at  ($ \okk *(RA) + \xkk *(RB) + \ykk *(RC) + \zkk *(RD) $) ;
    \node[color=\coliii,below] (Nxyz) at (RAxyz) {$\aa_{(\the\x,\the\y,\z)}$};
    \fill[color=\coliii] (RAxyz) circle (1pt);
  }

  \coordinate (eps0) at ($1/\kk*(0.05,-0.05,0)$); 
  \coordinate (eps) at ($1/\kk*(0,0.05,-0.05)$); 
  \pgfmathparse{2/\kk}\let\xkk\pgfmathresult
  \pgfmathparse{1/\kk}\let\ykk\pgfmathresult
  \pgfmathparse{0/\kk}\let\zkk\pgfmathresult
  \pgfmathparse{1-\xkk-\ykk-\zkk}\let\okk\pgfmathresult
  \coordinate (A210) at ($ \okk *(RA) + \xkk *(RB) + \ykk *(RC) + \zkk *(RD) $);
  \pgfmathparse{2/\kk}\let\xkk\pgfmathresult
  \pgfmathparse{0/\kk}\let\ykk\pgfmathresult
  \pgfmathparse{1/\kk}\let\zkk\pgfmathresult
  \pgfmathparse{1-\xkk-\ykk-\zkk}\let\okk\pgfmathresult
  \coordinate (A201) at ($ \okk *(RA) + \xkk *(RB) + \ykk *(RC) + \zkk *(RD) $);
  \draw[color=\colo,dashed,line width=1.4pt,->,>=latex] (RB) -- ($(A210) $);
  \draw[color=\coli,dashed,line width=0.7pt,->,>=latex] (A210) -- ($(A201) + (eps)$);
  \pgfmathparse{1/\kk}\let\xkk\pgfmathresult
  \pgfmathparse{2/\kk}\let\ykk\pgfmathresult
  \pgfmathparse{0/\kk}\let\zkk\pgfmathresult
  \pgfmathparse{1-\xkk-\ykk-\zkk}\let\okk\pgfmathresult
  \coordinate (A120) at ($ \okk *(RA) + \xkk *(RB) + \ykk *(RC) + \zkk *(RD) $);
  \pgfmathparse{1/\kk}\let\xkk\pgfmathresult
  \pgfmathparse{0/\kk}\let\ykk\pgfmathresult
  \pgfmathparse{2/\kk}\let\zkk\pgfmathresult
  \pgfmathparse{1-\xkk-\ykk-\zkk}\let\okk\pgfmathresult
  \coordinate (A102) at ($ \okk *(RA) + \xkk *(RB) + \ykk *(RC) + \zkk *(RD) $);
  \draw[color=\coli,dashed,line width=0.7pt,->,>=latex] ($(A201) + (eps)$) -- ($(A120) - (eps)$);
  \draw[color=\colii,dashed,line width=0.7pt,->,>=latex] (A120) -- ($(A102) + (eps)$);
  \coordinate (RCD) at ($1/2 *(RC) + 1/2 *(RD)$);
  \draw[color=\colii,dashed,line width=0.7pt,->,>=latex] ($(A102) + (eps)$) -- ($(RC) - (eps)$);
  \draw[color=\coliii,dashed,line width=2.1pt,->,>=latex] (RC) -- ($(RD) + (eps)$);

  \coordinate (A) at (0,3.2,-0.2);
  \coordinate (B) at  ($ (A) + (0,0.9,-0.2) $);
  \coordinate (C) at  ($ (A) + (-1.8,0.1,-0.4) $);
  \coordinate (D) at  ($ (A) + (0.05,0.3,1.1) $);
  \draw (A) node[left] {$\vv_0$} ;  
  \draw (B) node[above right=2pt] {$\vv_1$} ; 
  \draw (C) node[above right=2pt] {$\vv_2$} ; 
  \draw (D) node[right=1.5pt] {$\vv_3$} ; 
  \draw[line width=1.0pt,rounded corners=0.5pt] (A) -- (B) -- (D) -- cycle;
  \draw[line width=1.0pt,rounded corners=0.5pt] (A) -- (C) -- (B) -- cycle;
  \draw[line width=1.0pt,rounded corners=0.5pt] (A) -- (C) -- (D) -- cycle;

  \coordinate (E) at ($ 0.3*(B) - 0.45*(C) + 1.15*(D) $) ;
  \coordinate (F) at ($ -0.4*(B) + 0.25*(C) + 1.15*(D) $) ;
  \coordinate (G) at ($ -0.15*(B) + 1.3*(C) - 0.15*(D) $) ;
  \coordinate (H) at ($ 1.3*(B) - 0.15*(C) - 0.15*(D) $) ;

  \draw[color=\coliii,line width=0.2pt,fill=\coliii!40,fill
  opacity=\opacityo]  (E) -- (F) -- (G) -- (H) -- cycle;

  \node [color=\coliii] (KK3) at ($ (D) + (0,0,0.25) $) {$\Hplv_0$};
  \node[color=\colk] (K3) at ($ (D) + (0,0.1,-1.3) $) {$K^3$};

  \pgfmathparse{1-1/\kk}\let\kkp\pgfmathresult
  \coordinate (B1) at ($ \kkp*(A) + 1/\kk*(B) $) ;
  \coordinate (C1) at ($ \kkp*(A) + 1/\kk*(C) $) ;
  \coordinate (D1) at ($ \kkp*(A) + 1/\kk*(D) $) ;
  \pgfmathparse{1-2/\kk}\let\kkp\pgfmathresult
  \coordinate (B2) at ($ \kkp*(A) + 2/\kk*(B) $) ;
  \coordinate (C2) at ($ \kkp*(A) + 2/\kk*(C) $) ;
  \coordinate (D2) at ($ \kkp*(A) + 2/\kk*(D) $) ;

  \draw[color=\colk]   (B1) -- (C1) -- (D1) -- cycle;
  \draw[color=\colk]  (B2) -- (C2) -- (D2) -- cycle;
  \draw[color=\coliii,line width=1pt,fill=\coliii!40,fill opacity=\opacity]  (B) -- (C) -- (D) -- cycle;

  \fill[color=\colk] (A) circle (0.6pt);
  \foreach \x in {1,0}  {
    \pgfmathparse{1-\x}\let\YY\pgfmathresult
    \foreach \y in {0,...,\YY} {
      \pgfmathparse{1-\x-\y}\let\z\pgfmathresult
      \pgfmathparse{\x/\kk}\let\xkk\pgfmathresult
      \pgfmathparse{\y/\kk}\let\ykk\pgfmathresult
      \pgfmathparse{\z/\kk}\let\zkk\pgfmathresult
      \pgfmathparse{1-\xkk-\ykk-\zkk}\let\okk\pgfmathresult
      \fill[color=\colk,fill opacity=\opacityi]
      ($ \okk*(A) + \xkk*(B) + \ykk*(C) + \zkk*(D) $) circle (0.6pt);
    }
  }
  \foreach \x in {2,1,...,0}  {
    \pgfmathparse{2-\x}\let\YY\pgfmathresult
    \foreach \y in {0,...,\YY} {
      \pgfmathparse{2-\x-\y}\let\z\pgfmathresult
      \pgfmathparse{\x/\kk}\let\xkk\pgfmathresult
      \pgfmathparse{\y/\kk}\let\ykk\pgfmathresult
      \pgfmathparse{\z/\kk}\let\zkk\pgfmathresult
      \pgfmathparse{1-\xkk-\ykk-\zkk}\let\okk\pgfmathresult
      \fill[color=\colk,fill opacity=\opacityi]
      ($ \okk*(A) + \xkk*(B) + \ykk*(C) + \zkk*(D) $) circle (0.6pt);
    }
  }
  \newcount\x
  \foreach \y in {0}  {
    \foreach \z in {0}  {
      \pgfmathsetcount{\x}{\kk-\y-\z} 
      \pgfmathparse{\x/\kk}\let\xkk\pgfmathresult
      \pgfmathparse{\y/\kk}\let\ykk\pgfmathresult
      \pgfmathparse{\z/\kk}\let\zkk\pgfmathresult
      \pgfmathparse{1-\xkk-\ykk-\zkk}\let\okk\pgfmathresult
      \coordinate (Axyz) at  ($ \okk*(A) + \xkk*(B) + \ykk*(C) + \zkk*(D) $) ;
      \fill[color=\colo] (Axyz) circle (1pt);
      \node[color=\colo,below] (Nxyz) at (Axyz) {$\aa_{(\the\x,\y,\z)}$};
    }
  }
  \newcount\x
  \foreach \z in {0,1}  {
    \pgfmathsetcount{\y}{1-\z} 
    \pgfmathsetcount{\x}{\kk-\y-\z} 
    \pgfmathparse{\x/\kk}\let\xkk\pgfmathresult
    \pgfmathparse{\y/\kk}\let\ykk\pgfmathresult
    \pgfmathparse{\z/\kk}\let\zkk\pgfmathresult
    \pgfmathparse{1-\xkk-\ykk-\zkk}\let\okk\pgfmathresult
    \coordinate (Axyz) at  ($ \okk*(A) + \xkk*(B) + \ykk*(C) + \zkk*(D) $) ;
    \node[color=\coli,below] (Nxyz) at (Axyz) {$\aa_{(\the\x,\the\y,\z)}$};
    \fill[color=\coli] (Axyz) circle (1pt);
  }
  \newcount\x
  \foreach \z in {0,1,2}  {
    \pgfmathsetcount{\y}{2-\z} 
    \pgfmathsetcount{\x}{\kk-\y-\z} 
    \pgfmathparse{\x/\kk}\let\xkk\pgfmathresult
    \pgfmathparse{\y/\kk}\let\ykk\pgfmathresult
    \pgfmathparse{\z/\kk}\let\zkk\pgfmathresult
    \pgfmathparse{1-\xkk-\ykk-\zkk}\let\okk\pgfmathresult
    \coordinate (Axyz) at  ($ \okk*(A) + \xkk*(B) + \ykk*(C) + \zkk*(D) $) ;
    \node[color=\colii,below] (Nxyz) at (Axyz) {$\aa_{(\the\x,\the\y,\z)}$};
    \fill[color=\colii] (Axyz) circle (1pt);
  }
  \newcount\x
  \foreach \z in {0,1,...,3}  {
    \pgfmathsetcount{\y}{3-\z} 
    \pgfmathsetcount{\x}{\kk-\y-\z} 
    \pgfmathparse{\x/\kk}\let\xkk\pgfmathresult
    \pgfmathparse{\y/\kk}\let\ykk\pgfmathresult
    \pgfmathparse{\z/\kk}\let\zkk\pgfmathresult
    \pgfmathparse{1-\xkk-\ykk-\zkk}\let\okk\pgfmathresult
    \coordinate (Axyz) at  ($ \okk*(A) + \xkk*(B) + \ykk*(C) + \zkk*(D) $) ;
    \node[color=\coliii,below] (Nxyz) at (Axyz) {$\aa_{(\the\x,\the\y,\z)}$};
    \fill[color=\coliii] (Axyz) circle (1pt);
  }

  \coordinate (eps0) at ($1/\kk*(0.05,-0.05,0)$); 
  \coordinate (eps) at ($1/\kk*(0,0.05,-0.05)$); 
  \pgfmathparse{2/\kk}\let\xkk\pgfmathresult
  \pgfmathparse{1/\kk}\let\ykk\pgfmathresult
  \pgfmathparse{0/\kk}\let\zkk\pgfmathresult
  \pgfmathparse{1-\xkk-\ykk-\zkk}\let\okk\pgfmathresult
  \coordinate (A210) at ($ \okk*(A) + \xkk*(B) + \ykk*(C) + \zkk*(D) $);
  \pgfmathparse{2/\kk}\let\xkk\pgfmathresult
  \pgfmathparse{0/\kk}\let\ykk\pgfmathresult
  \pgfmathparse{1/\kk}\let\zkk\pgfmathresult
  \pgfmathparse{1-\xkk-\ykk-\zkk}\let\okk\pgfmathresult
  \coordinate (A201) at ($ \okk*(A) + \xkk*(B) + \ykk*(C) + \zkk*(D) $);
  \draw[color=\colo,dashed,line width=1.4pt,->,>=latex] (B) -- ($(A210) $);
  \draw[color=\coli,dashed,line width=0.7pt,->,>=latex] (A210) -- ($(A201) + (eps)$);
  \pgfmathparse{1/\kk}\let\xkk\pgfmathresult
  \pgfmathparse{2/\kk}\let\ykk\pgfmathresult
  \pgfmathparse{0/\kk}\let\zkk\pgfmathresult
  \pgfmathparse{1-\xkk-\ykk-\zkk}\let\okk\pgfmathresult
  \coordinate (A120) at ($ \okk*(A) + \xkk*(B) + \ykk*(C) + \zkk*(D) $);
  \pgfmathparse{1/\kk}\let\xkk\pgfmathresult
  \pgfmathparse{0/\kk}\let\ykk\pgfmathresult
  \pgfmathparse{2/\kk}\let\zkk\pgfmathresult
  \pgfmathparse{1-\xkk-\ykk-\zkk}\let\okk\pgfmathresult
  \coordinate (A102) at ($ \okk*(A) + \xkk*(B) + \ykk*(C) + \zkk*(D) $);
  \draw[color=\coli,dashed,line width=0.7pt,->,>=latex] ($(A201) + (eps)$) -- ($(A120) - (eps)$);
  \draw[color=\colii,dashed,line width=0.7pt,->,>=latex] (A120) -- ($(A102) + (eps)$);
  \coordinate (CD) at ($1/2*(C) + 1/2*(D)$);
  \draw[color=\colii,dashed,line width=0.7pt,->,>=latex] ($(A102) + (eps)$) -- ($(C) - (eps)$);
  \draw[color=\coliii,dashed,line width=1.8pt,->,>=latex] (C) -- ($(D) + (eps)$);

  \tikzstyle{fl}=[->,>=latex,very thick];
  \node (K2DD) at  ($ (K2) + (0,-0.1,-0.8) $) {} ;
  \node (K3D) at  ($ (D) + (0,0.15,-1.) $) {};
  \draw[fl] (K2DD) to[bend right] (K3D);
  \node (geomap) at  ($ (RA) + (0,0.,-0.8) $) {$\Injvo$};
  \node (K2DD) at  ($ (K2) + (0,0.05,0.05) $) {} ;
  \node (K3DR) at  ($ (RD) + (0,-0.2,-0.2) $) {};
  \draw[fl] (K2DD) to[bend left] (K3DR);
  \node (geomap) at  ($ 0.5*(AA) +  0.5*(RA) + (0,0.15,0.75) $) {$\InjhHdo$};
  \node (K3DR) at  ($ (RD) + (0,0.25,-0.15) $) {};
  \node (K3DD) at  ($ (D) + (0,-0.1,0.1) $) {};
  \draw[fl] (K3DR) to[bend left] (K3DD);
  \node (geomap) at  ($ 0.5*(RA) +  0.5*(A) + (0,0.3,1.) $) {$\Tgeomv$};

\end{tikzpicture}

%% file: fig_TetToTet_k3_perm_trsp0.tex
\begin{tikzpicture}[scale=4,math3d] 

  \def\kk{3}

  \def\colk{black}
  \def\colo{magenta}
  \def\coli{darkgreen}
  \def\colii{red}
  \def\coliii{blue}

  \def\opacity{0.4}
  \def\opacityi{0.6}
  \def\opacityo{0.2}

  \coordinate (A) at (0,0,0);
  \coordinate (B) at  ($ (A) + (1,0,0) $);
  \coordinate (C) at  ($ (A) + (0,1,0) $);
  \coordinate (D) at  ($ (A) + (0,0,1) $);
  \draw (A) node[left] {$\hvv_0$} ;  
  \draw (B) node[above left] {$\hvv_1$} ; 
  \draw (C) node[above=2pt] {$\hvv_2$} ; 
  \draw (D) node[right=1.5pt] {$\hvv_3$} ; 
  \draw[line width=1.0pt,rounded corners=0.5pt] (A) -- (B) -- (D) -- cycle;
  \draw[line width=1.0pt,fill=\coliii!40,fill opacity=\opacity,rounded corners=0.5pt] (A) -- (C) -- (B) -- cycle;
  \draw[line width=1.0pt,rounded corners=0.5pt] (B) -- (C) -- (D) -- cycle;
  \draw[line width=1.6pt,rounded corners=0.5pt] (A) -- (C) -- (D) -- cycle;

  \def\hsq{12}
  \pgfmathparse{1-1/\hsq}\let\hhsq\pgfmathresult
  \pgfmathparse{1-2/\hsq}\let\hhhsq\pgfmathresult
  \def\hsqB{8}
  \pgfmathparse{1-1/\hsqB}\let\hhsqB\pgfmathresult
  \pgfmathparse{1-1/\hsq-1/\hsqB}\let\hhhsqB\pgfmathresult
  \coordinate (hBB) at  ($ \hhsqB*(A) + 1/\hsqB*(B) $);
  \coordinate (hCC) at  ($ \hhsq*(A) + 1/\hsq*(C) $);
  \coordinate (hDD) at  ($ \hhsq*(A) + 1/\hsq*(D) $);
  \coordinate (hBC) at  ($ \hhhsqB*(A) + 1/\hsqB*(B) + 1/\hsq*(C) $);
  \coordinate (hBD) at  ($ \hhhsqB*(A) + 1/\hsqB*(B) + 1/\hsq*(D) $);
  \coordinate (hCD) at  ($ \hhhsq*(A) + 1/\hsq*(C) + 1/\hsq*(D) $);
  \draw[line width=0.4pt, color=\colk] (hBB) -- (hBC) -- (hCC);
  \draw[line width=0.4pt, color=\colk] (hBB) -- (hBD) -- (hDD);
  \draw[line width=0.4pt, color=\colk] (hCC) -- (hCD) -- (hDD);

  \pgfmathparse{1-1/\kk}\let\kkp\pgfmathresult
  \coordinate (B1) at ($ \kkp*(A) + 1/\kk*(B) $) ;
  \coordinate (C1) at ($ \kkp*(A) + 1/\kk*(C) $) ;
  \coordinate (D1) at ($ \kkp*(A) + 1/\kk*(D) $) ;
  \pgfmathparse{1-2/\kk}\let\kkp\pgfmathresult
  \coordinate (B2) at ($ \kkp*(A) + 2/\kk*(B) $) ;
  \coordinate (C2) at ($ \kkp*(A) + 2/\kk*(C) $) ;
  \coordinate (D2) at ($ \kkp*(A) + 2/\kk*(D) $) ;


  \coordinate (E) at ($ -0.15*(A) - 0.15*(B) + 1.3*(C) $) ;
  \coordinate (F) at ($ -2.6*(A) + 1.3*(B) + 1.3*(C) $) ;
  \coordinate (G) at ($ -0.15*(A) + 1.3*(B) - 0.15*(C) $) ;
  \coordinate (H) at ($ 1.4*(A) - 0.2*(B) - 0.2*(C) $) ;

  \draw[color=\coliii,line width=0.2pt,fill=\coliii!40,fill
  opacity=\opacityo]  (E) -- (F) -- (G) -- (H) -- cycle;

  \node [color=\coliii] (K3) at ($ (A) + (0.15,0.15,0) $) {} ; 
  \node [color=\coliii] (K33) at ($ (F) - (0.3,0.3,0) $) {$\hHpl_d$};
  \node [color=\colk] (K33) at ($ -0.3*(C) + 0.6*(D) $) {$\hK^d$};

  \fill[color=\colk,fill opacity=1.0] (A) circle (0.6pt);
  \foreach \x in {3,2,...,0}  {
    \pgfmathparse{\kk-\x}\let\YY\pgfmathresult
    \foreach \y in {0,...,\YY} {
      \pgfmathparse{\kk-\x-\y}\let\z\pgfmathresult
      \fill[color=\colk,fill opacity=1.0] ($ (A) + 1/\kk*(\x,\y,\z) $) circle (0.6pt);
    }
  }

  \coordinate (nA) at (0,1.8,0 );
  \coordinate (nB) at  ($ (nA) + (0,0.9,-0.4) $);
  \coordinate (nC) at  ($ (nA) + (-1.7,0.3,-0.4) $);
  \coordinate (nD) at  ($ (nA) + (0.05,0,1.3) $);
  \draw (nA) node[below=2.5pt] {$\vv_0=\TgeomTranspov(\hvv_3)$} ;
  \draw (nB) node[below] {$\vv_1=\TgeomTranspov(\hvv_1)$} ;
  \draw (nC) node[above] {$\vv_2=\TgeomTranspov(\hvv_2)$} ;
  \draw (nD) node[right=1.5pt] {$\vv_3=\TgeomTranspov(\hvv_0)$} ;
  \draw[line width=1.0pt,rounded corners=0.5pt] (nA) -- (nB) -- (nD) -- cycle;
  \draw[line width=1.0pt,rounded corners=0.5pt] (nA) -- (nC) -- (nB) -- cycle;
  \draw[line width=1.0pt,fill=\coliii!40,fill opacity=\opacity,rounded corners=0.5pt] (nB) -- (nC) -- (nD) -- cycle;

  \coordinate (nE) at ($ 0.15*(nB) - 0.35*(nC) + 1.25*(nD) $) ;
  \coordinate (nF) at ($ -0.25*(nB) + 0.2*(nC) + 1.15*(nD) $) ;
  \coordinate (nG) at ($ -0.15*(nB) + 1.3*(nC) - 0.15*(nD) $) ;
  \coordinate (nH) at ($ 1.3*(nB) - 0.15*(nC) - 0.15*(nD) $) ;

  \draw[color=\coliii,line width=0.2pt,fill=\coliii!40,fill
  opacity=\opacityo]  (nE) -- (nF) -- (nG) -- (nH) -- cycle;

  \node [color=\coliii] (nK3) at ($ (nD) + (0,0.1,0.2) $) {$\Hplv_0$};

  \pgfmathparse{1-1/\kk}\let\kkp\pgfmathresult
  \coordinate (nB1) at ($ \kkp*(nA) + 1/\kk*(nB) $) ;
  \coordinate (nC1) at ($ \kkp*(nA) + 1/\kk*(nC) $) ;
  \coordinate (nD1) at ($ \kkp*(nA) + 1/\kk*(nD) $) ;
  \pgfmathparse{1-2/\kk}\let\kkp\pgfmathresult
  \coordinate (nB2) at ($ \kkp*(nA) + 2/\kk*(nB) $) ;
  \coordinate (nC2) at ($ \kkp*(nA) + 2/\kk*(nC) $) ;
  \coordinate (nD2) at ($ \kkp*(nA) + 2/\kk*(nD) $) ;


  \node [color=\coliii] (nK33) at ($ (nD) + (0,0.55,-0.5) $) {};  
  \node [color=\colk] (nK333) at ($ 0.6*(nA) + 0.4*(nD) + (0,-0.15,0) $) {$K$};

  \foreach \x in {3,2,1,...,0}  {
    \pgfmathparse{3-\x}\let\YY\pgfmathresult
    \foreach \y in {0,...,\YY} {
      \pgfmathparse{3-\x-\y}\let\z\pgfmathresult
      \pgfmathparse{\x/\kk}\let\xkk\pgfmathresult
      \pgfmathparse{\y/\kk}\let\ykk\pgfmathresult
      \pgfmathparse{\z/\kk}\let\zkk\pgfmathresult
      \pgfmathparse{1-\xkk-\ykk-\zkk}\let\okk\pgfmathresult
      \fill[color=\colk,fill opacity=1.0]
      ($ \okk*(nA) + \xkk*(nB) + \ykk*(nC) + \zkk*(nD) $) circle (0.6pt);
    }
  }
  \newcount\x
  \pgfmathsetcount{\y}{0}
  \foreach \k in {0,...,\kk}  {
    \foreach \x in {0,...,\k} {
      \pgfmathparse{\k-\x-\y}\let\z\pgfmathresult
      \pgfmathparse{\x/\kk}\let\xkk\pgfmathresult
      \pgfmathparse{\y/\kk}\let\ykk\pgfmathresult
      \pgfmathparse{\z/\kk}\let\zkk\pgfmathresult
      \pgfmathparse{1-\xkk-\ykk-\zkk}\let\okk\pgfmathresult
      \fill[color=\colk,fill opacity=1.0]
      ($ \okk*(nA) + \xkk*(nB) + \ykk*(nC) + \zkk*(nD) $) circle (0.6pt);
    }
  }

  \tikzstyle{fl}=[->,>=latex,very thick];
  \node (K2DD) at  ($  1/2*(C) + 1/2*(D) $) {} ;
  \node (nK3D) at  ($ 0.7*(nD) + 0.3*(nA) $) {};
  \draw[fl] (K2DD) to[bend left] (nK3D);
  \node (fkd1) at  ($ (C) + (0,0.1,0.8) $) {$\TgeomTranspov$};
\end{tikzpicture}

%% file: fig_TFE.tex
\begin{tikzpicture}[scale=4,math3d] 

  \def\kk{3}

  \def\colk{black}
  \def\colo{magenta}
  \def\coli{darkgreen}
  \def\colii{red}
  \def\coliii{blue}

  \def\bcAAX{0.2}
  \def\bcCCX{0.5}
  \def\bcDDX{0.3}
  \def\bcAX{\bcAAX}
  \def\bcCX{\bcCCX}
  \def\bcDX{\bcDDX}

  \def\bcAAK{0.2}
  \def\bcCCK{0.2}
  \def\bcDDK{0.6}
  \def\bcAK{\bcAAK}
  \def\bcCK{\bcCCK}
  \def\bcDK{\bcDDK}

  \coordinate (AA) at (0,0,0);
  \coordinate (CC) at  ($ (AA) + (0,1,0) $);
  \coordinate (DD) at  ($ (AA) + (0,0,1) $);
  \draw (AA) node[color=\coliii,above left] {\large{$\hvv_0$}} ;
  \draw (CC) node[color=\coliii,above=1.5pt] {\large{$\hvv_1$}} ;
  \draw (DD) node[color=\coliii,right=1.5pt] {\large{$\hvv_2$}} ;
  \draw[line width=1.6pt,rounded corners=0.5pt] (AA) -- (CC) -- (DD) -- cycle;

  \node[color=\colk] (KK2) at ($ \bcAAK*(AA) + \bcCCK*(CC)+ \bcDDK*(DD) + (0,0,-0.15) $) {$\hK$};

  \def\hsq{12}
  \pgfmathparse{1-1/\hsq}\let\hhsq\pgfmathresult
  \pgfmathparse{1-2/\hsq}\let\hhhsq\pgfmathresult
  \coordinate (hCC) at  ($ \hhsq*(AA) + 1/\hsq*(CC) $);
  \coordinate (hDD) at  ($ \hhsq*(AA) + 1/\hsq*(DD) $);
  \coordinate (hCD) at  ($ \hhhsq*(AA) + 1/\hsq*(CC) + 1/\hsq*(DD) $);
  \draw[line width=0.4pt, color=\colk] (hCC) -- (hCD) -- (hDD);

  \coordinate (XX) at ($ \bcAAX*(AA) + \bcCCX*(CC)+ \bcDDX*(DD) $) ;
  \fill[color=\coliii] (XX) circle (0.8pt);
  \node[color=\coliii,below] (NXX) at (XX) {\large{$\hxx =\left(\Tgeom{\vvin}\right)^{-1}(\xxin)$}};

  \fill[color=\coliii] (AA) circle (0.8pt);
  \node[color=\coliii,below] (Nxyz) at (AA) {$(0.,0.)$};
  \fill[color=\coliii] (CC) circle (0.8pt);
  \node[color=\coliii,below] (Nxyz) at (CC) {$(1.,0.)$};
  \fill[color=\coliii] (DD) circle (0.8pt);
  \node[color=\coliii,left] (Nxyz) at (DD) {$(0.,1.)$};

  \coordinate (A) at (0,1.6,0.2);
  \coordinate (C) at  ($ (A) + (0,1.2,-0.2) $);
  \coordinate (D) at  ($ (A) + (0,0.5,0.75) $);
  \draw (A) node[color=\colii,below=2pt] {\large{$\vvout_0=\Tgeom{\vvout}(\hvv_0)$}} ;  
  \draw (C) node[color=\colii,below left=2pt] {\large{$\vvout_1=\Tgeom{\vvout}(\hvv_1)$}} ; 
  \draw (D) node[color=\colii,above=1.5pt] {\large{$\vvout_2=\Tgeom{\vvout}(\hvv_2)$}} ; 
  \draw[line width=1.6pt,rounded corners=0.5pt] (A) -- (C) -- (D) -- cycle;

  \node [color=\colk] (K2) at ($ \bcAK*(A) + \bcCK*(C)+ \bcDK*(D) $) {$\Kout$};

  \coordinate (X) at ($ \bcAX*(A) + \bcCX*(C)+ \bcDX*(D) $) ;
  \fill[color=\colii] (X) circle (0.8pt);
  \node[color=\colii,below] (NXX) at (X) {\large{$\xxout=\Tgeom{\vvout}(\hxx)$}};

  \fill[color=\colii] (A) circle (0.8pt);
  \fill[color=\colii] (C) circle (0.8pt);
  \fill[color=\colii] (D) circle (0.8pt);

  \coordinate (AAA) at (0,-0.8,0.7);
  \coordinate (CCC) at  ($ (AAA) + (0,-0.5,0.55) $);
  \coordinate (DDD) at  ($ (AAA) + (0,-0.35,-0.75) $);
  \draw (AAA) node[color=\colii,right] {\large{$\vvin_0$}} ;  
  \draw (CCC) node[color=\colii,right=2pt] {\large{$\vvin_1$}} ; 
  \draw (DDD) node[color=\colii,left=1.5pt] {\large{$\vvin_2$}} ; 
  \draw[line width=1.6pt,rounded corners=0.5pt] (AAA) -- (CCC) -- (DDD) -- cycle;

  \node [color=\colk] (K2) at ($ \bcAK*(AAA) + \bcCK*(CCC)+ \bcDK*(DDD) $) {$\Kin$};

  \coordinate (XXX) at ($ \bcAX*(AAA) + \bcCX*(CCC)+ \bcDX*(DDD) $) ;
  \fill[color=\colii] (XXX) circle (0.8pt);
  \node[color=\colii,below] (NXXX) at (XXX) {\large{$\xxin$}};

  \fill[color=\colii] (AAA) circle (0.8pt);
  \fill[color=\colii] (CCC) circle (0.8pt);
  \fill[color=\colii] (DDD) circle (0.8pt);

  \tikzstyle{fl}=[->,>=latex,very thick];
  \node (KK2D) at  ($ (XX) + (0,0.0,0.05) $) {} ;
  \node (K2D) at  ($ (X) + (0,-0.05,0.) $) {};
  \draw[fl] (KK2D) to[bend left] (K2D);
  \node (geomap) at  ($ (CC) + (0,0.1,0.65) $) {\large{$\Tgeom{\vvout}$}};

  \node (KKK2D) at  ($ (XXX) + (0,0.0,0.05) $) {} ;
  \draw[fl] (KKK2D) to[bend left] (KK2D);
  \node (ggeomap) at  ($ 0.5*(AA) + 0.5*(AAA) + (0,-0.1,0.62) $) {\large{$\left(\Tgeom{\vvin}\right)^{-1}$}};

\end{tikzpicture}

%% file: fig_Tet-Tria_k3.tex
\begin{tikzpicture}[scale=4,math3d] 

  \def\kk{3}

  \def\colk{black}
  \def\colo{magenta}
  \def\coli{darkgreen}
  \def\colii{red}
  \def\coliii{blue}

  \def\opacity{0.4}
  \def\opacityi{0.7}
  \def\opacityii{0.6}

  \coordinate (AA) at (0,0,-0.25);
  \coordinate (CC) at ($(AA) + (0,1,0)$);
  \coordinate (DD) at ($(AA) + (0,0,1)$);
  \draw (AA) node[above left] {$\hvv_0$};
  \draw (CC) node[above=1.5pt] {$\hvv_1=(1.,0.)$};
  \draw (DD) node[right=1.5pt] {$\hvv_2=(0.,1.)$};
  \draw[line width=1.0pt,rounded corners=0.5pt] (AA) -- (CC) -- (DD) -- cycle;
  \node[color=\colk] (K2) at ($(AA) + (0,0.65,0.65)$) {$\hK_{2}$};

  \def\hsq{15}
  \pgfmathparse{1-1/\hsq}\let\hhsq\pgfmathresult
  \pgfmathparse{1-2/\hsq}\let\hhhsq\pgfmathresult
  \coordinate (hCC) at  ($ \hhsq*(AA) + 1/\hsq*(CC) $);
  \coordinate (hDD) at  ($ \hhsq*(AA) + 1/\hsq*(DD) $);
  \coordinate (hCD) at  ($ \hhhsq*(AA) + 1/\hsq*(CC) + 1/\hsq*(DD) $);
  \draw[line width=0.4pt, color=\colk] (hCC) -- (hCD) -- (hDD);
  
  \fill[color=\colo] (AA) circle (0.8pt);
  \node[color=\colo,below] (Nxyz) at (AA) {$\haa_{(0,0)}$};
  \newcount\y
  \foreach \x in {1,0} {
    \pgfmathsetcount{\y}{1-\x} 
    \coordinate (Axyz) at ($(AA) + 1/\kk*(0,\x,\y)$);
    \node[color=\coli,below] (Nxyz) at (Axyz) {$\haa_{(\x,\the\y)}$};
    \fill[color=\coli] (Axyz) circle (0.8pt);
  }
  \newcount\y
  \foreach \x in {2,1,...,0} {
    \pgfmathsetcount{\y}{2-\x} 
    \coordinate (Axyz) at ($(AA) + 1/\kk*(0,\x,\y)$);
    \node[color=\colii,below] (Nxyz) at (Axyz) {$\haa_{(\x,\the\y)}$};
    \fill[color=\colii] (Axyz) circle (0.8pt);
  }
  \newcount\y
  \foreach \x in {3,2,...,0} {
    \pgfmathsetcount{\y}{3-\x} 
    \coordinate (Axyz) at ($(AA) + 1/\kk*(0,\x,\y)$);
    \node[color=\coliii,below] (Nxyz) at (Axyz) {$\haa_{(\x,\the\y)}$};
    \fill[color=\coliii] (Axyz) circle (0.8pt);
  }

  \coordinate (eps) at ($1/\kk*(0,0.05,-0.05)$); 
  \coordinate (A10) at ($(AA) + 1/\kk*(0,1,0)$);
  \coordinate (A01) at ($(AA) + 1/\kk*(0,0,1)$);
  \coordinate (A20) at ($(AA) + 1/\kk*(0,2,0)$);
  \coordinate (A02) at ($(AA) + 1/\kk*(0,0,2)$);

  \node[color=\colo,below=12pt] (Nxyz) at (AA) {$\calCdo$};
  \node[color=\coli,below=12pt] (Nxyz) at (A10) {$\calCdi$};
  \node[color=\colii,below=12pt] (Nxyz) at (A20) {$\calCdii$};
  \node[color=\coliii,below=12pt] (Nxyz) at (CC) {$\calCdiii$};
  
  \draw[color=\coli,line width=0.7pt] (A10) -- ($(A01) + (eps)$);
  \draw[color=\colii,line width=0.7pt] (A20) -- ($(A02) + (eps)$);
  \draw[color=\coliii,dashed,line width=1.4pt] (CC) -- ($(DD) + (eps)$);


  \coordinate (A) at (0,2.0,0);
  \coordinate (B) at ($(A) + (1,0,0)$);
  \coordinate (C) at ($(A) + (0,1,0)$);
  \coordinate (D) at ($(A) + (0,0,1)$);
  \draw (A) node[left] {$\hvv_0$}; 
  \draw (B) node[below=12pt] {$\hvv_1=(1.,0.,0.)$}; 
  \draw (C) node[above=2pt] {$\hvv_2=(0.,1.,0.)$}; 
  \draw (D) node[right=1.5pt] {$\hvv_3=(0.,0.,1.)$}; 
  \draw[line width=1.0pt,rounded corners=0.5pt] (A) -- (B) -- (D) -- cycle;
  \draw[line width=1.0pt,rounded corners=0.5pt] (A) -- (C) -- (B) -- cycle;
  \draw[line width=1.0pt,rounded corners=0.5pt] (B) -- (C) -- (D) -- cycle;
  \draw[line width=1.6pt,rounded corners=0.5pt] (A) -- (C) -- (D) -- cycle;

  \node[color=\colk] (K3) at ($(A) + (0,-0.3,0.7)$) {$\hK_{3}$};

  \pgfmathparse{1-1/\kk}\let\kkp\pgfmathresult
  \coordinate (B1) at ($\kkp*(A) + 1/\kk*(B)$);
  \coordinate (C1) at ($\kkp*(A) + 1/\kk*(C)$);
  \coordinate (D1) at ($\kkp*(A) + 1/\kk*(D)$);
  \pgfmathparse{1-2/\kk}\let\kkp\pgfmathresult
  \coordinate (B2) at ($\kkp*(A) + 2/\kk*(B)$);
  \coordinate (C2) at ($\kkp*(A) + 2/\kk*(C)$);
  \coordinate (D2) at ($\kkp*(A) + 2/\kk*(D)$);

  \draw[color=\coli,fill=\coli!40,fill opacity=\opacity] (B1) -- (C1) -- (D1) -- cycle;
  \draw[color=\colii,fill=\colii!20,fill opacity=\opacity] (B2) -- (C2) -- (D2) -- cycle;
  \draw[color=\coliii,fill=\coliii!20,fill opacity=\opacity] (B) -- (C) -- (D) -- cycle;

  \fill[color=\colo,fill opacity=\opacityii] (A) circle (0.8pt);
  \foreach \x in {1,0} {
    \pgfmathparse{1-\x}\let\YY\pgfmathresult
    \foreach \y in {0,...,\YY} {
      \pgfmathparse{1-\x-\y}\let\z\pgfmathresult
      \fill[color=\coli,fill opacity=\opacityii] ($(A) + 1/\kk*(\x,\y,\z)$) circle (0.8pt);
    }
  }
  \foreach \x in {2,1,...,0} {
    \pgfmathparse{2-\x}\let\YY\pgfmathresult
    \foreach \y in {0,...,\YY} {
      \pgfmathparse{2-\x-\y}\let\z\pgfmathresult
      \fill[color=\colii,fill opacity=\opacityii] ($(A) + 1/\kk*(\x,\y,\z)$) circle (0.8pt);
    }
  }
  \newcount\z
  \foreach \x in {3,2,...,0} {
    \pgfmathparse{3-\x}\let\YY\pgfmathresult
    \foreach \y in {0,...,\YY} {
      \pgfmathsetcount{\z}{3-\x-\y} 
      \coordinate (Axyz) at ($(A) + 1/\kk*(\x,\y,\z)$);
      \node[color=\coliii,below=0.15] (Nxyz) at (Axyz) {$\haa_{(\x,\y,\the\z)}$};
      \fill[color=\coliii] (Axyz) circle (0.8pt);
    }
  }


\end{tikzpicture}

%% file: fig_Tet_k3_slice.tex
\begin{tikzpicture}[scale=4,math3d] 

        \def\kk{3}

        \def\colk{black}
        \def\colo{magenta}
        \def\coli{darkgreen}
        \def\colii{red}
        \def\coliii{blue}

        \def\opacity{0.3}
        \def\opacityi{0.7}
        \def\opacityii{0.6}

	\coordinate (A) at (0,0,0);
	\coordinate (B) at  ($ (A) + (1,0,0) $);
	\coordinate (C) at  ($ (A) + (0,1,0) $);
	\coordinate (D) at  ($ (A) + (0,0,1) $);
        \draw (A) node[left] {$\hvv_0$} ;  
        \draw (B) node[below=11pt] {$\hvv_1$} ; 
        \draw (C) node[above=2pt] {$\hvv_2$} ; 
        \draw (D) node[right=1.5pt] {$\hvv_3$} ; 
	\draw[line width=1.0pt,rounded corners=0.5pt] (A) -- (B) -- (D) -- cycle;
	\draw[line width=1.0pt,rounded corners=0.5pt] (A) -- (C) -- (B) -- cycle;
	\draw[line width=1.0pt,rounded corners=0.5pt] (B) -- (C) -- (D) -- cycle;
	\draw[line width=1.6pt,rounded corners=0.5pt] (A) -- (C) -- (D) -- cycle;

        \node[color=\colk] (K3) at ($ (A) + (0,-0.3,0.7) $) {$\hK_{3}$};

        \pgfmathparse{1-1/\kk}\let\kkp\pgfmathresult
        \coordinate (B1) at ($ \kkp*(A) + 1/\kk*(B) $) ;
        \coordinate (C1) at ($ \kkp*(A) + 1/\kk*(C) $) ;
        \coordinate (D1) at ($ \kkp*(A) + 1/\kk*(D) $) ;
        \pgfmathparse{1-2/\kk}\let\kkp\pgfmathresult
        \coordinate (B2) at ($ \kkp*(A) + 2/\kk*(B) $) ;
        \coordinate (C2) at ($ \kkp*(A) + 2/\kk*(C) $) ;
        \coordinate (D2) at ($ \kkp*(A) + 2/\kk*(D) $) ;

        \coordinate (eps) at  ($ 1/\kk*(0,0.05,-0.05) $); 
	\coordinate (A210) at  ($ (A) + 1/\kk*(2,1,0) $);
	\coordinate (A201) at  ($ (A) + 1/\kk*(2,0,1) $);
	\coordinate (A120) at  ($ (A) + 1/\kk*(1,2,0) $);
	\coordinate (A102) at  ($ (A) + 1/\kk*(1,0,2) $);
        \coordinate (CD) at ($ 1/2*(C) + 1/2*(D) $);

        \draw[color=\colk,fill=\colo!10,fill opacity=\opacity]   (A) -- (C) -- (D) -- cycle;
        \draw[color=\colk,fill=\colo!10,fill opacity=\opacity]  (B1) -- (A120) -- (A102) -- cycle;
        \draw[color=\colk,fill=\colo!10,fill opacity=\opacity]  (B2) -- (A210) -- (A201) -- cycle;

        \fill[color=\colk,fill opacity=\opacityii] (A) circle (0.6pt);
        \foreach \x in {1,0}  {
          \pgfmathparse{1-\x}\let\YY\pgfmathresult
          \foreach \y in {0,...,\YY} {
            \pgfmathparse{1-\x-\y}\let\z\pgfmathresult
            \fill[color=\colk,fill opacity=\opacityii] ($ (A) + 1/\kk*(\x,\y,\z) $) circle (0.6pt);
          }
        }
        \foreach \x in {2,1,...,0}  {
          \pgfmathparse{2-\x}\let\YY\pgfmathresult
          \foreach \y in {0,...,\YY} {
            \pgfmathparse{2-\x-\y}\let\z\pgfmathresult
            \fill[color=\colk,fill opacity=\opacityii] ($ (A) + 1/\kk*(\x,\y,\z) $) circle (0.6pt);
          }
        }
        \draw[color=\colo,dashed,line width=2.5pt,<->,>=latex]  ($ (C) - (eps) $) -- ($ (D) + (eps) $);

        \draw[color=\colo,dashed,line width=2.5pt,<->,>=latex] ($ (A210) - (eps) $) -- ($ (A201) + (eps) $);
        \draw[color=\colo,dashed,line width=2.5pt,<->,>=latex] ($ (A120) - (eps) $) -- ($ (A102) + (eps) $);
        \newcount\z
        \foreach \x in {3,2,...,0}  {
          \pgfmathparse{3-\x}\let\YY\pgfmathresult
          \foreach \y in {0,...,\YY} {
            \pgfmathsetcount{\z}{3-\x-\y} 
            \coordinate (Axyz) at  ($ (A) + 1/\kk*(\x,\y,\z) $) ;
            \node[color=\coliii,below=-0.1] (Nxyz) at (Axyz) {$\haa_{(\x,\y,\the\z)}$};
            \fill[color=\coliii] (Axyz) circle (0.8pt);
          }
        }

        \draw (B) node[color=\colo,above left] {\tt mapF (insertF0 3) (CSdk (d-1) 0)} ; 
        \draw (A201) node[color=\colo,above left] {\tt mapF (insertF0 2) (CSdk (d-1) 1)} ; 
        \draw (A102) node[color=\colo,above left] {\tt mapF (insertF0 1) (CSdk (d-1) 2)} ; 
        \draw (D) node[color=\colo,above left] {\tt mapF (insertF0 0) (CSdk (d-1) 3)} ;

\end{tikzpicture}

%% file: fig_Tet-Tria_k3_grsymlex.tex
\begin{tikzpicture}[scale=4,math3d] 

  \def\kk{3}

  \def\colk{black}
  \def\colo{magenta}
  \def\coli{darkgreen}
  \def\colii{red}
  \def\coliii{blue}

  \def\opacity{0.4}
  \def\opacityi{0.7}
  \def\opacityii{0.6}

  \coordinate (AA) at (0,0,-0.25);
  \coordinate (CC) at ($(AA) + (0,1,0)$);
  \coordinate (DD) at ($(AA) + (0,0,1)$);
  \draw (AA) node[above left] {$\hvv_0$};
  \draw (CC) node[above=1.5pt] {$\hvv_1=(1.,0.)$};
  \draw (DD) node[right=1.5pt] {$\hvv_2=(0.,1.)$};
  \draw[line width=1.0pt,rounded corners=0.5pt] (AA) -- (CC) -- (DD) -- cycle;
  \node[color=\colk] (K2) at ($(AA) + (0,0.65,0.65)$) {$\hK_{2}$};

  \def\hsq{12}
  \pgfmathparse{1-1/\hsq}\let\hhsq\pgfmathresult
  \pgfmathparse{1-2/\hsq}\let\hhhsq\pgfmathresult
  \coordinate (hCC) at  ($ \hhsq*(AA) + 1/\hsq*(CC) $);
  \coordinate (hDD) at  ($ \hhsq*(AA) + 1/\hsq*(DD) $);
  \coordinate (hCD) at  ($ \hhhsq*(AA) + 1/\hsq*(CC) + 1/\hsq*(DD) $);
  \draw[line width=0.4pt, color=\colk] (hCC) -- (hCD) -- (hDD);

  \fill[color=\colo] (AA) circle (0.8pt);
  \node[color=\colo,below] (Nxyz) at (AA) {$\haa_{(0,0)}$};
  \newcount\y
  \foreach \x in {1,0} {
    \pgfmathsetcount{\y}{1-\x} 
    \coordinate (Axyz) at ($(AA) + 1/\kk*(0,\x,\y)$);
    \node[color=\coli,below] (Nxyz) at (Axyz) {$\haa_{(\x,\the\y)}$};
    \fill[color=\coli] (Axyz) circle (0.8pt);
  }
  \newcount\y
  \foreach \x in {2,1,...,0} {
    \pgfmathsetcount{\y}{2-\x} 
    \coordinate (Axyz) at ($(AA) + 1/\kk*(0,\x,\y)$);
    \node[color=\colii,below] (Nxyz) at (Axyz) {$\haa_{(\x,\the\y)}$};
    \fill[color=\colii] (Axyz) circle (0.8pt);
  }
  \newcount\y
  \foreach \x in {3,2,...,0} {
    \pgfmathsetcount{\y}{3-\x} 
    \coordinate (Axyz) at ($(AA) + 1/\kk*(0,\x,\y)$);
    \node[color=\coliii,below] (Nxyz) at (Axyz) {$\haa_{(\x,\the\y)}$};
    \fill[color=\coliii] (Axyz) circle (0.8pt);
  }

  \coordinate (eps0) at ($1/\kk*(0,-0.05,0)$); 
  \coordinate (eps) at ($1/\kk*(0,0.05,-0.05)$); 
  \node[color=\colo,below=12pt] (Nxyz) at (AA) {$\calCdo$};
  \coordinate (A10) at ($(AA) + 1/\kk*(0,1,0)$);
  \coordinate (A01) at ($(AA) + 1/\kk*(0,0,1)$);
  \draw[color=\colo,dashed,line width=1.4pt,->,>=latex] (AA) -- ($(A10) + (eps0)$);
  \draw[color=\coli,dashed,line width=1pt,->,>=latex] (A10) -- ($(A01) + (eps)$);
  \node[color=\coli,below=12pt] (Nxyz) at (A10) {$\calCdi$};
  \coordinate (A20) at ($(AA) + 1/\kk*(0,2,0)$);
  \coordinate (A02) at ($(AA) + 1/\kk*(0,0,2)$);
  \draw[color=\coli,dashed,line width=1pt,->,>=latex] ($(A01) + (eps)$) -- ($(A20) - (eps)$);
  \draw[color=\colii,dashed,line width=1pt,->,>=latex] (A20) -- ($(A02) + (eps)$);
  \node[color=\colii,below=12pt] (Nxyz) at (A20) {$\calCdii$};
  \coordinate (A30) at ($(AA) + 1/\kk*(0,3,0)$);
  \coordinate (CD) at ($1/2*(CC) + 1/2*(DD)$);
  \draw[color=\colii,dashed,line width=1pt,->,>=latex] ($(A02) + (eps)$) -- ($(A30) - (eps)$);
  \draw[color=\coliii,dashed,line width=1.4pt,->,>=latex] (A30) -- ($(DD) + (eps)$);
  \node[color=\coliii,below=12pt] (Nxyz) at (A30) {$\calCdiii$};

  \coordinate (A) at (0,2.0,0);
  \coordinate (B) at ($(A) + (1,0,0)$);
  \coordinate (C) at ($(A) + (0,1,0)$);
  \coordinate (D) at ($(A) + (0,0,1)$);
  \draw (A) node[left] {$\hvv_0$}; 
  \draw (B) node[below=12pt] {$\hvv_1=(1.,0.,0.)$}; 
  \draw (C) node[above=2pt] {$\hvv_2=(0.,1.,0.)$}; 
  \draw (D) node[right=1.5pt] {$\hvv_3=(0.,0.,1.)$}; 
  \draw[line width=1.0pt,rounded corners=0.5pt] (A) -- (B) -- (D) -- cycle;
  \draw[line width=1.0pt,rounded corners=0.5pt] (A) -- (C) -- (B) -- cycle;
  \draw[line width=1.0pt,rounded corners=0.5pt] (B) -- (C) -- (D) -- cycle;
  \draw[line width=1.6pt,rounded corners=0.5pt] (A) -- (C) -- (D) -- cycle;

  \node[color=\colk] (K3) at ($(A) + (0,-0.3,0.7)$) {$\hK_{3}$};

  \pgfmathparse{1-1/\kk}\let\kkp\pgfmathresult
  \coordinate (B1) at ($\kkp*(A) + 1/\kk*(B)$);
  \coordinate (C1) at ($\kkp*(A) + 1/\kk*(C)$);
  \coordinate (D1) at ($\kkp*(A) + 1/\kk*(D)$);
  \pgfmathparse{1-2/\kk}\let\kkp\pgfmathresult
  \coordinate (B2) at ($\kkp*(A) + 2/\kk*(B)$);
  \coordinate (C2) at ($\kkp*(A) + 2/\kk*(C)$);
  \coordinate (D2) at ($\kkp*(A) + 2/\kk*(D)$);

  \draw[color=\coli,fill=\coli!40,fill opacity=\opacity] (B1) -- (C1) -- (D1) -- cycle;
  \draw[color=\colii,fill=\colii!20,fill opacity=\opacity] (B2) -- (C2) -- (D2) -- cycle;
  \draw[color=\coliii,fill=\coliii!20,fill opacity=\opacity] (B) -- (C) -- (D) -- cycle;

  \fill[color=\colo,fill opacity=\opacityii] (A) circle (0.8pt);
  \foreach \x in {1,0} {
    \pgfmathparse{1-\x}\let\YY\pgfmathresult
    \foreach \y in {0,...,\YY} {
      \pgfmathparse{1-\x-\y}\let\z\pgfmathresult
      \fill[color=\coli,fill opacity=\opacityii] ($(A) + 1/\kk*(\x,\y,\z)$) circle (0.8pt);
    }
  }
  \foreach \x in {2,1,...,0} {
    \pgfmathparse{2-\x}\let\YY\pgfmathresult
    \foreach \y in {0,...,\YY} {
      \pgfmathparse{2-\x-\y}\let\z\pgfmathresult
      \fill[color=\colii,fill opacity=\opacityii] ($(A) + 1/\kk*(\x,\y,\z)$) circle (0.8pt);
    }
  }
  \newcount\z
  \foreach \x in {3,2,...,0} {
    \pgfmathparse{3-\x}\let\YY\pgfmathresult
    \foreach \y in {0,...,\YY} {
      \pgfmathsetcount{\z}{3-\x-\y} 
      \coordinate (Axyz) at ($(A) + 1/\kk*(\x,\y,\z)$);
      \node[color=\coliii,below=-0.15] (Nxyz) at (Axyz) {$\haa_{(\x,\y,\the\z)}$};
      \fill[color=\coliii] (Axyz) circle (0.8pt);
    }
  }

  \coordinate (eps0) at ($1/\kk*(0.05,-0.05,0)$); 
  \coordinate (eps) at ($1/\kk*(0,0.05,-0.05)$); 
  \coordinate (A210) at ($(A) + 1/\kk*(2,1,0)$);
  \coordinate (A201) at ($(A) + 1/\kk*(2,0,1)$);
  \draw[color=\coliii,dashed,line width=1.4pt,->,>=latex] (B) -- ($(A210) + (eps0)$);
  \draw[color=\coliii,dashed,line width=0.7pt,->,>=latex] (A210) -- ($(A201) + (eps)$);
  \coordinate (A120) at ($(A) + 1/\kk*(1,2,0)$);
  \coordinate (A102) at ($(A) + 1/\kk*(1,0,2)$);
  \draw[color=\coliii,dashed,line width=0.7pt,->,>=latex] ($(A201) + (eps)$) -- ($(A120) - (eps)$);
  \draw[color=\coliii,dashed,line width=0.7pt,->,>=latex] (A120) -- ($(A102) + (eps)$);
  \coordinate (CD) at ($1/2*(C) + 1/2*(D)$);
  \draw[color=\coliii,dashed,line width=0.7pt,->,>=latex] ($(A102) + (eps)$) -- ($(C) - (eps)$);
  \draw[color=\coliii,dashed,line width=1.4pt,->,>=latex] (C) -- ($(D) + (eps)$);

\end{tikzpicture}

%% file: fig_SegTriaTet_1to4.tex
  \begin{tikzpicture}[scale=2,math3d] 
        \def\kkmax{4}

        \def\colo{magenta}
        \def\coli{green}
        \def\colii{red}
        \def\coliii{blue}

	\coordinate (AA) at ($ (0,0,1.5) $);
	\coordinate (CC) at  ($ (AA) + (0,1,0) $);
	\draw[line width=1.0pt] (AA) -- (CC);
        \coordinate (Axyz) at  ($ 0.5*(AA) + 0.5*(CC) $) ;
        \fill[color=\colii] (Axyz) circle (1.5pt);
        \node[color=\colii,below] (Nxyz) at (Axyz) {{\tiny ${(0)}$}};  
        \foreach \K in {1,...,\kkmax}  {
	  \coordinate (AA) at ($ (0,0,1.5) + 1.5*(0,\K,0) $);
	  \coordinate (CC) at  ($ (AA) + (0,1,0) $);
	  \draw[line width=1.0pt] (AA) -- (CC);
          
          \foreach \x in {0,1,...,\K}  {
            \coordinate (Axyz) at  ($ (AA) + 1/\K*(0,\x,0) $) ;
            \ifthenelse{\x<\K}
                       { \fill[color=\coliii] (Axyz) circle (1.5pt);
                         \node[color=\coliii,below] (Nxyz) at (Axyz) {{\tiny ${(\x)}$}}; }
                       { \fill[color=\colii] (Axyz) circle (1.5pt);
                 \node[color=\colii,below] (Nxyz) at (Axyz) {{\tiny ${(\x)}$}}; } ; 
          }
        }
          
	\coordinate (AA) at ($ (0,0,0) $);
	\coordinate (CC) at  ($ (AA) + (0,1,0) $);
	\coordinate (DD) at  ($ (AA) + (0,0,1) $);
	\draw[line width=1.0pt,rounded corners=0.5pt] (AA) -- (CC) -- (DD) -- cycle;
        \coordinate (Axyz) at  ($ 1/3*(AA) + 1/3*(CC) + 1/3*(DD) $) ;
        \fill[color=\colii] (Axyz) circle (1.5pt);
        \node[color=\colii,below] (Nxyz) at (Axyz) {{\tiny ${(0,0)}$}};  
        \foreach \K in {1,...,\kkmax}  {
	  \coordinate (AA) at ($ (0,0,0) + 1.5*(0,\K,0) $);
	  \coordinate (CC) at  ($ (AA) + (0,1,0) $);
	  \coordinate (DD) at  ($ (AA) + (0,0,1) $);
	  \draw[line width=1.0pt,rounded corners=0.5pt] (AA) -- (CC) -- (DD) -- cycle;

          \foreach \k in {0,1,...,\K}  {
            \newcount\y
            \foreach \x in {\k,...,0}  {
              \pgfmathsetcount{\y}{\k-\x} 
              \coordinate (Axyz) at  ($ (AA) + 1/\K*(0,\x,\y) $) ;
              \ifthenelse{\k<\K}
                         {\fill[color=\coliii] (Axyz) circle (1.5pt);
                         \node[color=\coliii,below] (Nxyz) at (Axyz) {{\tiny ${(\x,\the\y)}$}}; }
                         {\fill[color=\colii] (Axyz) circle (1.5pt);
                           \node[color=\colii,below] (Nxyz) at (Axyz) {{\tiny ${(\x,\the\y)}$}}; } ; 
            }
          }
        }
        
	\coordinate (A) at ($ (0,0,-2) $);
	\coordinate (B) at  ($ (A) + (1,0,0) $);
	\coordinate (C) at  ($ (A) + (0,1,0) $);
	\coordinate (D) at  ($ (A) + (0,0,1) $);
	\draw[line width=1.0pt,rounded corners=0.5pt] (A) -- (B) -- (D) -- cycle;
	\draw[line width=1.0pt,rounded corners=0.5pt] (A) -- (C) -- (B) -- cycle;
	\draw[line width=1.0pt,rounded corners=0.5pt] (B) -- (C) -- (D) -- cycle;
	\draw[line width=1.6pt,rounded corners=0.5pt] (A) -- (C) -- (D) -- cycle;
        \coordinate (Axyz) at  ($ 1/4*(A) + 1/4*(B) + 1/4*(C) + 1/4*(D) $) ;
        \fill[color=\colii] (Axyz) circle (1.5pt);
        \node[color=\colii,above] (Nxyz) at (Axyz) {{\tiny ${(0,0,0)}$}};  

        \foreach \K in {1,...,\kkmax}  {
	  \coordinate (A) at ($ (0,0,-2) + 1.5*(0,\K,0) $);
	  \coordinate (B) at  ($ (A) + (1,0,0) $);
	  \coordinate (C) at  ($ (A) + (0,1,0) $);
	  \coordinate (D) at  ($ (A) + (0,0,1) $);
	  \draw[line width=1.0pt,rounded corners=0.5pt] (A) -- (B) -- (D) -- cycle;
	  \draw[line width=1.0pt,rounded corners=0.5pt] (A) -- (C) -- (B) -- cycle;
	  \draw[line width=1.0pt,rounded corners=0.5pt] (B) -- (C) -- (D) -- cycle;
	  \draw[line width=1.6pt,rounded corners=0.5pt] (A) -- (C) -- (D) -- cycle;

          \newcount\z
          \foreach \k in {0,1,...,\K}  {
            \foreach \x in {\k,...,0}  {
              \pgfmathparse{\k-\x}\let\YY\pgfmathresult
              \foreach \y in {0,...,\YY} {
                \pgfmathsetcount{\z}{\k-\x-\y} 
                \coordinate (Axyz) at  ($ (A) + 1/\K*(\x,\y,\z) $) ;
                \ifthenelse{\k<\K}
                           {\fill[color=\coliii] (Axyz) circle (1.5pt); }
                           {\fill[color=\colii] (Axyz) circle (1.5pt) ;
                             \node[color=\colii,below] (Nxyz) at  (Axyz) {{\tiny ${(\x,\y,\the\z)}$}}; };
              }
            }
          }
        }
  \end{tikzpicture}

%% file: fig_Tet_k3_k2.tex
\begin{tikzpicture}[scale=4,math3d] 

        \def\kk{3}

        \def\colo{magenta}
        \def\coli{darkgreen}
        \def\colii{red}
        \def\coliii{blue}

	\coordinate (A) at (0,0,0);
        \coordinate (B) at  ($ (A) + (1,0,0) $);
	\coordinate (C) at  ($ (A) + (0,1,0) $);
	\coordinate (D) at  ($ (A) + (0,0,1) $);
        \draw (A) node[left] {$\uhvv_0$} ;  
        \draw (B) node[right] {$\hvv_1$} ; 
        \draw (C) node[right=2pt] {$\hvv_2$} ; 
        \draw (D) node[right=1.5pt] {$\hvv_3$} ; 
	\draw[line width=1.0pt,rounded corners=0.5pt] (A) -- (B) -- (D) -- cycle;
	\draw[line width=1.0pt,rounded corners=0.5pt] (A) -- (C) -- (B) -- cycle;
	\draw[line width=1.6pt,rounded corners=0.5pt] (B) -- (C) --  (D) -- cycle; 

        \pgfmathparse{1-1/\kk}\let\kkp\pgfmathresult
        \coordinate (B1) at ($ \kkp*(A) + 1/\kk*(B) $) ;
        \coordinate (C1) at ($ \kkp*(A) + 1/\kk*(C) $) ;
        \coordinate (D1) at ($ \kkp*(A) + 1/\kk*(D) $) ;
        \coordinate (B2) at ($ \kkp*(A) + 2/\kk*(B) $) ;
        \coordinate (C2) at ($ \kkp*(A) + 2/\kk*(C) $) ;
        \coordinate (D2) at ($ \kkp*(A) + 2/\kk*(D) $) ;

        \draw[color=\coli,line width=0.8pt]   (B1) -- (C1) -- (D1) -- cycle;
        \draw[color=\colii,line width=0.8pt]  (B2) -- (C2) -- (D2) -- cycle;
        \draw[color=\coliii,line width=0.4pt]  (B) -- (C) -- (D) -- cycle;

        \coordinate (Bp) at  (B2) ;
        \coordinate (Cp) at  (C2) ;
        \coordinate (Dp) at  (D2) ;
        \draw (Bp) node[right=1.5pt] {$\uhvv_1$} ; 
        \draw (Cp) node[above right] {$\uhvv_2$} ; 
        \draw (Dp) node[above right] {$\uhvv_3$} ; 
        
        \fill[color=\colo] (A) circle (1.0pt);
        \newcount\YY
        \newcount\z
        \foreach \x in {1,0}  {
          \pgfmathsetcount{\YY}{1-\x}
          \foreach \y in {0,...,\YY} {
            \pgfmathsetcount{\z}{1-\x-\y}
            \coordinate (Axyz) at ($(A) + 1/\kk*(\x,\y,\z)$);
            \draw[color=\coli,below] (Axyz) node {$\uhaa_{(\x,\y,\the\z)}$};
            \fill[color=\coli] (Axyz) circle (1.0pt);
          }
        }
        \newcount\zint
        \foreach \x in {2,1,...,0}  {
          \pgfmathparse{2-\x}\let\YY\pgfmathresult
          \foreach \y in {0,...,\YY} {
            \pgfmathsetcount{\zint}{2-\x-\y} 
            \pgfmathparse{2-\x-\y}\let\z\pgfmathresult
            \pgfmathparse{\x/\kk}\let\xkk\pgfmathresult
            \pgfmathparse{\y/\kk}\let\ykk\pgfmathresult
            \pgfmathparse{\z/\kk}\let\zkk\pgfmathresult
            \pgfmathparse{1-\xkk-\ykk-\zkk}\let\okk\pgfmathresult
            \coordinate (Axyz) at  ($ \okk*(A) + \xkk*(B) + \ykk*(C) + \zkk*(D) $) ; 
            \node[color=\colii,below] (Nxyz) at (Axyz) {$\uhaa_{(\x,\y,\the\zint)}$};
            \fill[color=\colii] (Axyz) circle (1.0pt);
          }
        }
        \newcount\zint
        \foreach \x in {3,2,...,0}  {
          \pgfmathparse{3-\x}\let\YY\pgfmathresult
          \foreach \y in {0,...,\YY} {
            \pgfmathsetcount{\zint}{3-\x-\y} 
            \pgfmathparse{3-\x-\y}\let\z\pgfmathresult
            \pgfmathparse{\x/\kk}\let\xkk\pgfmathresult
            \pgfmathparse{\y/\kk}\let\ykk\pgfmathresult
            \pgfmathparse{\z/\kk}\let\zkk\pgfmathresult
            \pgfmathparse{1-\xkk-\ykk-\zkk}\let\okk\pgfmathresult
            \coordinate (Axyz) at  ($ \okk*(A) + \xkk*(B) + \ykk*(C) + \zkk*(D) $) ;
            \fill[color=\coliii] (Axyz) circle (0.6pt);
          }
        }

\end{tikzpicture}